\documentclass[aps,prd,twocolumn,preprintnumbers,superscriptaddress,nofootinbib]{revtex4-1}

\usepackage{graphicx,color,hyperref}
\usepackage{amsmath}
\usepackage{amsfonts}
\usepackage{amssymb}
\usepackage{braket}
\usepackage{mathtools}
\usepackage{graphicx}
\usepackage{float}
\usepackage{subfigure}
\usepackage[english]{babel}
\usepackage{hyperref}

\hypersetup{
colorlinks=true,linkcolor=red,anchorcolor=blue,citecolor=blue, filecolor=blue,urlcolor=red,bookmarksnumbered=true, pdfview=FitB
}

\newcommand{\bea}{\begin {eqnarray}}
\newcommand{\eea}{\end{eqnarray}}

\begin{document}

\title{Nucleon structure from basis light-front quantization}

\author{Siqi Xu }
\email{xsq234@impcas.ac.cn} 
\affiliation{Institute of Modern Physics, Chinese Academy of Sciences, Lanzhou, Gansu, 730000, China}
\affiliation{School of Nuclear Physics, University of Chinese Academy of Sciences, Beijing, 100049, China}
\affiliation{CAS Key Laboratory of High Precision Nuclear Spectroscopy, Institute of Modern Physics, Chinese Academy of Sciences, Lanzhou 730000, China}

\author{Chandan Mondal}
\email{mondal@impcas.ac.cn} 
\affiliation{Institute of Modern Physics, Chinese Academy of Sciences, Lanzhou, Gansu, 730000, China}
\affiliation{School of Nuclear Physics, University of Chinese Academy of Sciences, Beijing, 100049, China}
\affiliation{CAS Key Laboratory of High Precision Nuclear Spectroscopy, Institute of Modern Physics, Chinese Academy of Sciences, Lanzhou 730000, China}

\author{Jiangshan Lan}
\email{jiangshanlan@impcas.ac.cn}
\affiliation{Institute of Modern Physics, Chinese Academy of Sciences, Lanzhou, Gansu, 730000, China}
\affiliation{School of Nuclear Physics, University of Chinese Academy of Sciences, Beijing, 100049, China}
\affiliation{CAS Key Laboratory of High Precision Nuclear Spectroscopy, Institute of Modern Physics, Chinese Academy of Sciences, Lanzhou 730000, China}
\affiliation{Lanzhou University, Lanzhou 730000, China}

\author{Xingbo Zhao}
\email{xbzhao@impcas.ac.cn} 
\affiliation{Institute of Modern Physics, Chinese Academy of Sciences, Lanzhou, Gansu, 730000, China}
\affiliation{School of Nuclear Physics, University of Chinese Academy of Sciences, Beijing, 100049, China}
\affiliation{CAS Key Laboratory of High Precision Nuclear Spectroscopy, Institute of Modern Physics, Chinese Academy of Sciences, Lanzhou 730000, China}

\author{Yang Li}
\email{leeyoung1987@ustc.edu.cn}
\affiliation{Department of Modern Physics, University of Science and Technology of China, Hefei 230026, China}
\affiliation{Department of Physics and Astronomy, Iowa State University, Ames, Iowa 50011, USA}


\author{James P. Vary}
\email{jvary@iastate.edu}
\affiliation{Department of Physics and Astronomy, Iowa State University, Ames, Iowa 50011, USA} 

\collaboration{BLFQ Collaboration}

\date{\today}


\begin{abstract}
We produce the light-front wave functions (LFWFs) of the nucleon from a basis light-front approach in the leading Fock sector representation. We solve for the mass eigenstates from a light-front effective Hamiltonian, which includes a confining potential adopted from light-front holography in the transverse direction, a longitudinal confinement, and a one-gluon exchange interaction with fixed coupling.
We then employ the LFWFs to obtain the electromagnetic and axial form factors, the parton distribution functions (PDFs) and the generalized parton distribution functions for the nucleon. The electromagnetic and axial form factors of the proton agree with the experimental data, whereas the neutron form factors deviate somewhat from the experiments in the low momentum transfer region. The unpolarized, the helicity, and the transversity valence quark PDFs, after QCD scale evolution, are fairly consistent with the global fits to the data at the relevant experimental scales. The helicity asymmetry for the down quark also agrees well with the measurements, however, the asymmetry for the up quark shows a deviation from the data, especially in the small $x$ region.
We also find that the tensor charge agrees well with the extracted data and the lattice QCD predictions, while the axial charge is somewhat outside the experimental error bar. The electromagnetic radii of the proton, the magnetic radius of the neutron, and the axial radius are in excellent agreement with the measurements, while the neutron charge radius deviates from experiment.

\end{abstract}

\maketitle

\section{introduction}
One of the main goals in hadron physics is to
understand how nucleons and other hadrons are built up
from quarks and gluons. The subject has been investigated extensively with dedicated experiments and theoretical efforts for several decades.
One of the most powerful tools to investigate the structure of the nucleon is high energy electron scattering. 
The electromagnetic form factors (FFs) which can be probed through elastic scattering are among the most basic quantities containing information about the internal structure of the nucleon. The Fourier transform of the FFs gives information about spatial distributions such as the charge and the magnetization densities inside the nucleon. On the other hand,
from deep inelastic scattering (DIS) processes one can extract the parton distribution functions (PDFs), which encode the nonperturbative structure of the nucleon in terms of the distribution of longitudinal momentum carried by the quarks and gluons as its constituents.
Both the observables, FFs and PDFs, have taught us a great deal about the nucleon, but these quantities do not provide full three-dimensional structural information of the nucleon.
Meanwhile, it has become clear that the generalized parton distributions (GPDs), appearing in the description of hard exclusive reactions, like deeply virtual Compton scattering (DVCS) or vector meson production allow us to draw three-dimensional pictures of the nucleon. From GPDs we learn essential information about the distribution and orbital motion of the constituents. 

The matrix element of the electromagnetic current for the nucleon is parameterized by two independent FFs namely the Dirac and Pauli FFs. The nucleon FFs attract numerous dedicated experimental and theoretical efforts for several decades~\cite{Gao,Hyd,Punjabi,Pacetti:2015iqa,Punjabi:2015bba}. There are two well established methods to extract the
nucleon FFs from experiments. One is from unpolarized scattering cross section data by the Rosenbluth separation technique. In this method, one extracts the nucleon Sachs FFs, which are expressed in terms of the Dirac and Pauli FFs. The other technique uses either the target or the recoiled polarized proton along with the polarized lepton beam and is known as the polarization transfer
method in which the ratios of the Sachs FFs for the nucleon are measured. However, there are inconsistencies in the extraction of the data for the proton electric to magnetic Sachs FF ratio. In double polarization experiments~\cite{Gao:2000ne,Punjabi:2005wq,Gayou:2001qd,Puckett:2011xg,Puckett:2010ac}, the ratio decreases almost linearly for momentum transfer ($Q^2$) 
$>0.5\, \text{GeV}^2$
, whereas the results obtained from the Rosenbluth separation method~\cite{Hand:1963zz,Janssens:1965kd,Price:1971zk,Litt:1969my,Berger:1971kr,Bartel:1973rf,Borkowski:1974mb,Simon:1980hu,Walker:1993vj,Andivahis:1994rq,Christy:2004rc,Qattan:2004ht} remain constant in the spacelike region. 
The experimental data for the neutron FFs are not available in the large $Q^2=-q^2$ regime and predictions of the neutron FFs are even more challenging to explain using phenomenological models. Meanwhile, high precision measurements of both proton and neutron FFs are expected from ongoing and forthcoming experiments at Jefferson Lab~\cite{a,b,c,d,e,f,g}. The nucleon electromagnetic FFs
have been theoretically investigated using different approaches in Refs.~\cite{Alexandrou:2017ypw,Chambers:2017tuf,Alexandrou:2018sjm,Shintani:2018ozy,He:2017viu,Alarcon:2017lhg,Abidin:2009hr,Gutsche:2017lyu,Mondal:2016xpk,Sufian:2016hwn,Brodsky:2014yha,Mondal:2015uha,Chakrabarti:2013dda,Ye:2017gyb,Gutsche:2014yea,Gutsche:2013zia,Cloet:2012cy,Pasquini:2007iz}, whereas the flavor decomposition of the nucleon FFs has been reported in Refs.\cite{Cates:2011pz,Qattan:2012zf,Diehl:2013xca,Chakrabarti:2013dda}.

Unlike the nucleon electromagnetic FFs which  have been explored experimentally to a large extent, our information about the axial FFs is very limited.  Until now, there are only two sets of experiments that can be used to determine axial FFs: first, (anti)neutrino scattering off protons or nuclei and, second, charged pion electroproduction. We refer to the articles~\cite{Bernard:2001rs,Schindler:2006jq} for a review of
experimental data on the nucleon axial FFs. The axial FFs can be extracted
using various theoretical approaches~\cite{Tsushima:1988xv,JuliaDiaz:2004qr,Mamedov:2016ype,Liu:2016kpb,Anikin:2016teg,Adamuscin:2007fk,Aznauryan:2012ba,Ramalho:2017tga,Hashamipour:2019pgy}. In recent years, lattice QCD simulations of axial FFs have been reported for pion masses in the range $0.2-0.6$ GeV~\cite{Alexandrou:2013joa,Bhattacharya:2013ehc,Liang:2016fgy,Green:2017keo,Yao:2017fym,Abdel-Rehim:2015owa,Bali:2014nma,Bhattacharya:2016zcn}, while in a very recent study, the lattice QCD calculation of nucleon axial FF in $2+1$ flavor near the physical pion mass has been presented by the PACS Collaboration\cite{Ishikawa:2018rew}.


PDFs, which encode the nonpertubative
structure of the nucleon in terms of the number densities of its confined constituents, are functions of the light-front longitudinal momentum fraction ($x$) of the nucleon carried by the constituents. At leading twist, the complete spin structure of the nucleon is described in terms of three independent PDFs, namely, the unpolarized $f_1(x)$, the helicity $g_1(x)$, and the transversity $h_1(x)$. 

Precise knowledge of PDFs is needed for the analysis and interpretation of the scattering experiments in
the LHC era.  Global fitting collaborations such as  NNPDF \cite{Ball:2017nwa}, HERAPDF \cite{Alekhin:2017kpj}, MMHT \cite{Harland-Lang:2014zoa}, CTEQ \cite{Dulat:2015mca}, and MSTW \cite{Martin:2009iq} have made considerable efforts to determine PDFs and their uncertainties. Precise determination of the helicity PDF from polarized lepton-proton inclusive processes is now available \cite{Aidala:2012mv,Deur:2018roz}. Perturbative QCD (pQCD) indicates that the polarized to unpolarized PDF ratio for the up quark increases toward 1 as $x\to 1$~\cite{Brodsky:1994kg,Avakian:2007xa}. For the down quark, the ratio is found to remain negative  in the experimentally covered region of $x\lesssim 0.6$ \cite{Zheng:2003un,Zheng:2004ce,Parno:2014xzb,Dharmawardane:2006zd,Airapetian:2003ct,Airapetian:2004zf,Alekseev:2010ub}, without any indication of a sign reversal at large $x$-values. This is supported by the global pQCD analyses of the experimental data extracted to large $x$ \cite{deFlorian:2008mr,deFlorian:2009vb,Nocera:2014gqa,Jimenez-Delgado:2014xza,Ethier:2017zbq} as well as by the Dyson-Schwinger equation approach \cite{Roberts:2013mja}. However, a recent study from light-front holography predicts the sign reversal of the polarized down-quark distribution in the proton at large $x$ \cite{Liu:2019vsn}. Nucleon PDFs have also been investigated with lattice QCD using different approaches such as the path-integral formulation \cite{Liu:1999ak,Liu:1993cv}, the inversion method \cite{Horsley:2012pz,Chambers:2017dov}, pseudo-PDFs \cite{Radyushkin:2017cyf,Orginos:2017kos}, quasi-PDFs \cite{Ji:2013dva,Lin:2014zya,Alexandrou:2015rja,Alexandrou:2016jqi}, and lattice cross sections \cite{Ma:2017pxb}. The current status and challenges of lattice calculations of PDFs can be found in Ref. \cite{Lin:2017snn}.
 
While the unpolarized and the helicity PDFs have been investigated for decades and are well determined, much less information is available on the transversity PDF. Due to its chiral-odd nature, the distribution can only be accessed in a process wherein it couples to another chiral-odd quantity \cite{Jaffe:1991kp}. 

The $h_1$ distribution describes the correlation between the transverse polarization of the constituents and the transverse polarization
of the nucleon. An important approach for studying the $h_1$ distribution is to measure the Collins azimuthal asymmetries in semi-inclusive hadron production in deep
inelastic scattering \cite{Collins:1992kk}. Considerable efforts have been made to measure spin asymmetries by the HERMES Collaboration \cite{Airapetian:2004tw,Airapetian:2010ds}, JLab HALL A \cite{Qian:2011py}, and the COMPASS
Collaboration \cite{Adolph:2012sn} experiments whereas the azimuthal
angular asymmetries of two back-to-back hadrons produced in $e^+e^-$ annihilations have been measured by the BELLE and BABAR collaborations \cite{Abe:2005zx,Seidl:2008xc}. However, the extraction of the transversity PDF
requires knowledge of the chiral-odd Collins fragmentation
functions \cite{Collins:1992kk,Boer:2003ya,Bacchetta:2008wb,Courtoy:2012ry}. 
In the last few years, extensive efforts to extract the transversity distributions has been carried out using the data on
polarized single-hadron SIDIS and back-to-back emission of two hadrons in $e^+e^-$ annihilations \cite{Anselmino:2007fs,Anselmino:2008jk,Anselmino:2013vqa,Kang:2015msa,Radici:2015mwa,Radici:2016lam,Bacchetta:2011ip,Bacchetta:2012ty}. Recently, the extraction of the transversity PDF from a global analysis of electron-proton and proton-proton data has been reported in Ref. \cite{Radici:2018iag}.

In principle, the GPDs provide valuable information about the spin
and orbital angular momentum of the constituents, as well as the spatial structure of the nucleon. As opposed to ordinary PDFs, GPDs are functions of three variables, namely, the longitudinal momentum fraction $x$ of the partons, the square of the
total momentum transferred $(t)$ and the skewness $(\zeta)$, which corresponds to the longitudinal momentum transferred.  We refer to the articles~\cite{Diehl:2003ny,Belitsky:2005qn,Goeke:2001tz,Ji:1998pc,Burkardt:2002hr} for reviews on this subject. 

The GPDs reduce to the ordinary PDFs in the forward limit $(t=0)$. The first moments of the GPDs are related
to the FFs, while the second moments of the sum of
the GPDs are related to the angular momentum by a sum rule proposed by Ji~\cite{Ji:1996ek}. Being off-forward
matrix elements, the GPDs do not have probabilistic
interpretations. Having said that, the Fourier transforms of the GPDs with respect to the momentum transfer purely in the transverse direction $(\zeta=0)$ provide the impact parameter dependent GPDs, which have probabilistic
interpretation and satisfy the positivity condition~\cite{Burkardt:2002hr,Burkardt:2000za,Gockeler:2006zu,Diehl:2002he}. The impact parameter dependent GPDs give us the information about partonic distributions in  the transverse position space for a given longitudinal momentum $(x)$. 


Unlike the FFs and PDFs, it is very difficult to measure the GPDs accessible in DVCS scattering \cite{Ji:1996nm,Radyushkin:1997ki}. First experimental DVCS results in terms of the beam spin asymmetry have been reported by HERMES at DESY \cite{Airapetian:2001yk} and CLAS at JLab \cite{Stepanyan:2001sm}. Since then, many
more results are available from the experiments performed by  the H1, ZEUS and HERMES collaborations at DESY \cite{Adloff:1999kg,Adloff:2001cn,Breitweg:1998nh,Chekanov:2003ya,Aktas:2005ty,Airapetian:2006zr,Aaron:2007ab,Airapetian:2011uq,Airapetian:2008aa,Airapetian:2010aa}, COMPASS collaborations at CERN~\cite{dHose:2004usi} and the Hall A and Hall B/CLAS collaborations at JLab \cite{Camacho:2006qlk,Chen:2006na,Girod:2007aa,Mazouz:2007aa}. Exclusive production of $\rho^0$ meson \cite{Adolph} and $\omega$ meson \cite{Adolph:2016ehf}  by scattering muons off a transversely polarized proton has been measured in a recent COMPASS experiment. The target spin asymmetries extracted in those experiments agree well with GPD-based model calculations. 

There have also been proposals to measure the GPDs through diffractive double meson production~\cite{Enberg,Yu}. 
The role of the GPDs in leptoproduction of vector mesons~\cite{Goloskokov1} as well as in hard exclusive electroproduction of pseudoscalar mesons~\cite{Goloskokov2}  has been studied within the framework of the handbag approach. 
In parallel to the efforts to investigate the GPDs from experiments, several theoretical predictions for the GPDs have been reported by using different approaches such as light-front quantization~\cite{Tiburzi:2001ta,Tiburzi:2001je,Mukherjee:2002xi}, constituent quark models (CQM) \cite{Scopetta:2003et,Scopetta:2002xq,Boffi:2002yy,Pasquini1,Boffi:2003yj,Scopetta:2004wt}, bag models \cite{Ji:1997gm,Anikin:2001zv}, soliton models \cite{Goeke:2001tz,Petrov:1998kf,Penttinen:1999th}, light-front quark-diquark models~\cite{Mondal:2017wbf,Maji:2017ill,Chakrabarti:2015ama,Maji:2015vsa,Mondal:2015uha},  AdS/QCD~\cite{Vega,Vega2,CM1,Rinaldi:2017roc,Traini:2016jko,deTeramond:2018ecg}, etc.
The moments of the GPDs have been evaluated with lattice QCD~\cite{gock,Hagler:2009,Hagler:2004,Bratt:2010}. Recently, the first calculation of the $x$ dependence of the nucleon GPDs within lattice QCD has been presented in Ref.~\cite{Alexandrou:2020zbe}.

In this paper, we employ the theoretical framework of basis light-front quantization (BLFQ)~\cite{Vary:2009gt,Honkanen:2010rc,Zhao:2014xaa,Wiecki:2014ola,Li:2015zda,Li:2017mlw,Jia:2018ary} to study nucleon properties. We adopt an effective light-front Hamiltonian and solve for its mass eigenstates at the scales suitable for low-resolution probes. With quarks as the only explicit degrees of freedom, our Hamiltonian includes the holographic QCD confinement potential~\cite{Brodsky:2014yha} supplemented by the longitudinal confinement~\cite{Li:2017mlw}. Our Hamiltonian also incorporates the one gluon exchange (OGE) interactions~\cite{Li:2015zda} to account for the dynamical spin effects. By solving this Hamiltonian in the constituent valence quark Fock space, and fitting the quark masses, confining strengths, and coupling constant,  we obtain the nucleon light-front wave functions (LFWFs). We then employ the LFWFs to compute the electromagnetic and axial FFs, transverse densities, PDFs, GPDs, radii, axial and tensor charges of the nucleon. We compare our results with experiment and with other theoretical approaches.

The paper is organized as follows. A detailed description of the BLFQ formalism and the light-front effective Hamiltonian for the nucleon is discussed in Sec.~\ref{sec:BLFQ}. The numerical results of the various nucleon observables are presented in Sec.~\ref{sec:results}. We summarize our findings in Sec.~\ref{sec:con}.
\section{Basis Light-Front Quantization}\label{sec:BLFQ}
The theoretical framework of BLFQ~\cite{Vary:2009gt} has emerged as a promising tool to solve relativistic bound state problems in quantum field theories. BLFQ, as a nonperturbative approach, is based on the Hamiltonian formalism and incorporates the advantages of the light-front dynamics \cite{Brodsky:1997de}. 
This approach has been successfully applied to QED systems including the electron anomalous magnetic moment~\cite{Zhao:2014xaa,Honkanen:2010rc} and the strong coupling bound-state positronium problem \cite{Wiecki:2014ola}. It has also been employed to solve light mesons~\cite{Jia:2018ary,Lan:2019vui,Lan:2019rba}, heavy quarkonia~\cite{Li:2015zda,Li:2017mlw,Li:2018uif,Lan:2019img}, heavy-light mesons~\cite{Tang:2018myz,Tang:2019gvn}, and the proton \cite{Xu:2019xhk} as QCD bound states.

The structure of the bound states is encoded in the LFWFs, which are obtained from diagonalizing the light-front Hamiltonian ($H =P^-P^+$):
\begin{align}
H \ket{\beta}= M_h^2 \ket{\beta},
\end{align}
where the eigenvalues $M_h^2$
correspond to the mass-squared spectrum and the associated eigenvectors $\ket{\beta}$ encode
structural information of the bound states. In our approach, we define the light front coordinate variables as $x^{\pm}=x^0\pm x^3$ and the corresponding momentum variables as $p^{\mp}=p^0\mp p^3$.
In writing Eq. (1), we indicate that we will adopt a framework for solving the Hamiltonian eigenvalue problem in which the LFWFs are boost invariant in the longitudinal and the transverse directions.
The BLFQ approach employs a suite of analytical and numerical techniques for setting up and solving this eigenvalue problem in a convenient basis space~\cite{Li:2015zda,Li:2017mlw,Lan:2019vui}. 

In this paper, we solve the nucleon bound state problem in BLFQ using an effective light-front Hamiltonian ($H_{\rm{eff}}$) defined below.
At fixed light-front time, the nucleon state can be expressed schematically in terms of various quark ($q$), antiquark ($\bar q$) and gluon $(g)$ Fock components,
\begin{align}\label{Eq1}
|N\rangle=&\psi_{(3q)}|qqq\rangle+\psi_{(3q+q\bar q)}|qqqq\bar q\rangle \nonumber\\&+\psi_{(3q+1g)}|qqqg\rangle+\dots\, , 
\end{align}
where the $\psi_{(\dots)}$ correspond to the probability amplitudes to find the different parton configurations in the nucleon. Within BLFQ, each Fock component itself consists of an infinite number of basis states. For the purpose of numerical calculations, we employ both a Fock-sector truncation and limits on the basis states within each Fock component. Here, we consider only the leading Fock sector to describe the valence quark contribution to nucleon properties. 

The basis states of each Fock particle are expressed in terms of the longitudinal and transverse coordinates along with the helicity quantum numbers~\cite{Zhao:2013cma}.  We omit the color degree of freedom in the current approach since, for the pure valence sector treated as a color singlet, a color factor suffices when combined with the strength of our effective one-gluon exchange interaction (see below).
The longitudinal momentum of the particle is identified by the quantum number $k$. We confine the longitudinal coordinate $x^-$ to a box of length $2L$ with antiperiodic (periodic) boundary conditions for fermions (bosons)\footnote{Although we do not include dynamical gluons in the present work, we specify their corresponding modes as seen
for dynamical photons in Ref.~\cite{Zhao:2013cma}, for example.}. Thus, the  amplitude in longitudinal coordinate space is given by
\begin{eqnarray}
\Psi_k(x^-) = \frac{1}{2L} e^{i\frac{\pi}{L}kx^-},
\end{eqnarray} 
with the discretized longitudinal momentum 
$p^+=2\pi k/L$, where the dimensionless quantity $k=\frac{1}{2},\frac{3}{2},\frac{5}{2},\ldots$ for fermions, while for bosons $k=1,2,3,\ldots$. The zero mode for bosons is neglected.
All many-body basis states are selected to have the same total longitudinal momentum $P^+=\sum_ip_i^+$,
where the sum is over the particles in a particular basis state. We then parameterize $P^+$ using a dimensionless variable $K=\sum_i k_i$ such that 
$P^+=\frac{2\pi}{L}K$. For a given particle $i$, the longitudinal momentum fraction $x$ is now defined as
$
x_i=p_i^+/P^+=k_i/K.
$

 In the transverse direction, we emply the two dimensional harmonic oscillator (2D-HO) basis $\phi_{nm}(\vec{p}_\perp;b)$, which is characterized by two quantum numbers $n$ and $m$ corresponding to the radial excitation and angular momentum projection, respectively, of the particle. In momentum space, the orthonormalized 2D-HO wave functions are given by ~\cite{Vary:2009gt,Zhao:2013cma}
\begin{align}
\phi_{n,m}(\vec{p}_{\perp};b)
 =&\frac{\sqrt{2}}{b(2\pi)^{\frac{3}{2}}}\sqrt{\frac{n!}{(n+|m|)!}}e^{-\vec{p}_{\perp}^2/(2b^2)}\nonumber\\
  &\times\left(\frac{|\vec{p}_{\perp}|}{b}\right)^{|m|}L^{|m|}_{n}(\frac{\vec{p}_{\perp}^2}{b^2})e^{im\theta},\label{ho}
\end{align}
where $b$ is an HO basis scale parameter with the dimension of mass and $\vec{p}_{\perp}$ represents the transverse momentum of the particle; $L^{\alpha}_{n}(x)$ are the generalized  Laguerre polynomials and $\theta={\rm arg}(\vec{p}_{\perp}/b)$. For the spin degrees of freedom, the quantum number $\lambda$ is used to label the helicity of the particle. Thus, each single-particle basis state is identified using four quantum numbers, $\{x,n,m,\lambda\}$. In addition, we require that our many-body basis states have well defined values of the total angular momentum projection
$
M_J=\sum_i\left(m_i+\lambda_i\right).
$

Beyond the Fock space truncation, within each Fock-sector, further truncation is still needed to reduce the basis to a finite
dimension. We truncate the infinite basis by introducing a truncation parameter $K$ on the longitudinal direction  and in the transverse direction, we require the total transverse quantum number 
\begin{align}
N_\alpha=\sum_i (2n_i+| m_i |+1)
\end{align}
for multiparticle basis state satisfies $N_\alpha \le N_{\text{max}}$, where $N_{\text{max}}$ is a chosen truncation parameter in the transverse direction. Here, $K$ is the basis resolution in the longitudinal direction, whereas $N_{\rm{max}}$ controls the transverse momentum covered by the 2D-HO basis functions. The $N_{\rm max}$ truncation naturally provides ultraviolet (UV) and infrared (IR) cutoffs. In momentum space, the UV cutoff, $\Lambda_{\rm UV} \simeq b\sqrt{N_{\rm max}}$ and the IR cutoff, $\lambda_{\rm IR} \simeq b/\sqrt{N_{\rm max}}$. With increasing the $N_{\text{max}}$, the UV (IR) cutoff increases (decreases)~\cite{Zhao:2014xaa} and both UV and IR increase as the basis scale parameter $b$ increases. 

\subsection{Light-front effective Hamiltonian}
For the valence Fock sector of the baryon, we adopt a light-front effective Hamiltonian ($H_{\rm{\rm eff}}=P^-_{\rm{eff}}P^+$), which is given by
\begin{align}\label{hami}
H_{\rm{eff}} =\sum_{i} \frac{m_i^2+\vec{p}_{i\perp}^2}{x_i}+\frac{1}{2}\sum_{i,j} V_{i,j}^{\rm{conf}}+ \frac{1}{2}\sum_{i,j} V_{i,j}^{\rm{OGE}},
\end{align}
where $\sum_i x_i=1$; $m$ is the constituent mass of quark and  $i,j$ denote the index of particles in a Fock-sector. $V^{\rm conf}_{i,j}$ represents the confining potential, which includes both  the transverse and the longitudinal confinements. 

We generalize the soft-wall holographic confinement~\cite{Brodsky:2014yha} in the transverse direction, and also employ a complementary longitudinal confining potential that reproduces 3D confinement in the nonrelativistic limit~\cite{Li:2015zda}. 
For a many-body system the complete confining potential is then written as
\begin{align}
V^{\rm{conf}}_{i,j}=\kappa^4 \vec{r}_{ij\perp}^2+\frac{\kappa^4}{(m_i+m_j)^2}\partial_{x_i}(x_ix_j\partial_{x_j}),
\end{align}
where $\vec{r}_{ij\perp}=\sqrt{x_{i} x_{j}}(\vec{r}_{i\perp}-\vec{r}_{j\perp})$ is the relative coordinate, $\kappa$ is the strength of the confining potential in the transverse and longitudinal direction and $\partial_{x}\equiv (\partial/\partial x)_{r_{ij\perp}}$. Note that the form of the longitudinal confining potential for $q\bar q$ has been suggested in the relative momentum coordinate~\cite{Li:2015zda}. Here, the potential is approximately generalized for a many-body system in the single-particle momentum coordinate. We assume that the correction is small and the potential is much less sensitive to the choice of the momentum coordinate for a large basis. 

The last term in Eq.~(\ref{hami}) represents the one-gluon exchange (OGE) interaction: 
\begin{align}
V^{\rm{OGE}}_{i,j} = \frac{4\pi C_F \alpha_s}{Q^2_{ij}} \bar{u}_{s^{\prime}_i}(p^{\prime}_i)\gamma^{\mu}u_{s_i}(p_i)\bar{u}_{s^{\prime}_j}(p^{\prime}_j)\gamma_{\mu}u_{s_j}(p_j)
\end{align}
with fixed coupling constant $\alpha_s$. Here, $Q^2_{ij}=-(1/2)(p^{\prime}_i-p_i)^2-(1/2)(p_j-p^{\prime}_j)^2$ is the average of 4-momentum square carried by the exchanged gluon. In terms of kinematical variables,
\begin{align}
  Q^2_{ij} =& \frac{1}{2} \Bigg[\Big(\frac{\vec{p}_{i\perp}^2+m_i^2}{x_i}-\frac{\vec{p}_{i\perp}^{\prime 2}+m_i^2}{x^{\prime}_i}\nonumber\\ &
  -\frac{(\vec{p}_{i\perp}^2-\vec{p}_{i\perp}^{\prime 2})+\mu_g^2}{x_i-x^{\prime}_i}\Big) -(i\rightarrow j)\Bigg],
\end{align}
where $\mu_g$ is the gluon mass.
 The color factor $C_F=-2/3$, which implies that the OGE is an attractive potential. The spinor ${u}_{s_i}(p_i)$ is the solution of the free Dirac equation, with the subscripts representing the spin and $\vec{p}_{i\perp}$ is the momentum of the valence quark $i$. Implementing the OGE interaction, we naturally generate the dynamical spin structure in the LFWFs, which plays a crucial role in computing the spin dependent observables.

Here, we construct our basis using single-particle coordinates. The advantage of using these coordinates is that we can treat each particle in the Fock space on equal footing, and it enables dealing with symmetry among identical particles~\cite{Zhao:2014xaa}. On the other hand, $H_{\rm eff}$ incorporates the transverse center-of-mass (c.m.) motion and this is intertwined with intrinsic motion when working on a basis of single-particle states. In order to preserve boost invariance in the transverse direction, we introduce a constraint term 
\begin{align}
    \label{Hprime}
    H^{\prime}=\lambda_{L} ( H_{\rm c.m.}-2b^2 I)\, ,
\end{align}
into the effective Hamiltonian which effectively factorizes the transverse c.m. motion from the intrinsic motion. We subtract the zero-point energy $ 2b^2 $ and multiply by a Lagrange multiplier $\lambda_{L}$. $ \hat I $ indicates the unity operator and the c.m. motion is governed by~\cite{Wiecki:2014ola},
\begin{align}
    H_{\rm c.m.}=\left(\sum_i \vec{p}_{i\perp}\right)^2+b^4\left(\sum_i x_i  \vec{r}_{i\perp}\right)^2 . 
\end{align}
With $\lambda_L$ sufficiently large and positive, we are able to shift the excited states of c.m. motion to higher energy away from the low-lying states with LFWFs in factorized form. Therefore,
the effective Hamiltonian we diagonalize is
\begin{align}
  H_{\rm eff}^{\prime} = H_{\rm eff} - \left(\sum_i \vec{p}_{i\perp}\right)^2 + \lambda_L (H_{\rm c.m.} - 2b^2I).
\end{align}

Upon diagonalization of this light-front Hamiltonian matrix $H_{\rm eff}^{\prime}$ within the BLFQ basis, we produce the eigenvalues that correspond to the mass spectrum. We also produce the eigenvectors that correspond to the LFWFs in the BLFQ basis that encode the structural information of the systems. The lowest eigenstate is naturally identified as the nucleon state, denoted as $\ket{P, {\Lambda}}$, where the $\Lambda$ indicates the helicity of the nucleon. 
The resulting valence LFWF in momentum space is expressed as an expansion in the orthonormal basis set designed to preserve the symmetries of the effective Hamiltonian
\begin{align}
&\Psi^{\Lambda}_{\{x_i,\vec{p}_{i\perp},\lambda_i\}}=\braket{P, {\Lambda}|\{x_i,\vec{p}_{i\perp},\lambda_i\}}\nonumber \\&\quad\quad=\sum_{\{n_i,m_i\}}\big( \psi^{\Lambda}_{\{x_{i},n_{i},m_{i},\lambda_i\}} \prod_i \phi_{n_i,m_i}(\vec{p}_{i\perp};b) \big),\label{wavefunctions}
\end{align}
with $\psi^{\Lambda}_{\{x_{i},n_{i},m_{i},\lambda_i\}}=\braket{P, {\Lambda}|\{x_i,n_i,m_i,\lambda_i\}}$ as the LFWF in BLFQ. Note that the LFWFs should have parity symmetry ($P$), which is broken by the Fock-space truncation. However, one can use mirror parity $\hat{P}_x=\hat{R}_x(\pi)P$~\cite{Brodsky:2006ez} to replace the parity. Under the mirror parity transformation, our wave function follows the relation
\begin{align}
\psi^{\downarrow}_{\{x_i,n_i,m_i,\lambda_i\}}&=&(-1)^{\sum_{i}m_i+1}\psi^{\uparrow}_{\{x_i,n_i,-m_i,-\lambda_i\}},
\end{align}
where the arrow indicates the helicity
of the nucleon.

\begin{table}[htp]	
\centering
\caption{Model parameters for the basis truncations $N_{\rm{max}}=10$ and $K=16.5$.}\label{tab:parameter}
\begin{tabular}{|cccc|cc}
\hline\hline
$m_{\rm{q/k}}$ 	   ~&~   $m_{\rm{q/g}}$    ~&~ $\kappa$      	 ~&~ $\alpha_s$    \\
\hline
$0.3$ GeV ~&~	 $0.2$	GeV    ~&~ $0.34$ GeV ~&~ $1.1\pm 0.1$ \\
\hline\hline
\end{tabular}
\end{table} 
\begin{table}[htp]	
\centering
\caption{The dependence of the fitted coupling constant $\alpha_s$ on basis truncation parameters, $N_{\rm{max}}$ and $K$ with other parameters held fixed to the values quoted in Table I.}\label{tab:coupling}
\begin{tabular}{|cc|cc|cc}
\hline\hline
$[N_{\rm{max}},\,K]$  ~&~ $\alpha_s$ ~&~ $[N_{\rm{max}},\,K]$  ~&~ $\alpha_s$   \\
\hline
$[6,\,16.5]$ ~&~ $1.4\pm 0.20$ ~&~ $[10,\,10.5]$ ~&~ $1.0\pm 0.15$\\
$[8,\,16.5]$ ~&~ $1.2\pm 0.15$ ~&~ $[10,\,16.5]$ ~&~ $1.1\pm 0.10$\\
\hline\hline
\end{tabular}
\end{table} 
\section{Numerical results}\label{sec:results}
There are four parameters in our calculation: the quark mass in the kinetic energy ($m_{q/k}$), the quark mass in the OGE interaction ($m_{q/g}$), the strength of confining potential ($\kappa$), and the coupling constant ($\alpha_s$). 
We now sketch our reasoning for the flexibility in the choice of the vertex mass.  In particular, our approach features an effective OGE interaction that is important for short distance physics and approximately describes the processes where valence quarks emit and absorb a gluon. As it is an effective interaction, it accounts for fluctuations between the $\ket{qqq}$, $\ket{qqqg}$, and higher Fock sectors. According to the mass evolution in renormalization group theory, the dynamical OGE would also generate contributions to the quark mass arising from higher momentum scales leading to a decrease in the quark mass from the gluon dynamics. In turn, this leads to the suggestion that the mass in the OGE interaction would be lighter than the kinetic mass, which is associated with the long-range physics in our effective Hamiltonian. This treatment is also noticed and adopted in the literature~\cite{Brisudova:1994it,Burkardt:1998dd,Burkardt:1991tj}. 

 In our approach, we select the truncation parameters $N_{\rm{max}}=10$ and $K=16.5$. The model parameters are summarized in Table~\ref{tab:parameter}. 
 We set those parameters by fitting the nucleon mass and the flavor FFs~\cite{Cates:2011pz,Qattan:2012zf,Diehl:2013xca}. 
 We estimate an uncertainty on $\alpha_s$ that accounts for the model selections and major fitting uncertainties. As can be seen from Table~\ref{tab:coupling}, our estimated uncertainty for the coupling constant decreases with increasing the basis cutoffs $N_{\rm{max}}$. 
 Using those model parameters, we then present the nucleon electromagnetic FFs and their ratios, the axial FF as well as the leading twist PDFs and GPDs. We also predict the radii, axial and tensor charges of the nucleon.
\subsection{Electromagnetic form factors}
In the light-front framework, the Dirac and the Pauli FFs of the nucleon, $F_1(Q^2)$ and $F_2(Q^2)$ respectively, are identified with the helicity-conserving and helicity-flip matrix elements of the vector ($J^+\equiv\sum_q e_q \bar{\psi}_q\gamma^+\psi_q$) current:
\begin{align}\label{DFFs}
\braket{P+q,\uparrow|\frac{J^+(0)}{2P^+}|P,\uparrow} =& F_1(Q^2); \\
\braket{P+q,\uparrow|\frac{J^+(0)}{2P^+}|P,\downarrow} =& -(q^1-iq^2)\frac{F_2(Q^2)}{2M},\label{PFFs}
\end{align}
where $Q^2=-q^2$ is the square of the momentum transfer, $M$ is the mass of the nucleon, and $e_q$ is the charge of the individual quarks. Within the valence Fock-sector, the nucleon state with momentum  $P$ can be written in terms of three-particle LFWFs:
\begin{align}
\ket{P,{\Lambda}} =&  \int \prod_{i=1}^{3} \left[\frac{{\rm d}x_i{\rm d}^2 \vec{p}_{i\perp}}{\sqrt{x_i}16\pi^3}\right] \nonumber\\
& \times 16\pi^3\delta \left(1-\sum_{i=1}^{3} x_i\right) \delta^2 \left(\sum_{i=1}^{3}\vec{p}_{i\perp}\right) \nonumber\\
& 
\times \Psi^{\Lambda}_{\{x_i,\vec{p}_{i\perp},\lambda_i\}} \ket{\{x_iP^+,\vec{p}_{i\perp}+x_i\vec{P}_{\perp},\lambda_i\}};\label{wavefunction_expansion}
\end{align}
here $x_i=p_i^+/P^+$   and $\vec{p}_{i\perp}$ 
represents the relative transverse momentum of the $i$-th constituent.
 Substituting the nucleon states and the quark field operators ($\psi_q$ and $\bar{\psi}_q$) in Eqs.~(\ref{DFFs}) and (\ref{PFFs}) leads
to the flavor Dirac and Pauli form factors in terms of the overlap of the LFWFs~\cite{Brodsky:2000xy,Brodsky:2000xy}:
\begin{align}
F_1^q(Q^2)&= 
 \sum_{\{\lambda_i\}} \int \left[{\rm d}\mathcal{X} \,{\rm d}\mathcal{P}_\perp\right] \Psi^{\uparrow *}_{\{x^{\prime}_i,\vec{p}^{\prime}_{i\perp},\lambda_i\}}\Psi^{\uparrow}_{\{x_i,\vec{p}_{i\perp},\lambda_i\}} ;    \\
F_2^q(Q^2)&= -\frac{2M}{(q^1-iq^2)}
 \sum_{\{\lambda_i\}} \int \left[{\rm d}\mathcal{X} \,{\rm d}\mathcal{P}_\perp\right]\nonumber\\&\quad\quad\quad\quad\times \Psi^{\uparrow *}_{\{x^{\prime}_i,\vec{p}^{\prime}_{i\perp},\lambda_i\}}\Psi^{\downarrow}_{\{x_i,\vec{p}_{i\perp},\lambda_i\}}\, , 
\end{align}
where $x^{\prime}_1=x_1$ and $\vec{p}^{\prime}_{1\perp}=\vec{p}_{1\perp}+(1-x_1)\vec{q}_{\perp}$ for the struck quark, while $x^{\prime}_i={x_i}$ and $\vec{p}^{\prime}_{i\perp}=\vec{p}_{i\perp}-{x_i} \vec{q}_{\perp}$ for the spectators ($i=2,3$). Here, we use the abbreviation
\begin{align}
\left[{\rm d}\mathcal{X} \,{\rm d}\mathcal{P}_\perp\right]=&\prod_{i=1}^3 \left[\frac{{\rm d}x_i{\rm d}^2 \vec{p}_{i\perp}}{16\pi^3}\right]\nonumber\\&\times 16 \pi^3 \delta \left(1-\sum_{i=1}^{3} x_i\right) \delta^2 \left(\sum_{i=1}^{3}\vec{p}_{i\perp}\right) .  
\end{align}

We consider the frame where the momentum transfer occurs purely in the transverse direction, i.e., $q=(0,0,\vec{q}_{\perp})$,  thus $Q^2=-q^2={\vec{q}_{\perp}}^2$. Note that under the charge and isospin symmetry, $\langle p\mid \bar{u}\gamma^\mu u\mid p\rangle= \langle n\mid \bar{d}\gamma^\mu d\mid n\rangle$, where $p(n)$ represents the proton (neutron) and $u(d)$ denotes the up (down) quark field. 
The Dirac FFs follow the normalizations
\begin{align}
&F^{\rm{u/p}}_{1}(0)=2,\quad F^{\rm{d/p}}_{1}(0)=1 ; \nonumber\\
&F^{\rm{u/n}}_{1}(0)=1,\quad F^{\rm{d/n}}_{1}(0)=2,
\end{align}
while, the  Pauli FFs at $Q^2=0$ provide the anomalous magnetic moment:
\begin{align}
&F^{\rm{u/p}}_{2}(0)=\kappa_u,\quad F^{\rm{d/p}}_{2}(0)=\kappa_d ; \nonumber\\
&F^{\rm{u/n}}_{2}(0)=\kappa_d,\quad F^{\rm{d/n}}_{2}(0)=\kappa_u,
\end{align}
where $\kappa_{u(d)}$ is the anomalous magnetic moment of the up (down) quark in the proton.

\begin{figure}[htbp]
\centering
\includegraphics[width=\columnwidth]{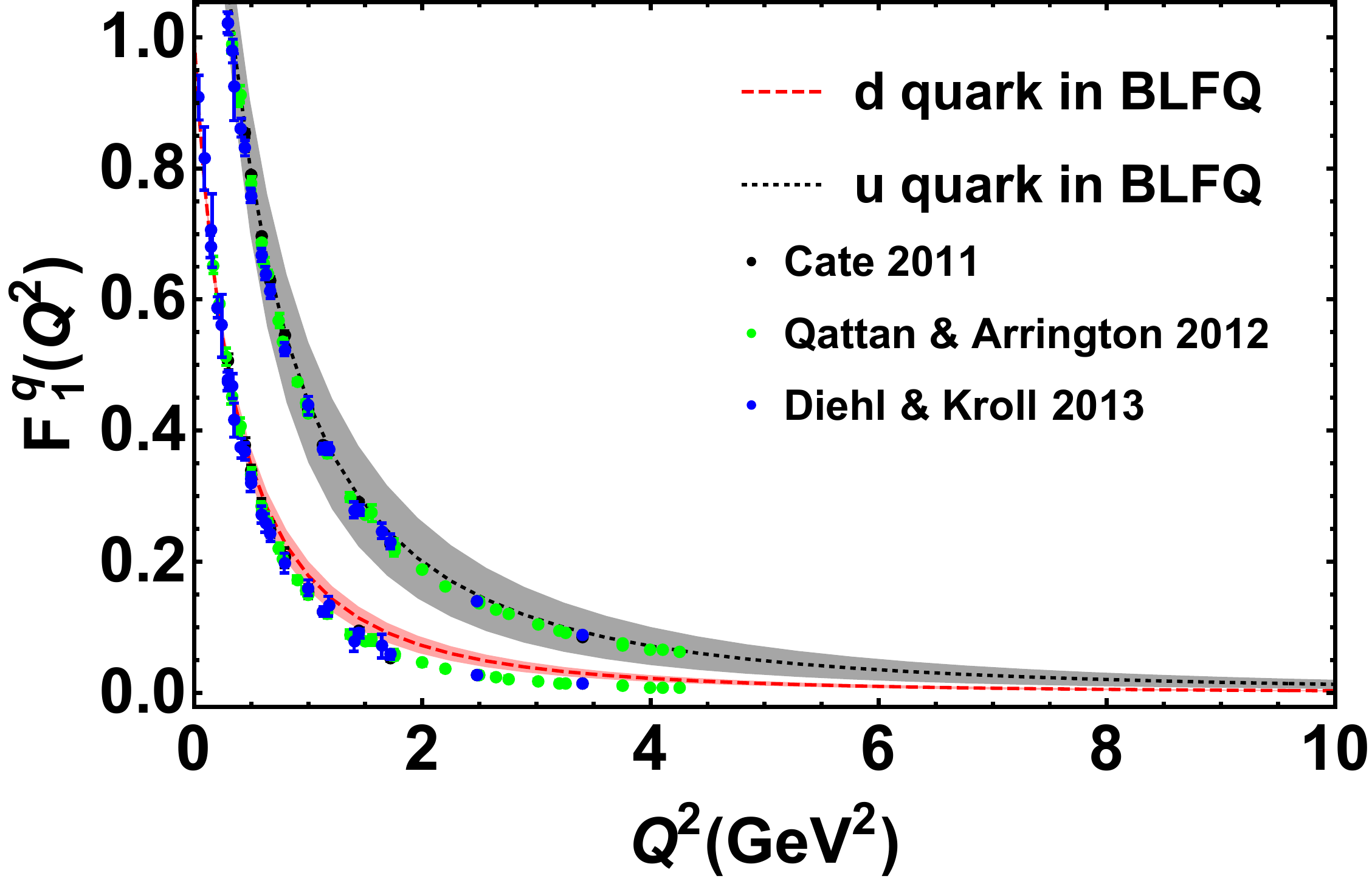}
\includegraphics[width=\columnwidth]{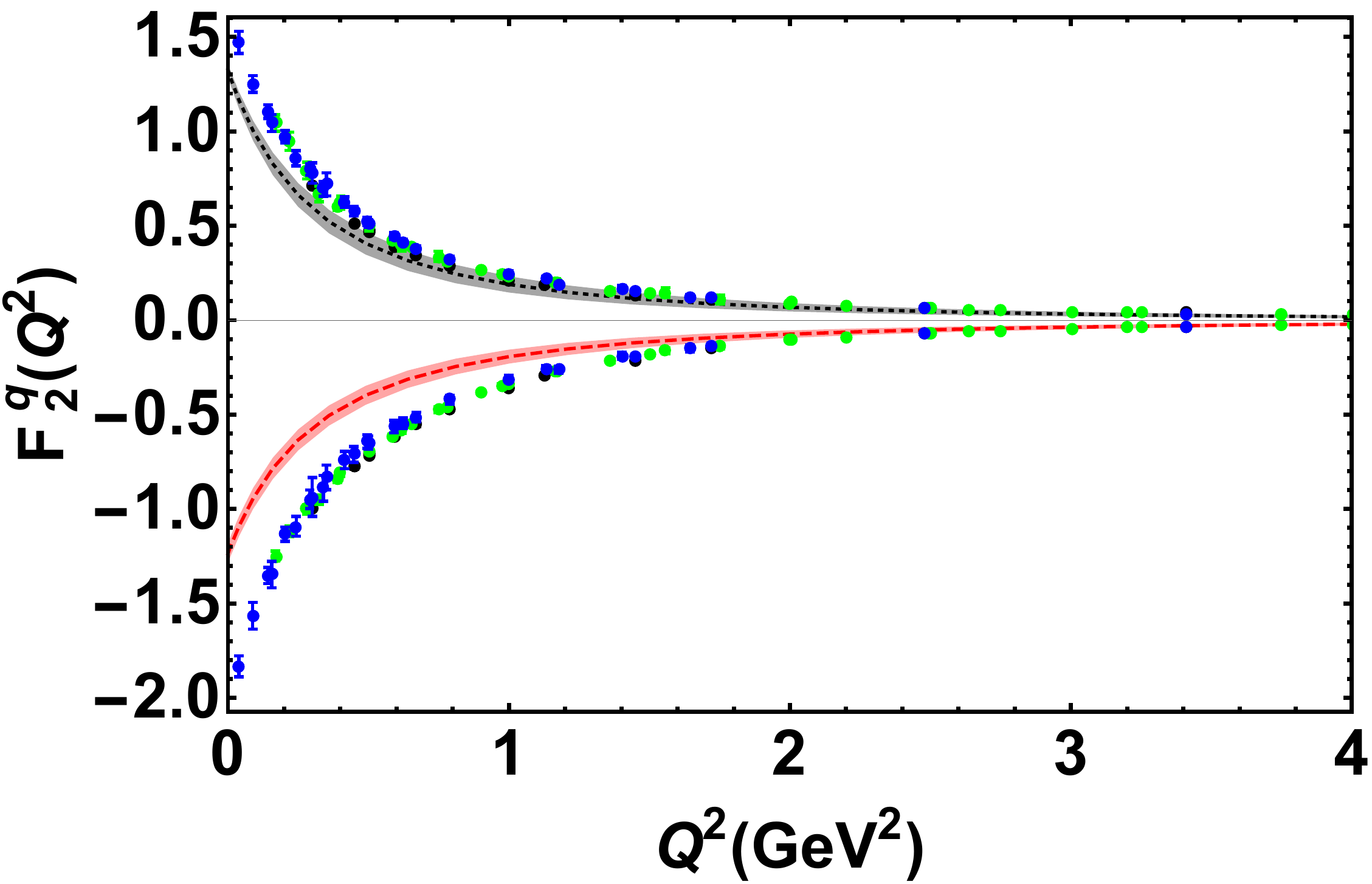}
\caption{the Dirac FFs (upper panel) and the Pauli FFs (lower panel) for the up and down quark. The gray and red bands are BLFQ results for the up and the down quark, respectively, reflecting our $\alpha_s$ uncertainty of $10\%$. The experimental data are taken from Refs.~\cite{Cates:2011pz,Qattan:2012zf,Diehl:2013xca}}
\label{flavor}
\end{figure}
\begin{figure*}[htbp]
\centering
\subfigure[]{
\begin{minipage}[t]{0.45\linewidth}
\centering
\includegraphics[width=\columnwidth]{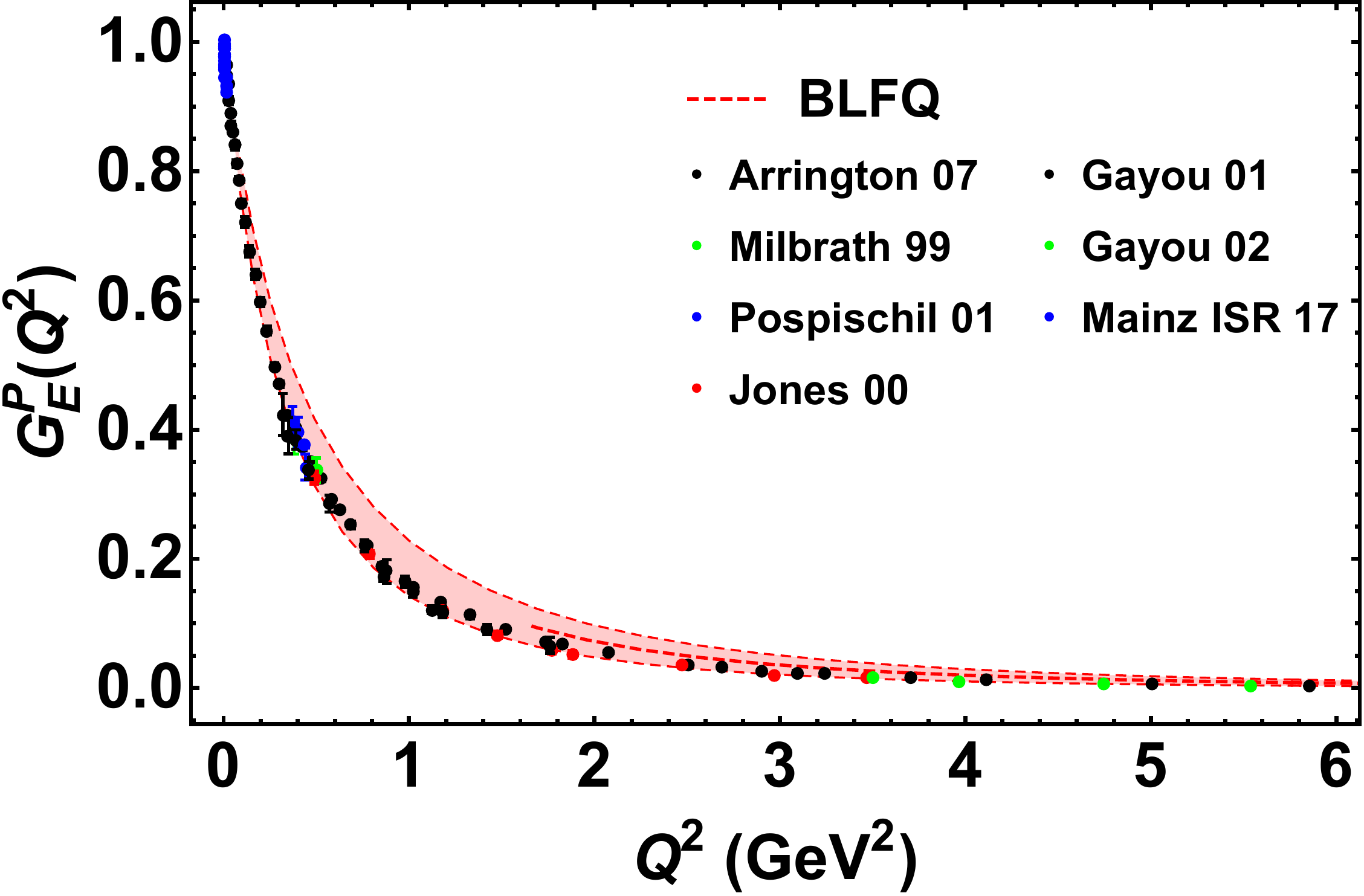}
\end{minipage}
\label{Gep}
}
\subfigure[]{
\begin{minipage}[t]{0.45\linewidth}
\centering
\includegraphics[width=\columnwidth]{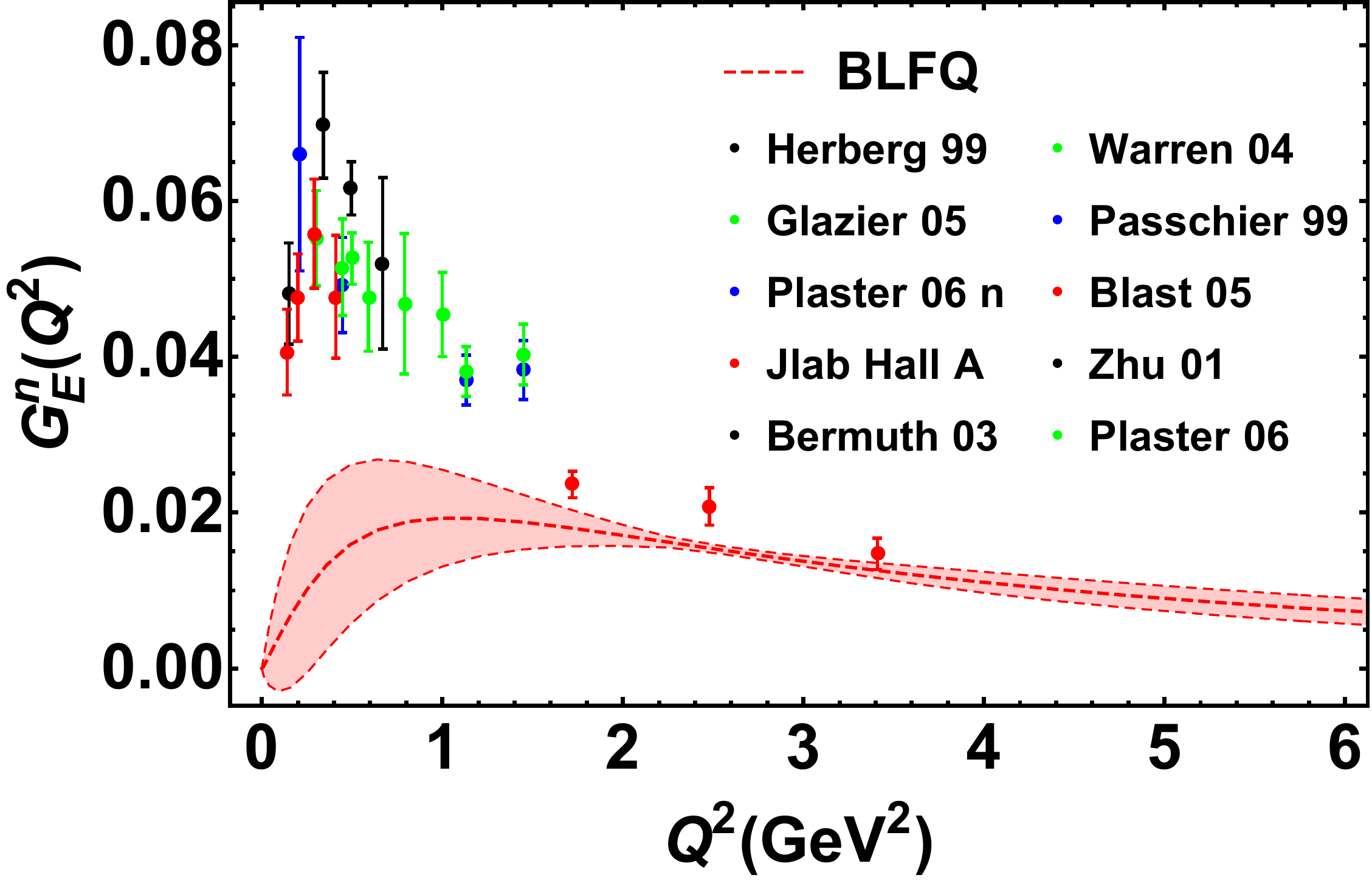}
\end{minipage}
\label{Gen}
}
\subfigure[]{
\begin{minipage}[t]{0.45\linewidth}
\centering
\includegraphics[width=\columnwidth]{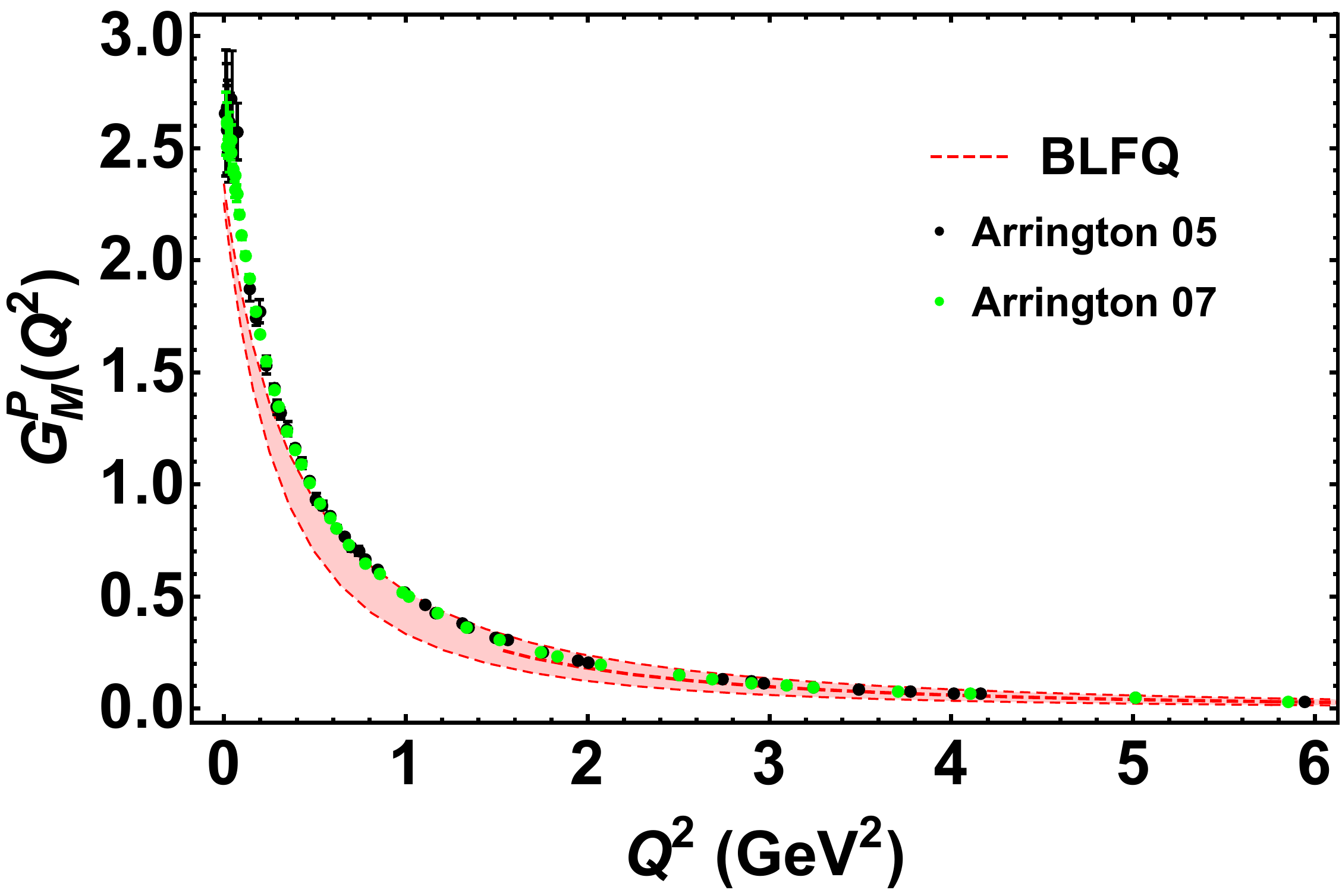}
\end{minipage}
\label{Gmp}
}
\subfigure[]{
\begin{minipage}[t]{0.45\linewidth}
\centering
\includegraphics[width=\columnwidth]{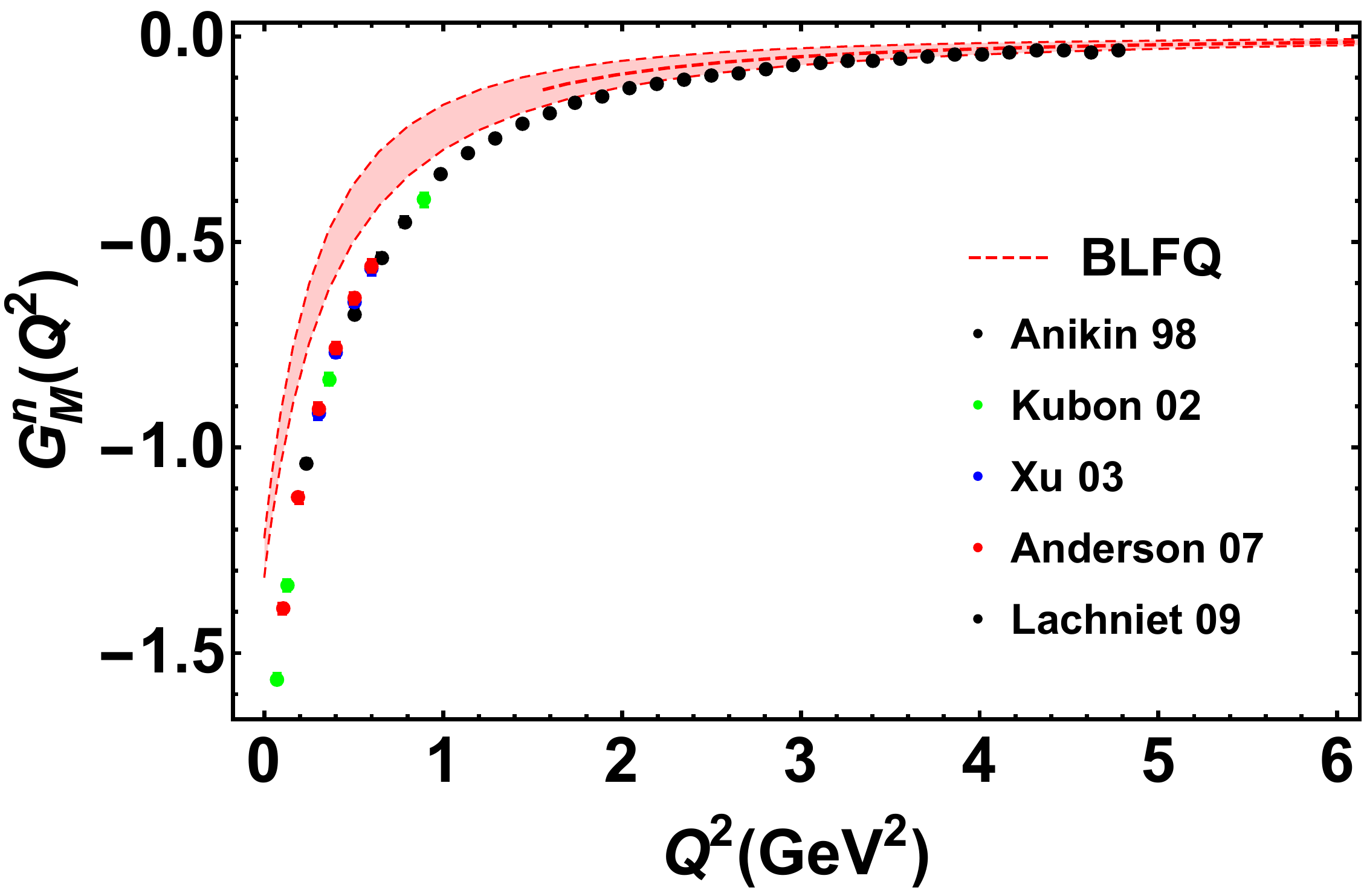}
\end{minipage}
\label{Gmn}
}
\caption{The Sachs FFs for the proton and the neutron. The red bands are BLFQ results.
The data are taken from Refs.~\cite{Gayou:2001qt,Jones:1999rz,Arrington:2007ux,Gayou:2001qd,Pospischil:2001pp,Milbrath:1997de} for $G^{p}_E(Q^2)$, Refs.~\cite{Geis:2008aa,Warren:2003ma,Riordan:2010id,Bermuth:2003qh,Bermuth:2003qh,Plaster:2005cx,Glazier:2004ny,Schiavilla:2001qe} for $G^{n}_E(Q^2)$, Ref.~\cite{Arrington:2004ae,Arrington:2007ux} for $G^{p}_M(Q^2)$ and Ref.~\cite{Lachniet:2008qf,Anderson:2006jp,Xu:2002xc,Kubon:2001rj,Anklin:1998ae} for $G^{n}_M(Q^2)$.}
\label{sach}
\end{figure*}
\begin{figure*}[htbp]
\centering
\subfigure[]{
\begin{minipage}[t]{0.45\linewidth}
\centering
\includegraphics[width=\columnwidth]{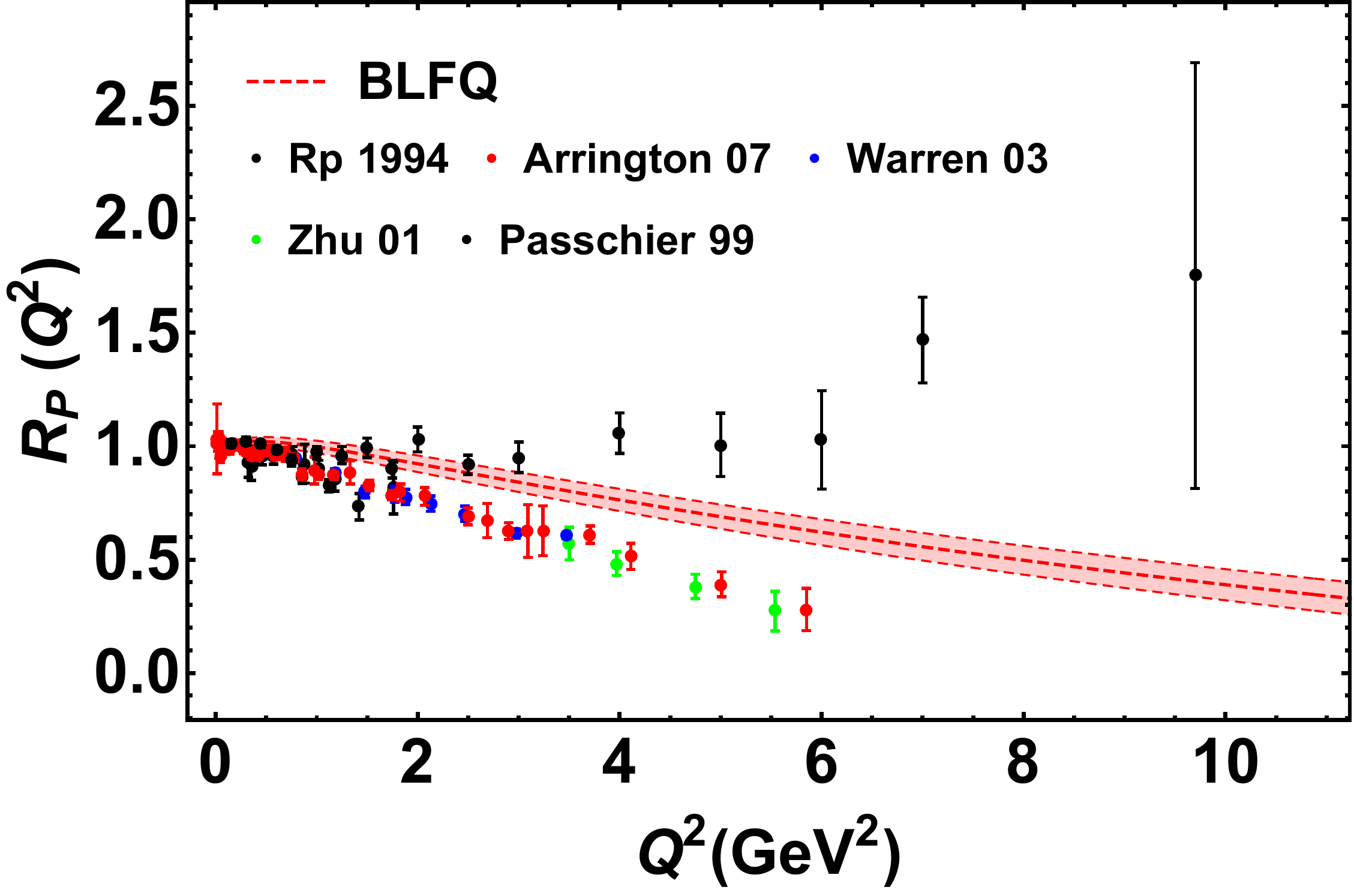}
\end{minipage}
\label{R_proton}
}
\subfigure[]{
\begin{minipage}[t]{0.45\linewidth}
\centering
\includegraphics[width=\columnwidth]{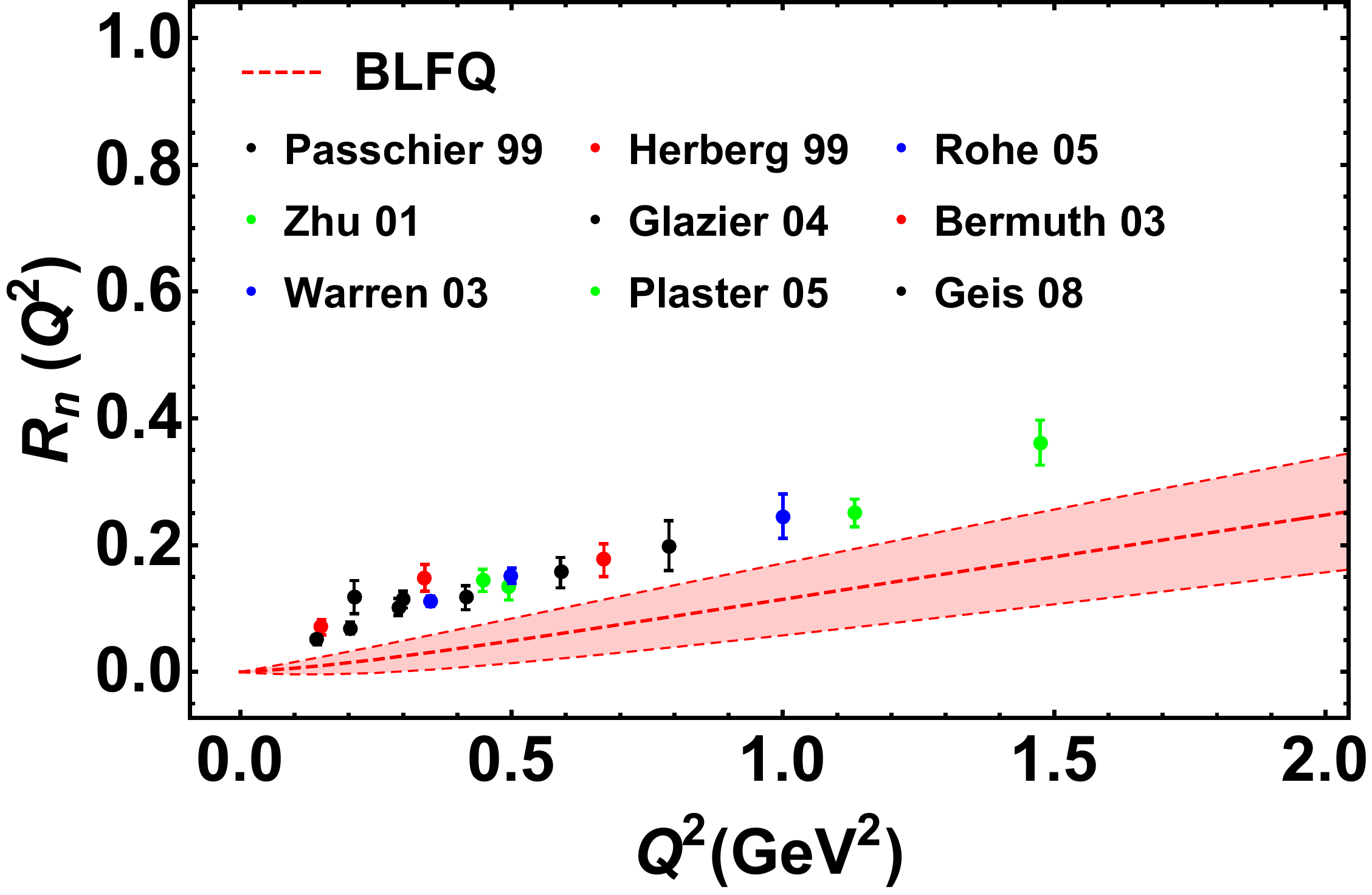}
\end{minipage}
\label{R_neutron}
}
\caption{Ratio of Sachs form factor $R^i=\mu_i G^i_E/G^i_M$. The red bands are BLFQ results. The data are taken from Refs.~\cite{Gayou:2001qd,Pospischil:2001pp,Milbrath:1997de,Punjabi:2005wq,Ron:2011rd,Walker:1993vj,Zhan:2011ji,MacLachlan:2006vw,Paolone:2010qc} for proton and Refs.~\cite{Geis:2008aa,Warren:2003ma,Bermuth:2003qh,Plaster:2005cx,Glazier:2004ny,Passchier:1999cj,Zhu:2001md,Herberg:1999ud,Riordan:2010id} for neutron.}
\label{ratio_sach}
\end{figure*}
\begin{figure*}[htbp]
\centering
\subfigure[]{
\begin{minipage}[t]{0.45\linewidth}
\centering
\includegraphics[width=\columnwidth]{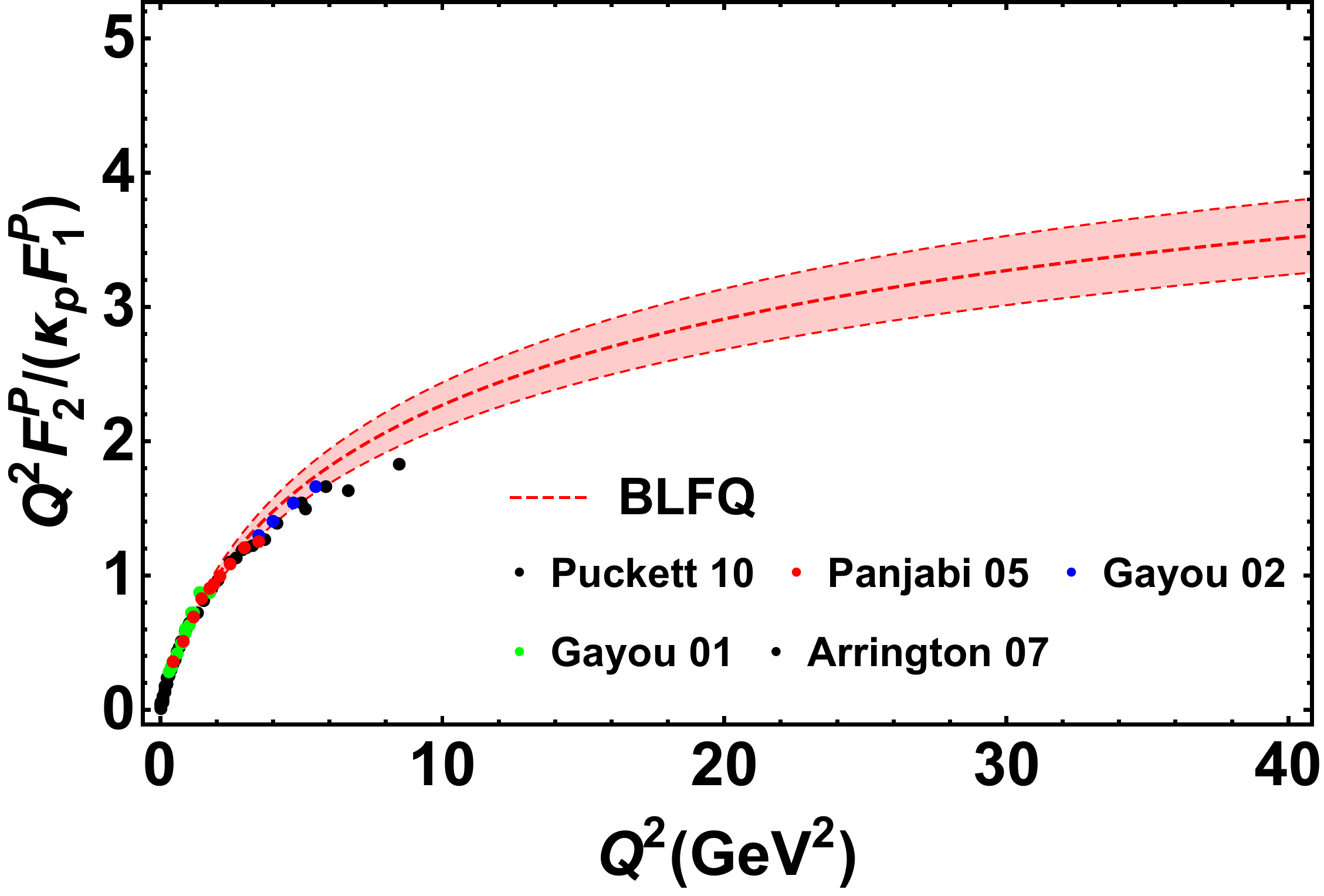}
\end{minipage}
\label{p1}
}
\subfigure[]{
\begin{minipage}[t]{0.45\linewidth}
\centering
\includegraphics[width=\columnwidth]{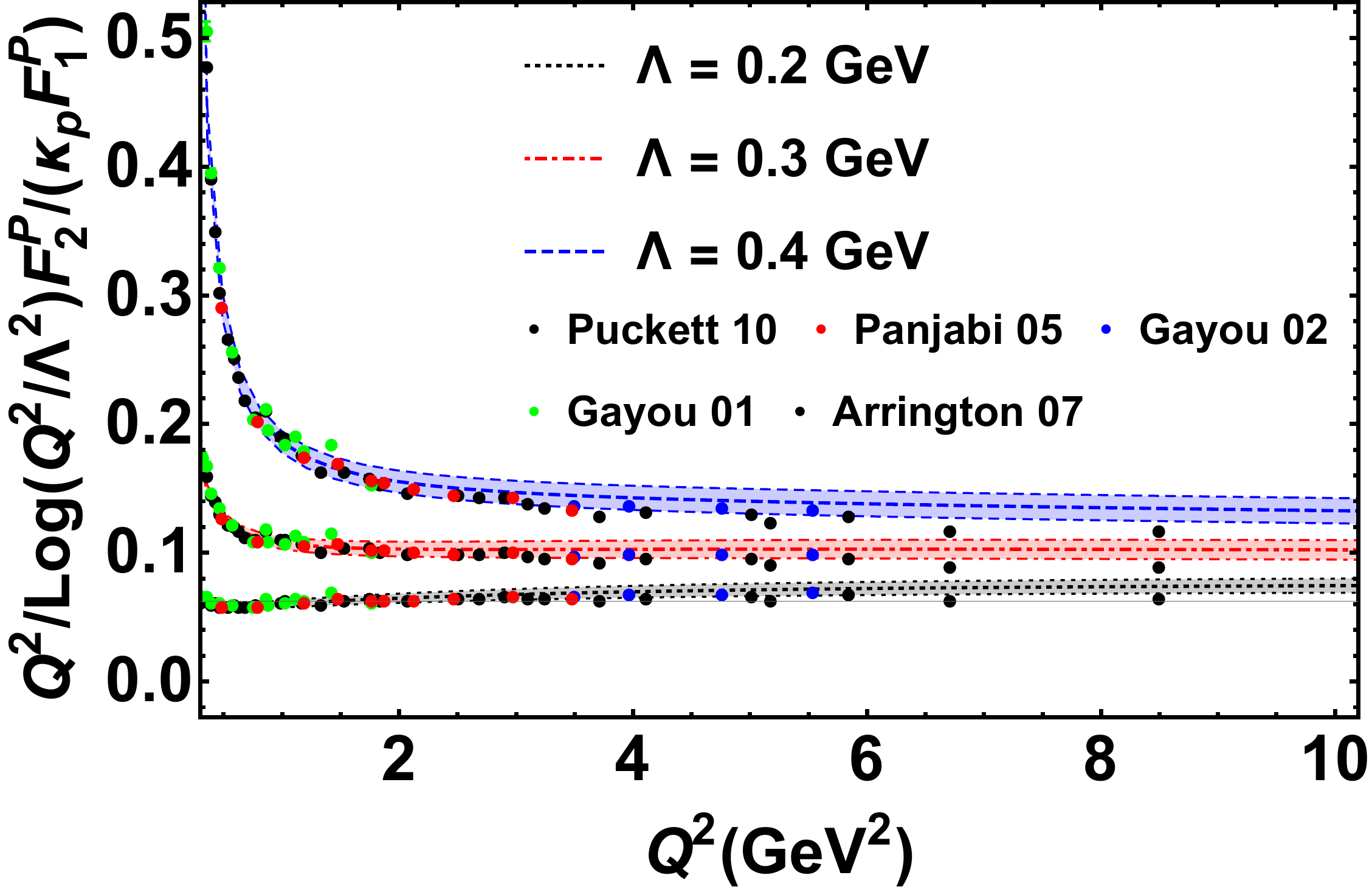}
\end{minipage}
\label{p2}
}
\caption{The ratio of the
proton Pauli form factor $F_2^p$ and the Dirac form factor $F_1^p$; (a) $Q^2 F_2^p/(\kappa_p F_1^p)$ and (b) $(Q^2/{\rm Log}^2[Q^2/\Lambda^2]) F_2^p/(\kappa_p F_1^p)$ for three different values of $\Lambda=0.2,~0.3$ and $0.4$ GeV. The data are taken from Refs.~\cite{Gayou:2001qt,Jones:1999rz,Gayou:2001qd,Punjabi:2005wq,Puckett:2010ac}.}
\label{proton_ratio}
\end{figure*}

We compute the Dirac and the Pauli FFs for up and down quarks using the LFWFs defined in Eq.~(\ref{wavefunctions}). The flavor FFs are shown in Fig.~\ref{flavor}, where we compare the BLFQ results with the extracted data obtained from the decomposition of the nucleon FFs adopting isospin symmetry~\cite{Cates:2011pz,Qattan:2012zf,Diehl:2013xca}. 
%
The flavor Dirac FFs within the proton agree well with the extracted data. However, the Pauli FFs are seen to deviate from the data at the low $Q^2$ region by amounts that are larger for the down quark than the up quark. It should be noted that these results are obtained within leading Fock representation, while the higher Fock components $|qqqg\rangle$ and $|qqqq\bar{q}\rangle$ are anticipated to have significant effects on the Pauli FFs \cite{Sufian:2016hwn}. With the inclusion of dynamical gluons and sea quarks, the quark spin contribution may be suppressed, and the orbital angular momentum can play an enhanced role and tune the S and P wave contributions that could increase the Pauli FFs of the down and up quarks.
%
%
In our approach, we obtain the anomalous magnetic moment of the up (down) and the down (up) quark in the proton (neutron), $\kappa_u=1.481\pm 0.029$ and $\kappa_d=-1.367\pm 0.025$ (in units of the nuclear magneton, $\mu_{\rm N}$), respectively, whereas the extracted values from the experimental data of the nucleon anomalous magnetic moments are: $\kappa_u^{\rm exp}=2\kappa_p+\kappa_n=1.673$ and $\kappa_d^{\rm exp}=\kappa_p+2\kappa_n=-2.033$~\cite{Cates:2011pz}.

We evaluate the nucleon Sach's FFs, which are expressed in terms of Dirac and Pauli FFs as
\begin{align}
G_{\rm E}^{\rm N}(Q^2)=&F_1^{{\rm N}}(Q^2) - \frac{Q^2}{4M^2} F_2^{{\rm N}}(Q^2); \\
G_{\rm M}^{\rm N}(Q^2)=&F_1^{{\rm N}}(Q^2) + F_2^{\rm N}(Q^2),
\end{align}
where $F_{1(2)}^{\rm N}=\sum_{\rm f} e_{\rm f} F_{1(2)}^{{\rm f}}$ is the Dirac (Pauli) FF of the nucleon. Here, ${\rm N}$ stands for the nucleon and $e_{\rm f}$ symbolizes the quark charge of flavor ${\rm f}$. The electric and the magnetic Sach's FFs of the proton are shown in Figs.~\ref{Gep} and \ref{Gmp}. Overall, we find good agreement between our approach and experiment for the proton electric FF. The magnetic FF of the proton is also in reasonable agreement with the data at large $Q^2$, while at low $Q^2$ this FF shows a small deviation from the data. 
The neutron Sach's FFs are displayed in Figs.~\ref{Gen} and \ref{Gmn}. We find that at large $Q^2$ regime, the neutron magnetic FF is consistent with experimental data while, at low $Q^2$, the magnitude of this FF falls below the data. The prediction for
the neutron's charge FF falls well below the data at low $Q^2$ where both experimental and theoretical uncertainties are large.  
The discrepancies between theory and experiment for the neutron FFs are not too surprising because
the down quark FF, $F_2^{\rm d}$ (Fig.~\ref{flavor}) shows a deviation from the data in this model.

Another critical measurement is the ratio of the nucleon Sach's FFs. As remarked before, there are inconsistencies in the extraction of the data for the proton electric to magnetic Sachs FF ratio. The ratio decreases almost linearly as $Q^2$ increases 
($>0.5\, \text{GeV}^2$) in double polarization experiments~\cite{Gao:2000ne,Punjabi:2005wq,Gayou:2001qd,Puckett:2011xg}, while the results obtained from the Rosenbluth separation method~\cite{Hand:1963zz,Janssens:1965kd,Price:1971zk,Litt:1969my,Berger:1971kr,Bartel:1973rf,Borkowski:1974mb,Simon:1980hu,Walker:1993vj,Andivahis:1994rq,Christy:2004rc,Qattan:2004ht} remain constant in the spacelike region. Meanwhile, the data for the neutron indicate that the ratio increases more or less linearly with increasing $Q^2$. We compare our results for the ratio of the nucleon Sach's FFs: $R_{\rm N} = \mu_{\rm N} G^{\rm N}_{\rm E}/G^{\rm N}_{\rm M}$  with the experimental data in Fig~\ref{ratio_sach}. We notice that the ratio for the proton in our approach follows the trend of the data from double polarization experiments. 
The ratio for the neutron also exhibits the trend of the measured data, but the slope and magnitude are somewhat smaller than the data.

Experimental results on the proton Dirac form factor $F_1^p$~\cite{Arnold:1986nq,Sill:1992qw} have been found to be in reasonable agreement with a scaling prediction based on pQCD, $F_1^p\propto 1/Q^4$~\cite{Lepage:1979za}. However, it has been argued that pQCD is not applicable for exclusive processes at experimentally accessible values of $Q^2$~\cite{Isgur:1989cy}. Indeed, it has been noticed that the experimental data from Jefferson Lab~\cite{Gayou:2001qt,Jones:1999rz,Gayou:2001qd,Punjabi:2005wq,Puckett:2010ac} on the ratio, $F_2^p/F_1^p$ disagree with the suggested scaling $F_2^p/F_1^p \propto 1/Q^2$~\cite{Lepage:1979za}. Meanwhile, these same data  have been found to be in fair agreement with an updated pQCD prediction, $Q^2 F_2^p/F_1^p \propto {\rm Log}^2[Q^2/\Lambda^2]$~\cite{Belitsky:2002kj} even at large $Q^2$. Here, $\Lambda$ represents the QCD scale parameter. The scaling of $F_2^p/F_1^p$ in our approach is illustrated in Fig.~\ref{proton_ratio}. We observe that our result agrees reasonably well with the updated pQCD prediction, $Q^2 F_2^p/F_1^p \propto {\rm Log}^2[Q^2/\Lambda^2]$~\cite{Belitsky:2002kj}.

The magnetic moment of the nucleon is related to the nucleon magnetic Sach's FF at $Q^2=0$. 
In our approach, expressed in units of nuclear magnetons, we obtain the magnetic moment of the proton the neutron close to the recent lattice QCD results as shown in Table~\ref{tab:mu}~\cite{Alexandrou:2018sjm}. On the other hand, the experimental value of the magnetic moments is larger in magnitude by $11 \%$ for $\mu_p$ and by $36 \%$ for $\mu_n$~\cite{Tanabashi:2018oca}.

From the Sachs FFs, we can also
compute the electromagnetic radii of the nucleon, which are defined by
\begin{align}
\braket{r^2_{\rm E}}^{\rm N}=&-6 \frac{{\rm d}G_{\rm E}^{\rm N}(Q^2)}{{\rm d}Q^2}\bigg|_{Q^2=0}, \\
\braket{r^2_{\rm M}}^{\rm N}=&-\frac{6}{G_{\rm M}^{\rm N}(0)}\frac{{\rm d}G_{\rm M}^{\rm N}(Q^2)}{{\rm d}Q^2}\bigg|_{Q^2=0}.
\end{align}
The radii are presented in Table~\ref{tab:radii}. We compare the BLFQ results with the measured data~\cite{Tanabashi:2018oca} and the recent lattice results~\cite{Alexandrou:2018sjm}. We find a reasonable agreement with experiment within our uncertainties stemming from our uncertainty in $\alpha_s$.
\begin{table}[htp]	
\centering
\caption{The magnetic moment of the proton and the neutron in units of nuclear magnetons. We compare our results with the experimental data~\cite{Tanabashi:2018oca} and with lattice results~\cite{Alexandrou:2018sjm}.}
\label{tab:mu}
\begin{tabular}{|cccc|cc}
\hline\hline
Quantity  	   					~& BLFQ	          			~& Measurement\footnote{The uncertainty in the measured data of the proton and the neutron magnetic moments are ($8.2\times 10^{-10}$) and ($4.5\times 10^{-7}$), respectively.}       			~& Lattice   		\\
\hline
$\mu_{\rm{p}}$    		~& $2.443\pm0.027$	~& $2.79$		  	~& $2.43(9)$		\\
\hline
$\mu_{\rm{n}}$   ~& $-1.405\pm0.026$	  		~& $-1.91$ 	  	~& $-1.54(6)$		\\
\hline\hline
\end{tabular}
\end{table}
\begin{table}[htp]	
\centering
\caption{The electromagnetic radii of the proton and the neutron. We compare our results with the experimental data~\cite{Tanabashi:2018oca} and with lattice results~\cite{Alexandrou:2018sjm}.}
\label{tab:radii}
\begin{tabular}{|cccc|cc}
\hline\hline
Quantity  	   					& BLFQ	          			& Measurement       			& Lattice   		\\
\hline
$r^{\rm{p}}_{\rm E} ~[\rm{fm}]$    		& $0.802^{+0.042}_{-0.040}$	& $0.833\pm 0.010$		  	& $0.742(13)$		\\
$r^{\rm{p}}_{\rm M} ~[\rm{fm}]$    		& $0.834^{+0.029}_{-0.029}$	& $0.851\pm 0.026$  		& $0.710(26)$		\\
\hline
$\braket{r^2_{\rm E}}^{\rm{n}} ~[\rm{fm}^2]$   & $-0.033\pm 0.198$	  		& $-0.1161\pm 0.0022$ 	  	& $-0.074(16)$		\\
$r^{\rm{n}}_{\rm M} ~[\rm{fm}]$    		& $0.861^{+0.021}_{-0.019}$	& $0.864^{+0.009}_{-0.008}$ & $0.716(29)$		\\
\hline\hline
\end{tabular}
\end{table}

\subsection{Axial form factor}
Another important quantity in our understanding of the nucleon structure is the nucleon isovector axial FF. At zero momentum transfer, the axial FF defines the nucleon axial charge $g_A$. The momentum-transfer dependence of the axial FF is relevant to experimental processes, for example, elastic neutrino-nucleon scattering. The precise understanding of such processes is essential to obtain the accuracy goals in resolving neutrino-oscillation parameters~\cite{Alvarez-Ruso:2017oui}. The isovector axial current of light quarks in QCD, 
$
A^\mu_a(z)=\bar{\psi}_q(z)\gamma^{\mu}\gamma^5\frac{\tau^a}{2}\psi_q(z)
$,
with $\tau^a$ ($a\equiv 1,~2,~3$) being the Pauli isospin matrices, is used to define the axial FF. In the isospin symmetry limit, the matrix element of this current between the nucleon states is parameterized as
\begin{align}
&\braket{P+q, \Lambda'|A^\mu_a(0)|P, \Lambda}
=\bar{u}(P+q, \Lambda')\big[\gamma^{\mu}G_{\rm A}(Q^2)\nonumber\\&\quad\quad\quad\quad\quad+\frac{q^{\mu}}{2M}G_{\rm P}(Q^2)\big]\gamma^5\frac{\tau^a}{2}u(P,\Lambda),
\end{align}
where $G_{\rm A}(Q^2)$ and $G_{\rm P}(Q^2)$ are called the nucleon axial and induced pseudoscalar FFs, respectively. The axial FF, which is our current focus, can be expanded in the regime of small $Q^2$ as
\begin{align}
G_{\rm A}(Q^2)=g_{\rm A}\Big[1+\frac{1}{6}\langle r_{\rm A}^2\rangle Q^2+\mathcal{O}(Q^4)\Big]\ ,
\end{align}
with $g_{\rm A} \equiv G_{\rm A}(0)$ and $\langle r_{\rm A}^2\rangle$ being the axial-vector charge and axial radius squared, respectively. 

In the light-front formalism, similar to the electromagnetic FFs, we obtain the axial FF in terms of the overlap of the LFWFs using the plus component of $A^\mu_a(0)$ and the nucleon state given in Eq.~(\ref{wavefunction_expansion})
\begin{align}
G^q_{\rm A}(Q^2)=&\sum_{\{\lambda_i\}} \int \left[{\rm d}\mathcal{X} \,{\rm d}\mathcal{P}_\perp\right]\nonumber\\&\times  \lambda_1 \,\Psi^{\uparrow *}_{\{x_i,\vec{p}^{\prime}_{i\perp},\lambda_i\}}\Psi^{\uparrow}_{\{x_i,\vec{p}_{i\perp},\lambda_i\}}.
\end{align}
Here $\lambda_1=1(-1)$ for the struck quark helicity. The $G^q_{\rm A}(Q^2)$ is the flavor axial FF. 

\begin{figure}[htbp]
\centering
\includegraphics[width=\columnwidth]{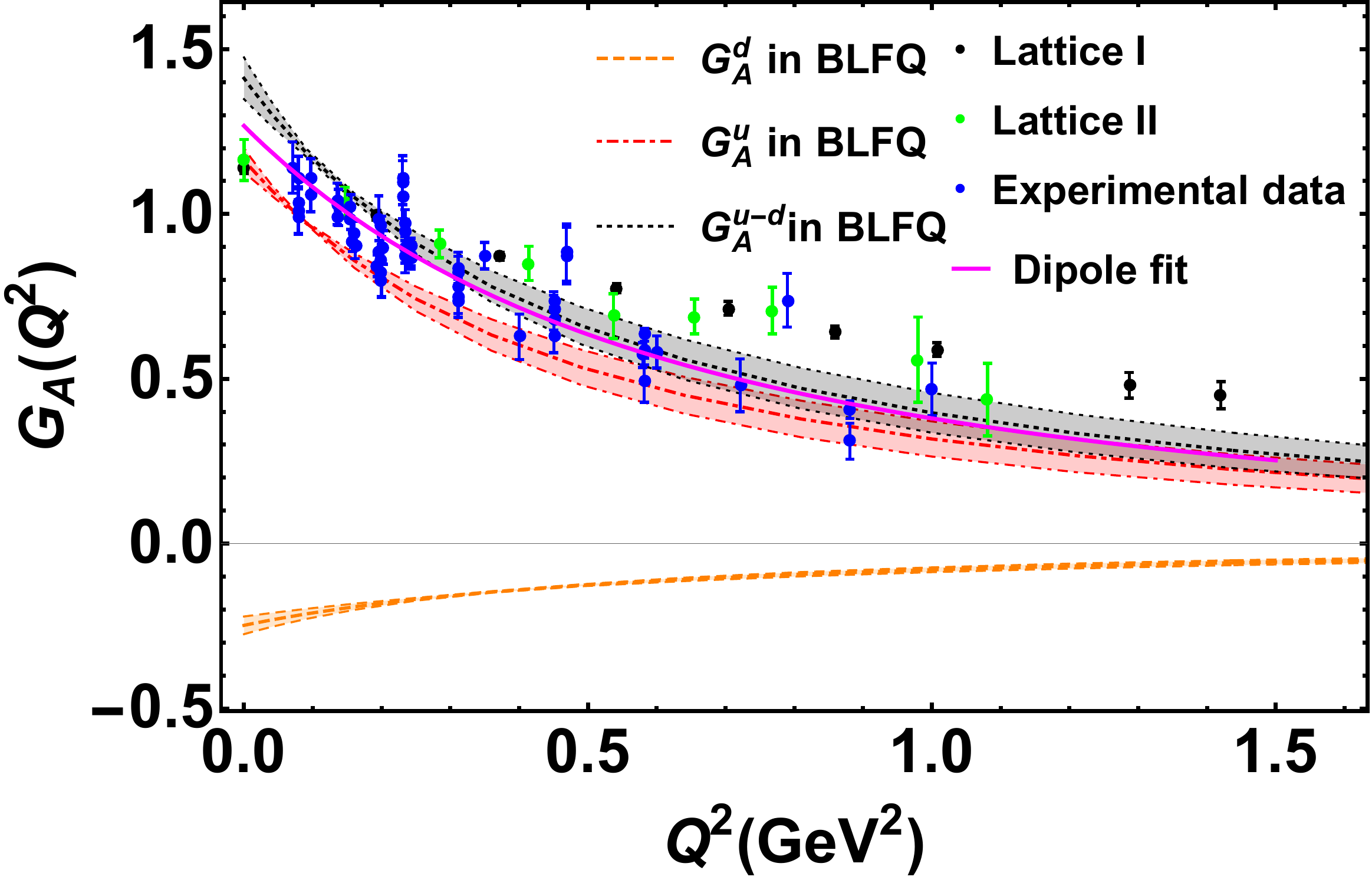}
\caption{ The axial form factors $G_{\rm A}=G_{\rm A}^u-G_A^d$ and $G_{\rm A}^u$, $G_{\rm A}^d$ as the function of $Q^2$. The gray band ($G_{\rm A}$), red band ($G_{\rm A}^u$), and orange band ($G_{\rm A}^d$) are the BLFQ results. The extracted data are taken from the Refs.~\cite{Bernard:2001rs,Hashamipour:2019pgy} and the lattice calculations from ~\cite{Alexandrou:2013joa}. The magenta line represents the dipole fit to the experimental data~\cite{Bernard:2001rs}.}
\label{AFF}
\end{figure}

In Fig.~\ref{AFF}, we show the axial FF, $G_{\rm A}=G_{\rm A}^u-G_{\rm A}^d$ as a function of $Q^2$, while the contribution from up and down quark in $G_{\rm A}(Q^2)$ is also presented. We compare our results with the available data from (anti)neutrino scattering off protons or nuclei and charged pion electroproduction experiments~\cite{Bernard:2001rs,Schindler:2006jq,Hashamipour:2019pgy} and the lattice result from the Ref.~\cite{Alexandrou:2013joa}.
Considering the experimental uncertainties and our current treatment of the BLFQ uncertainties, we find good agreement with experiment. Note that the experimental data can be well-described by the following dipole ansatz:
\begin{align}
G_{\rm A}(Q^2) = \frac{g_{\rm A}}{(1+Q^2/M_{\rm A}^2)^2}\,,\label{dipole_fit}
\end{align}
with $g_{\rm A}=1.2673\pm0.0035$ and the axial mass $M_{\rm A}=1.1~\rm{GeV}$~\cite{Bernard:2001rs}. We notice that our result is compatible with the dipole fit. 

At $Q^2=0$, the axial FF is identified as the axial charge, $g_{\rm A}=G_{\rm A}(0)$. Experimentally, the value of $g_A$ is very accurately determined through neutron $\beta$-decay, $g_{\rm A}=1.2723\pm 0.0023$~\cite{Tanabashi:2018oca}. We obtain $g_{\rm A}=1.41\pm 0.06$, somewhat larger than the extracted data and  the lattice QCD prediction as quoted in Table~\ref{tab:AFF}. The nucleon axial charge describes the difference of the spins carried by the up and down quarks in the proton. For the up quark, our prediction is overestimated, while for the down quark the value is underestimated in contrast with the extracted data and with lattice simulations. Summing over the flavors, we get the total spin contributed by the quarks to the proton spin. Within our model that includes only the leading Fock sector, we find that the quark spin contributes $\sim 91\%$ to the proton spin, while the contribution of quark spin occupies only $\sim 40\%$ as reported from the experiment~\cite{Leader:2010rb}. This evident discrepancy points to the need to extend our model to include the higher Fock sectors which have a significant effect on the quark contribution to the nucleon spin. With dynamical gluons and sea quarks, the quark spin contribution can be diminished and the orbital angular momentum can play a larger role in understanding the nucleon spin. At the same time, the gluon and sea quark contributions to the total spin will emerge.

\begin{figure*}[htbp]
  \centering
  \subfigure[]{
  \begin{minipage}[t]{0.45\linewidth}
  \centering
  \includegraphics[width=\columnwidth]{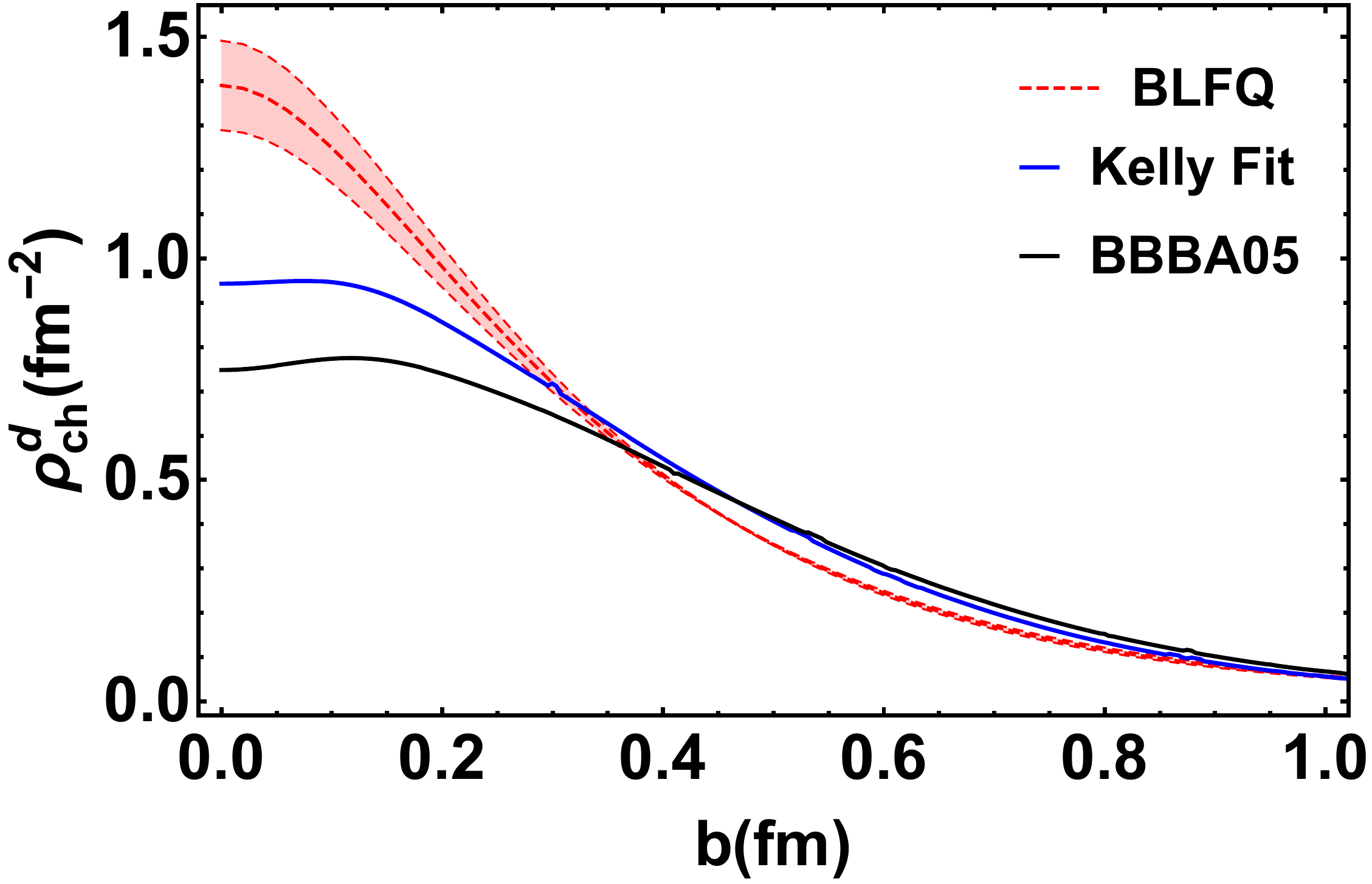}
  \end{minipage}
  \label{density_d_charge}
  }
  \subfigure[]{
  \begin{minipage}[t]{0.45\linewidth}
  \centering
  \includegraphics[width=\columnwidth]{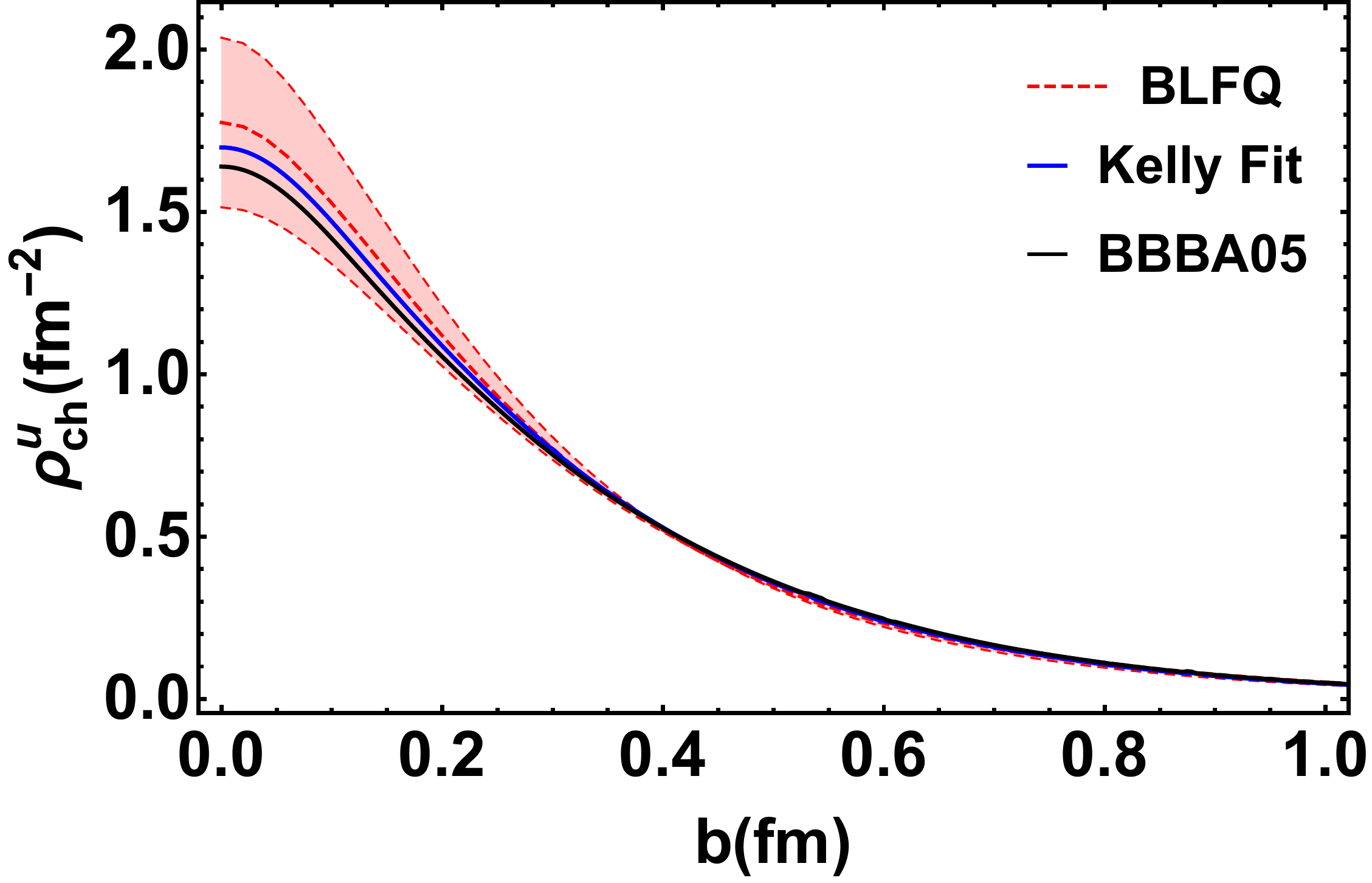}
  \end{minipage}
  \label{density_u_charge}
  }
  \subfigure[]{
  \begin{minipage}[t]{0.45\linewidth}
  \centering
  \includegraphics[width=\columnwidth]{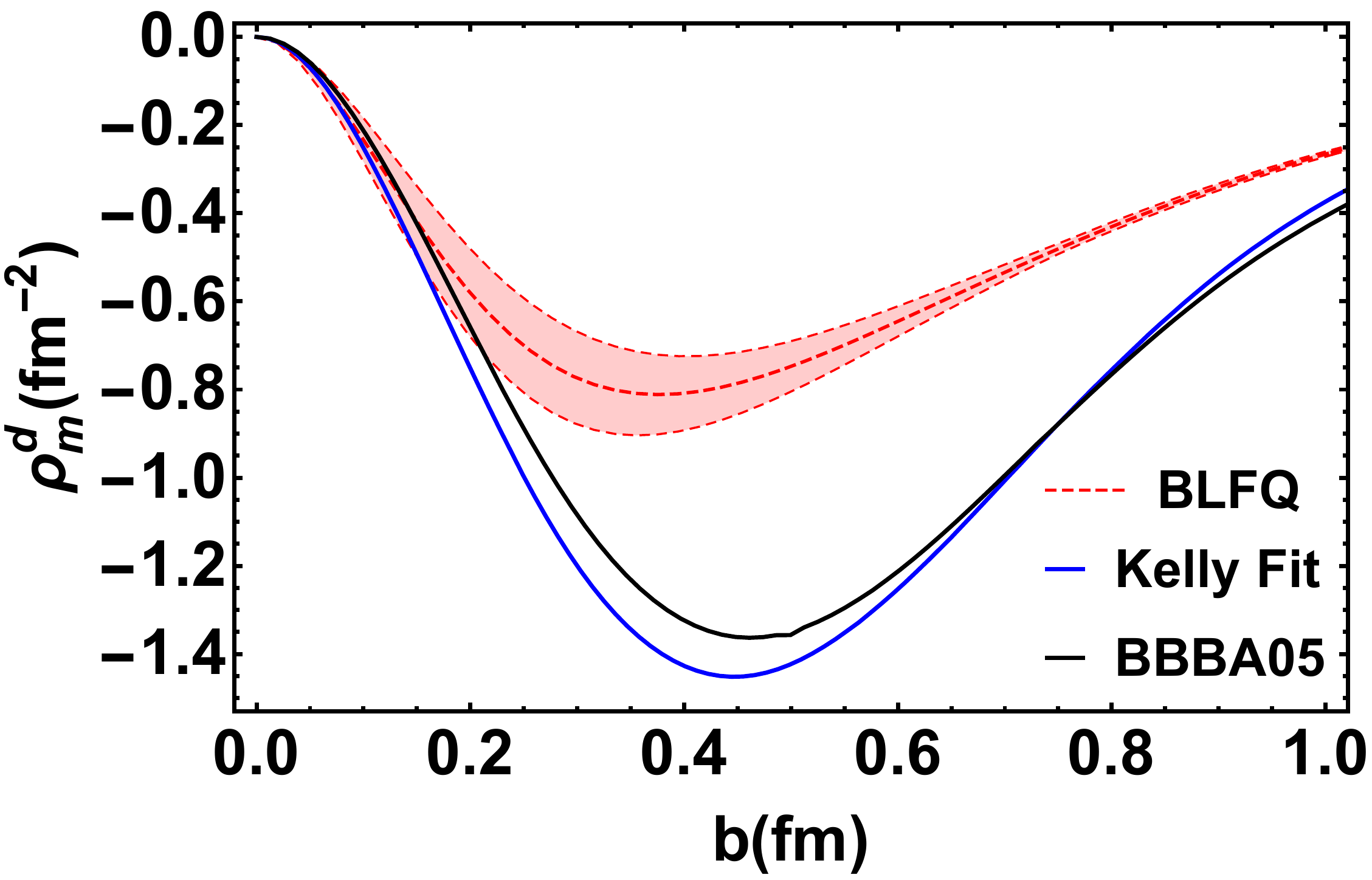}
  \end{minipage}
  \label{density_d_mag}
  }
  \subfigure[]{
  \begin{minipage}[t]{0.45\linewidth}
  \centering
  \includegraphics[width=\columnwidth]{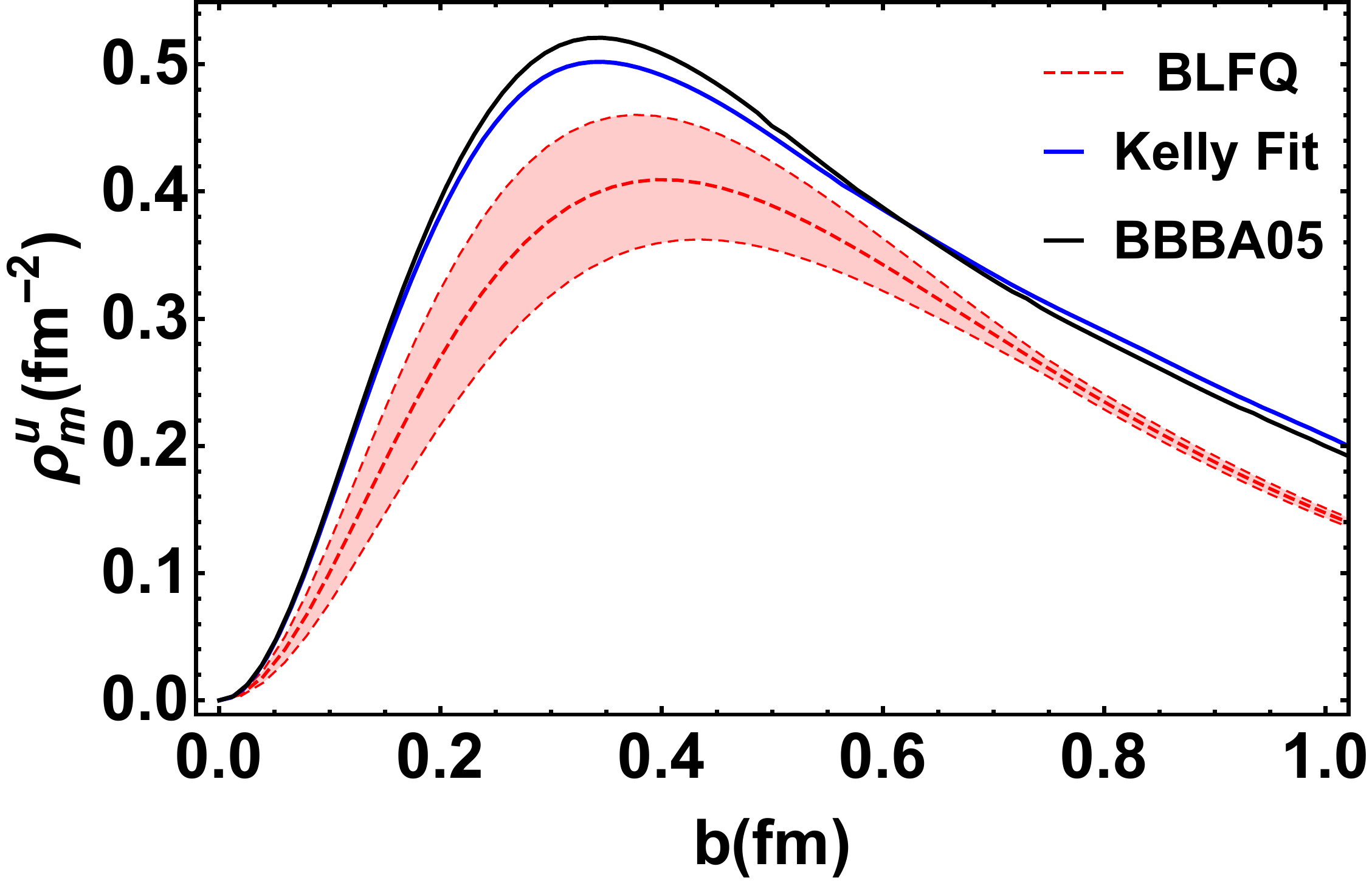}
  \end{minipage}
  \label{density_u_mag}
  }
  \caption{Quark transverse charge and anomalous magnetization densities for the unpolarized nucleon: (a),(c) for the down quark; (b),(d) for the up quark. The red bands correspond to the uncertainty range of the BLFQ results. The blue and black lines represent the parameterizations of Kelly~\cite{Kelly:2004hm}, and Bradford {\it et al}.~\cite{Bradford:2006yz}, respectively. }
  \label{density_quark}
\end{figure*}
\begin{figure*}[htbp]
  \centering
  \subfigure[]{
  \begin{minipage}[t]{0.45\linewidth}
  \centering
  \includegraphics[width=\columnwidth]{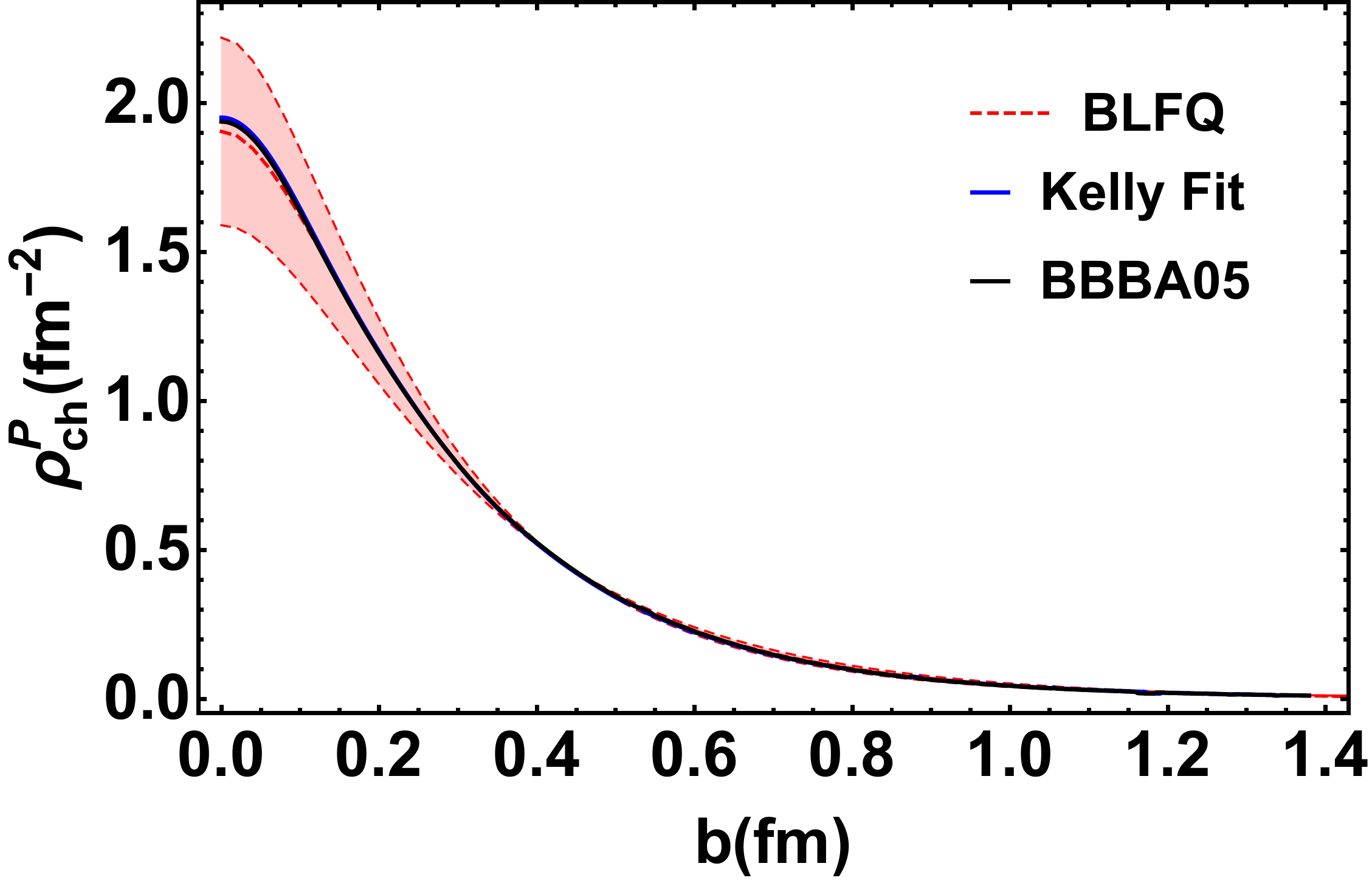}
  \end{minipage}
  \label{density_P_charge}
  }
  \subfigure[]{
  \begin{minipage}[t]{0.45\linewidth}
  \centering
  \includegraphics[width=\columnwidth]{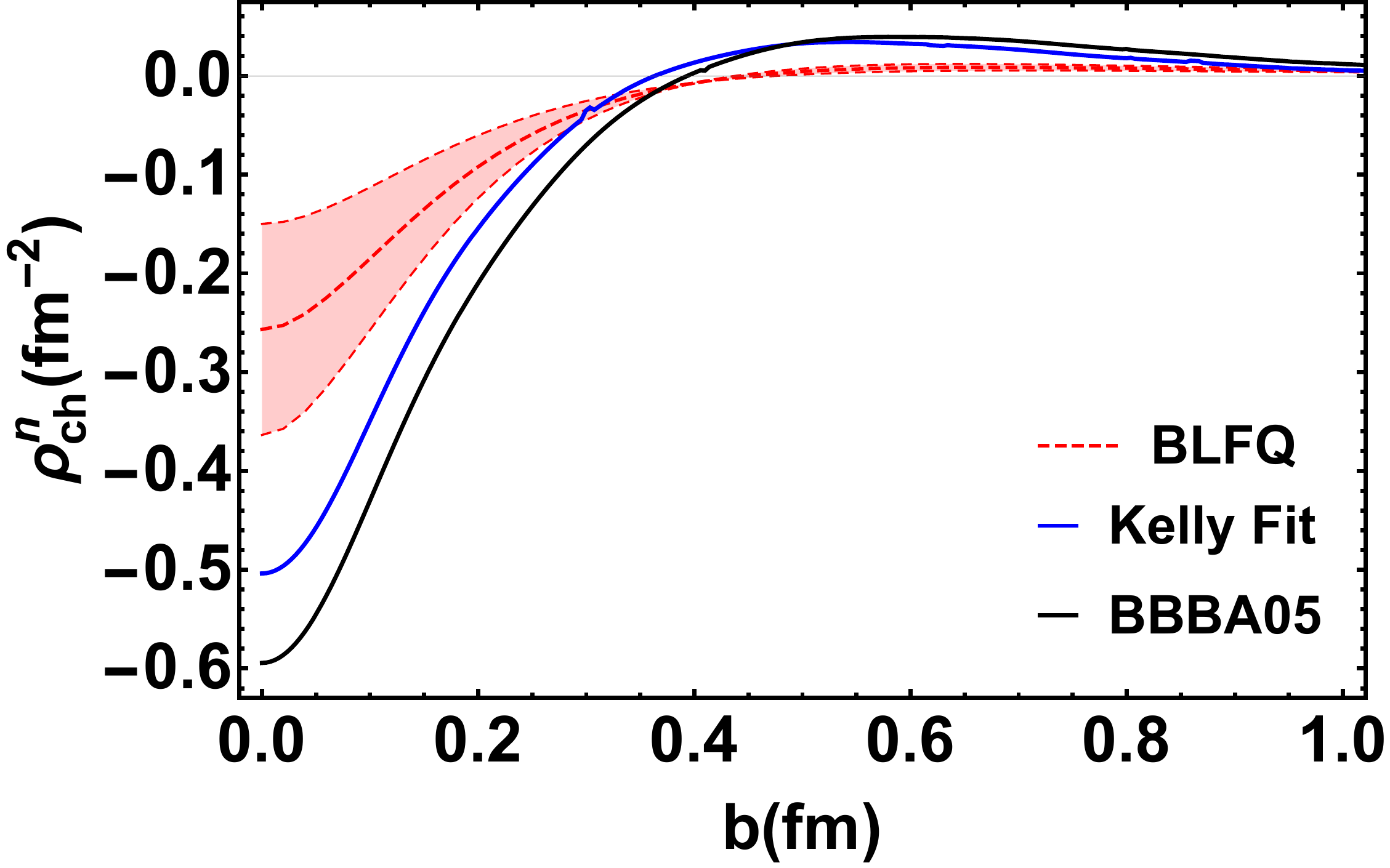}
  \end{minipage}
  \label{density_N_charge}
  }
  \subfigure[]{
  \begin{minipage}[t]{0.45\linewidth}
  \centering
  \includegraphics[width=\columnwidth]{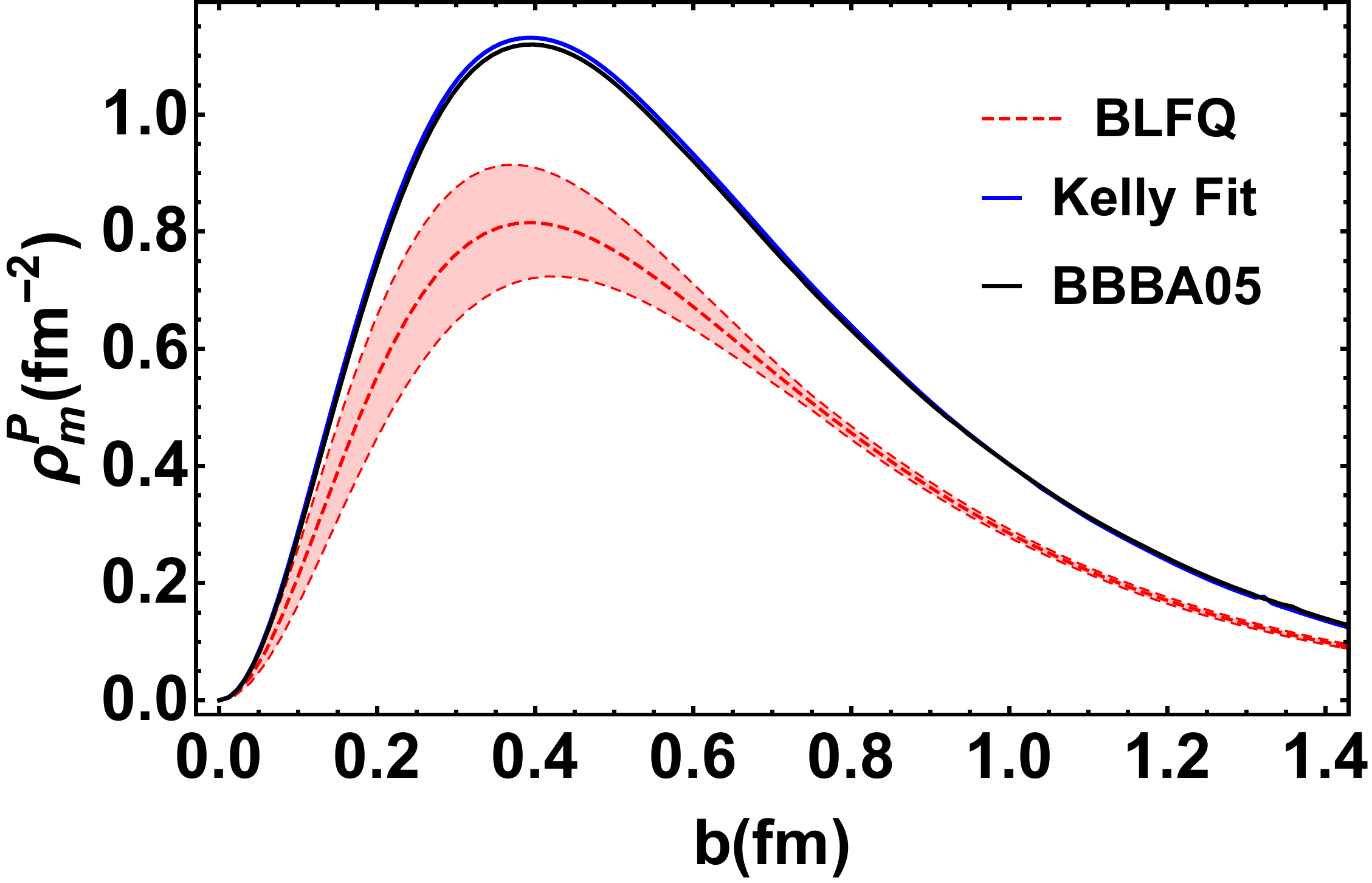}
  \end{minipage}
  \label{density_P_mag}
  }
  \subfigure[]{
  \begin{minipage}[t]{0.45\linewidth}
  \centering
  \includegraphics[width=\columnwidth]{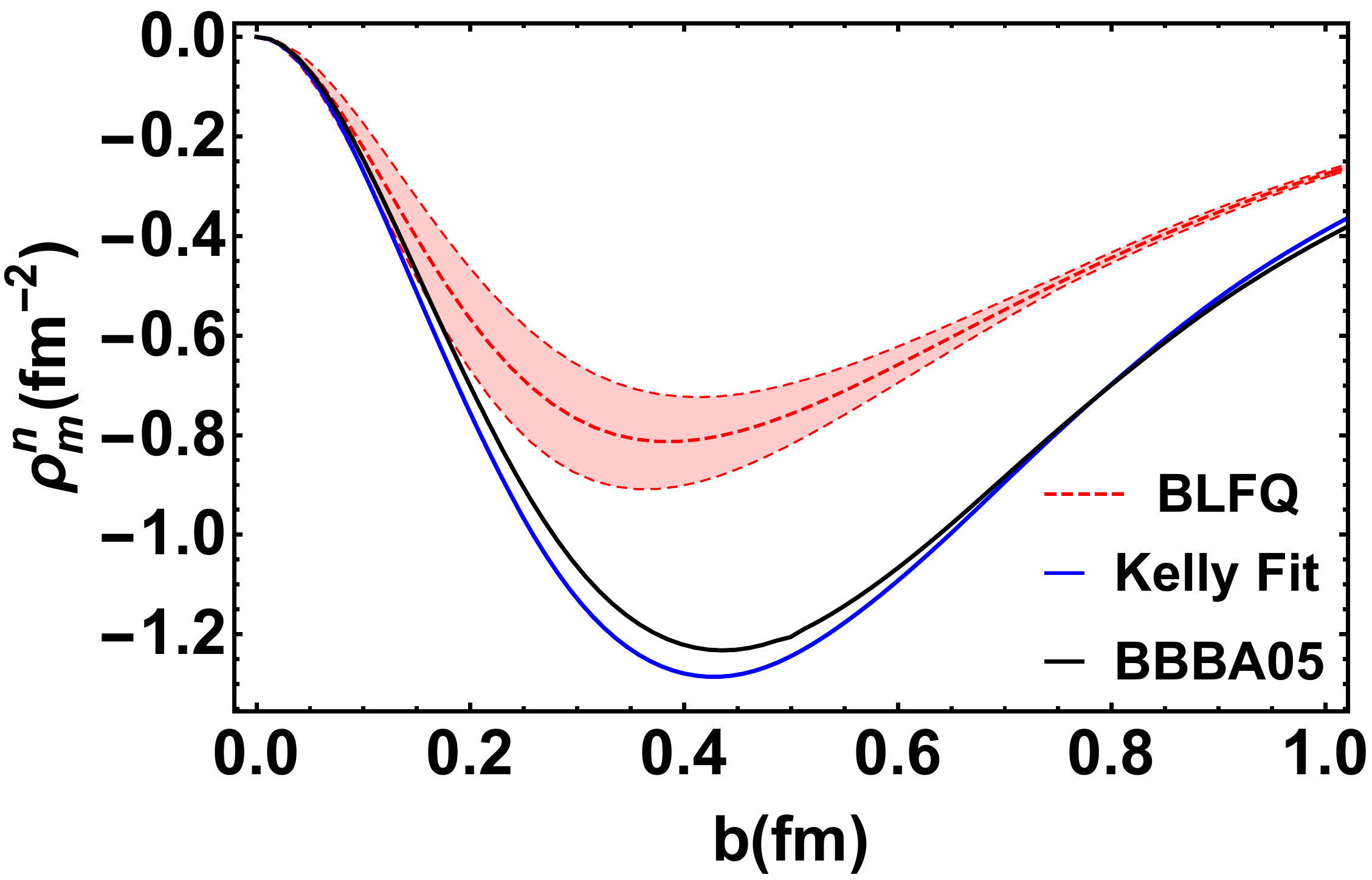}
  \end{minipage}
  \label{density_N_mag}
  }
  \caption{The nucleon transverse charge and anomalous magnetization densities for the unpolarized nucleon. (a),(c) for the proton.(b),(d) for the neutron. The red bands correspond to the uncertainty range of the BLFQ results. The blue and black lines represent the parameterizations of Kelly~\cite{Kelly:2004hm}, and Bradford {\it et al}.~\cite{Bradford:2006yz}, respectively. }
  \label{density_nucleon}
  \end{figure*}

%
\begin{table}[htp]	
\centering
\caption{The axial charge and axial radius. Our results are compared with the extracted data~\cite{Tanabashi:2018oca,Leader:2010rb,Hill:2017wgb} and recent lattice QCD calculation~\cite{Yao:2017fym,Alexandrou:2017oeh}.}
\label{tab:AFF}
\begin{tabular}{|cccc|cc}
\hline\hline
Quantity  	   	& BLFQ	          			& Extracted data       	& Lattice   	\\
\hline
$g_{\rm A}^u $    	& $1.16\pm 0.04$	  			& $0.82\pm 0.07$		    	~&~ $0.830(26)$	\\
$g_{\rm A}^d $    	& $-0.248\pm 0.027$	  		& $-0.45\pm 0.07$				~&~ $-0.386(16)$	\\
$g_{\rm A}^{u-d} $    & $1.41\pm 0.06$	  		  	& $1.2723\pm 0.0023$			~&~ $1.237(74)$	\\
\hline
$\sqrt{\braket{r_{\rm A}^2}} ~\rm{fm}$  & $0.680^{+0.070}_{-0.073}$	& $0.667\pm 0.12$  			& $0.512(34)$	\\
\hline\hline
\end{tabular}
\end{table}

We also calculate the axial radius which is defined as
\begin{align}
\braket{r_{\rm A}^2}=\frac{6}{g_{\rm A}}\frac{{\rm d}G_{\rm A}(Q^2)}{{\rm d}Q^2}\big|_{Q^2=0}.
\end{align}
We present the result of $\sqrt{\braket{r_A^2}}$ in Table~\ref{tab:AFF}. We find that the BLFQ result shows an excellent agreement with the extracted data from the analysis of neutrino-nucleon scattering experiments~\cite{Hill:2017wgb,Meyer:2016oeg}. It appears that the good agreement of this scalar observable, in contrast with the spin-sensitive observables, suggests the important role of the neglected higher Fock sectors will likely be seen more dramatically in those spin-sensitive observables.
\subsection{Transverse charge and magnetization densities}
Electric charge and magnetization densities in the transverse plane also give insights into the structure of the nucleon.
The charge density in the transverse plane of an unpolarized nucleon is defined as the two-dimensional Fourier transform of the Dirac FF~\cite{Miller:2007uy,Mondal:charge,Carlson:2007xd}
\begin{align}
\rho_{\rm fch} (b_{\perp}) = \int \frac{{\rm d}^2\vec{ q}_{\perp}}{(2\pi)^2} F_1(Q^2)e^{i\vec{ q}_{\perp}\cdot \vec{b}_{\perp}},
\end{align}
where $b_{\perp}=|\vec{b}_{\perp}|$ represents the impact parameter. It is worthwhile to mention that one has to distinguish the quark and the flavor charge densities, which we label as $\rho_{\rm ch}^q (b_{\perp})$ and $\rho_{\rm fch}^q (b_{\perp})$, respectively. The quark charge density is defined as the Fourier transform of a single quark Dirac FF in the nucleon. Due to the charge and isospin symmetry, the up and
down quark densities in the proton are the same as the down and up quark densities in the neutron. Under this symmetry, one has $\rho_{\rm ch}^d=\rho_{\rm fch}^d$ and $\rho_{\rm ch}^u=\rho_{\rm fch}^u/2$~\cite{Miller:2007uy,Mondal:charge}. Similarly, the magnetization density is defined as the Fourier transform of the Pauli FF,
\begin{align}
  \tilde{\rho}_{\rm fm} (b_{\perp}) = \int \frac{{\rm d}^2\vec{ q}_{\perp}}{(2\pi)^2} F_2(Q^2)e^{i\vec{ q}_{\perp}\cdot \vec{b}_{\perp}},
\end{align}
while the anomalous magnetization density is given by
\begin{align}
  \rho_{\rm fm} (b_{\perp}) = - b_{\perp} \frac{\partial \tilde{\rho}_{\rm fm}}{\partial b_{\perp}}.
\end{align}

The decomposition of the nucleon transverse densities can be defined in a similar fashion as the electromagnetic FFs. In terms of the two flavors, the nucleon charge/magnetization density can be decomposed as
\begin{align}
  \rho_{\rm ch/m}^{\rm p} (b_{\perp}) &= e_u \rho_{\rm fch/fm}^u (b_{\perp}) + e_d \rho_{\rm fch/fm}^d (b_{\perp}), \nonumber\\
  \rho_{\rm ch/m}^{\rm n} (b_{\perp}) &= e_u \rho_{\rm fch/fm}^d (b_{\perp}) + e_d \rho_{\rm fch/fm}^u (b_{\perp}),
\end{align}
where $e_u$ and $e_d$ are the charge of the up and down quarks. We can also express the nucleon transverse densities in the quark densities representation as
\begin{align}
  \rho_{\rm ch/m}^{\rm p} (b_{\perp}) &= 2e_u \rho_{\rm ch/m}^u (b_{\perp}) + e_d \rho_{\rm ch/m}^d (b_{\perp}), \nonumber\\
  \rho_{\rm ch/m}^{\rm n} (b_{\perp}) &= e_u \rho_{\rm ch/m}^d (b_{\perp}) + 2e_d \rho_{\rm ch/m}^u (b_{\perp}).
\end{align}
%
Note that these quantities are not directly measured in experiments. Meanwhile, an estimation of the nucleon charge and magnetization densities has been done from experimental data of electromagnetic FFs in Ref.~\cite{Kelly:2002if}. To gain an insight into the transverse structure of the nucleon, we evaluate the charge and anomalous magnetization densities within the BLFQ framework and compare with the
two different global parameterizations proposed by Kelly~\cite{Kelly:2004hm} and by Bradford \textit {et al.}~\cite{Bradford:2006yz}.

We show the charge and magnetization densities of the individual quarks in Fig~\ref{density_quark}. We observe that the up quark charge density is in good agreement with the global parameterizations of
Kelly and Bradford \textit {et al.}. However, the charge density for the down quark from BLFQ
deviates at small $b$ from those parameterizations. The qualitative behavior of the anomalous magnetization densities for both the quarks tracks the parameterizations but our results are smaller in magnitude. This may, as discussed above, be attributed to the neglected higher Fock sectors that may make important contributions to the magnetic FFs~\cite{Sufian:2016hwn}.

The transverse densities of the nucleon are presented in Fig.~\ref{density_nucleon}. We find that the proton charge density in our approach shows an excellent agreement with the global parameterizations. However, our current treatment predicts an insufficient magnitude for the neutron transverse charge density. Having said that, it is worth noting that the qualitative behavior of the neutron charge density agrees with the parameterizations. Due to the similar reason as mentioned before for the quarks, the anomalous magnetization densities for the nucleon in our model deviate from the global parameterizations.

\subsection{Parton distribution functions}
%
The PDF, the probability that a parton carries a certain fraction of the total light-front longitudinal momentum of a hadron, gives us information about the nonperturbative structure of hadrons. 
The quark PDF of the nucleon, which encodes the distribution of longitudinal momentum and polarization carried by the quark in the nucleon, is defined as
\begin{align}\label{defi_pdf}
\Phi^{\Gamma(q)}(x)=&\frac{1}{2}\int \frac{d z^-}{4\pi} e^{ip^+z^-/2}\nonumber\\
&\times \langle P,\Lambda|\bar{\psi}_{q}(0)\Gamma\psi_{q}(z^-)|P, \Lambda\rangle \bigg|_{z^+=\vec{z}_\perp=0}.
\end{align}
For different Dirac structures, we get different quark PDFs of the nucleon. For example, for $\Gamma=\gamma^+, \gamma^+ \gamma^5, i\sigma^{j+}\gamma^5$ we have the unpolarized PDF $f(x)$, helicity distribution $g_1(x)$ and transversity distribution $h_1(x)$ respectively. It is worth noting that we work in the light-front gauge $A^+ = 0$, so that the gauge link appearing in between the quark fields in Eq.~(\ref{defi_pdf}) is unity. We compute the quark PDFs from the eigenstates of our light front effective Hamiltonian, Eq.~(\ref{hami}), in the constituent valence quarks representation suitable for low-momentum scale applications.
\\

\subsubsection{Unpolarized PDFs and QCD evolution}
%
Following the two-point correlation function defined in Eq.~(\ref{defi_pdf}), 
in the LFWFs overlap representation, the
unpolarized PDFs in the valence sector at the initial scale ($\mu_0$) are obtained as
\begin{align}
f^q(x)=&\sum_{\{\lambda_i\}} \int \left[{\rm d}\mathcal{X} \,{\rm d}\mathcal{P}_\perp\right] \nonumber\\&\times\Psi^{\uparrow *}_{\{x_i,\vec{p}_{i\perp},\lambda_i\}}\Psi^{\uparrow}_{\{x_i,\vec{p}_{i\perp},\lambda_i\}} \delta \left(x-x_q\right).
\end{align}
Using the LFWFs within BLFQ given in Eq.~(\ref{wavefunctions}), we compute the unpolarized PDFs for the valence quarks, which are normalized as
\begin{align}
&\int_{0}^{1} dxf^u(x)=F^{\rm{u}}_1(0)=n_u, \nonumber\\
&\int_{0}^{1} dxf^d(x)=F^{\rm{d}}_1(0)=n_d,
\end{align}
with $n_q$ being the number of quarks of flavor $q$ in the nucleon.
Further, at the initial scale, the following momentum sum rule is satisfied by our PDFs,
\begin{align}
\int_{0}^{1} dx x(f^u(x)+f^d(x)) = 1. 
\end{align}

By performing the QCD evolution, one can obtain the valence quark PDFs at higher $\mu^2$ scales with the input PDFs at the initial scale. In this paper,
we set a 10$\%$ uncertainty in the initial scale. We interpret the initial scale associated with our LFWFs as the effective scale where the structures of the nucleon are described by the motion of the valence quarks only.
The scale evolution allows the constituent quarks to produce gluons, with the emitted gluons capable of generating quark-antiquark pairs as well as additional gluons. In this picture, the sea quark and gluon components are revealed at a higher scale through QCD evolution.

The QCD evolution is governed by the Dokshitzer-Gribov-Lipatov-Altarelli-Parisi (DGLAP) equations~\cite{Dokshitzer:1977sg,Gribov:1972ri,Altarelli:1977zs}. Here, we use the next-to-next-to-leading order (NNLO) DGLAP equations of QCD, to evolve our valence quark PDFs from our model scale $\mu_0^2$ to higher scale $\mu^2$. 
For this purpose, we utilize the Higher Order Perturbative Parton Evolution toolkit (HOPPET) to numerically solve the NNLO DGLAP equations~\cite{Salam:2008qg}.
We employ the condition that the running coupling constant $\alpha_s(\mu^2)$ saturates in the infrared at a cutoff value of max $\alpha_s\sim 1$~\cite{Lan:2019vui,Lan:2019rba}, consistent with our fit value discussed before. 
We adopt $\mu_0^2=0.195\pm 0.020$ GeV$^2$ for the initial scale of our PDFs, which we determine by matching the moment of the valence quark PDFs: $\braket{x}_q=\int_0^1 \,{\rm d}x\, x\, f^q(x)$ at $\mu^2=10$ GeV$^2$, with the result from the global fits $\braket{x}_u+\braket{x}_d=0.3742$~\cite{deTeramond:2018ecg}, after performing the QCD evolution of our valence quark PDFs. 


\begin{figure}[htbp]
\centering
\includegraphics[width=\columnwidth]{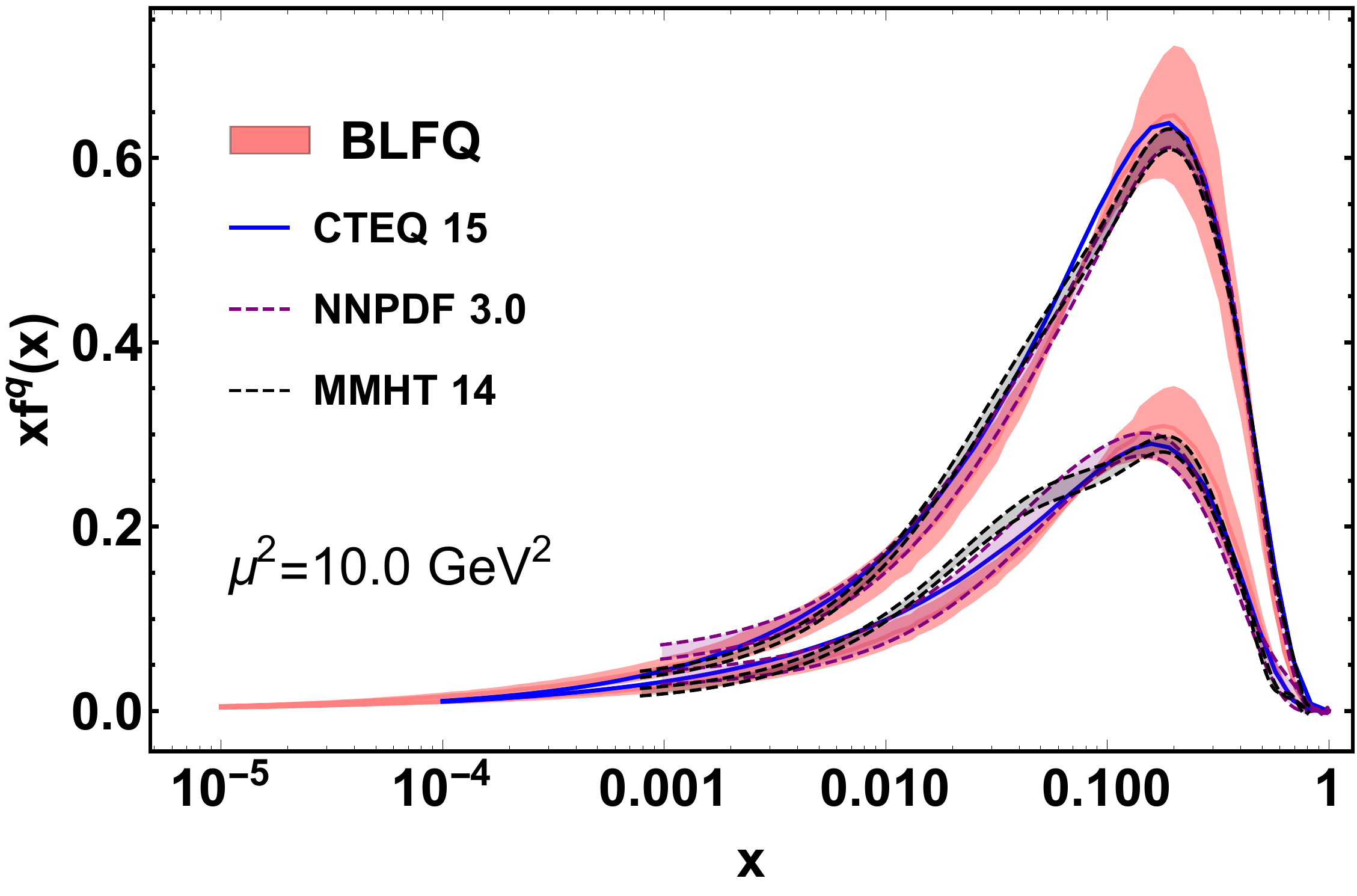}
\caption{The unpolarized proton's PDFs of the proton $xf^q(x)$ for the up (top curves) and down  (bottom curves) quarks. The light red bands correspond to the uncertainty in the BLFQ results. We compare our results with the global fits: CTEQ~\cite{Dulat:2015mca}(blue solid), NNPDF~\cite{Ball:2017nwa}(purple dashed), MMHT~\cite{Harland-Lang:2014zoa}(black dashed).}
\label{unpolarized_pdf}
\end{figure}

In Fig~\ref{unpolarized_pdf}, we show the unpolarized PDFs of the valence quarks at $\mu^2=10$ GeV$^2$ and compare our results with the global fits by CTEQ 2015 \cite{Dulat:2015mca}, NNPDF3.0 \cite{Ball:2017nwa} and MMHT14~\cite{Harland-Lang:2014zoa} collaborations. 
 The error bands in our evolved PDFs are due to the $10 \%$ uncertainties in the initial scale $\mu_0^2$ and the coupling constant $\alpha_s$. Our unpolarized valence PDFs for both the up and the down quarks agree well with the global fits.
According to the Drell-Yan-West relation~\cite{Drell-Yan,West}, at large $\mu^2$ the valence quark PDFs fall off at large $x$ as $(1-x)^p$, where $p$ is related to the number of valence quarks and for the nucleon $p=3$. In our treatment, we observe that the up quark unpolarized distribution falls off at large $x$ as $(1-x)^{2.99}$, while for the down quark the distribution goes as $(1-x)^{3.24}$. These are consistent with the Drell-Yan-West relation and support the perturbative QCD prediction~\cite{Brodsky:1994kg}.


\begin{figure}[htbp]
\centering
\includegraphics[width=\columnwidth]{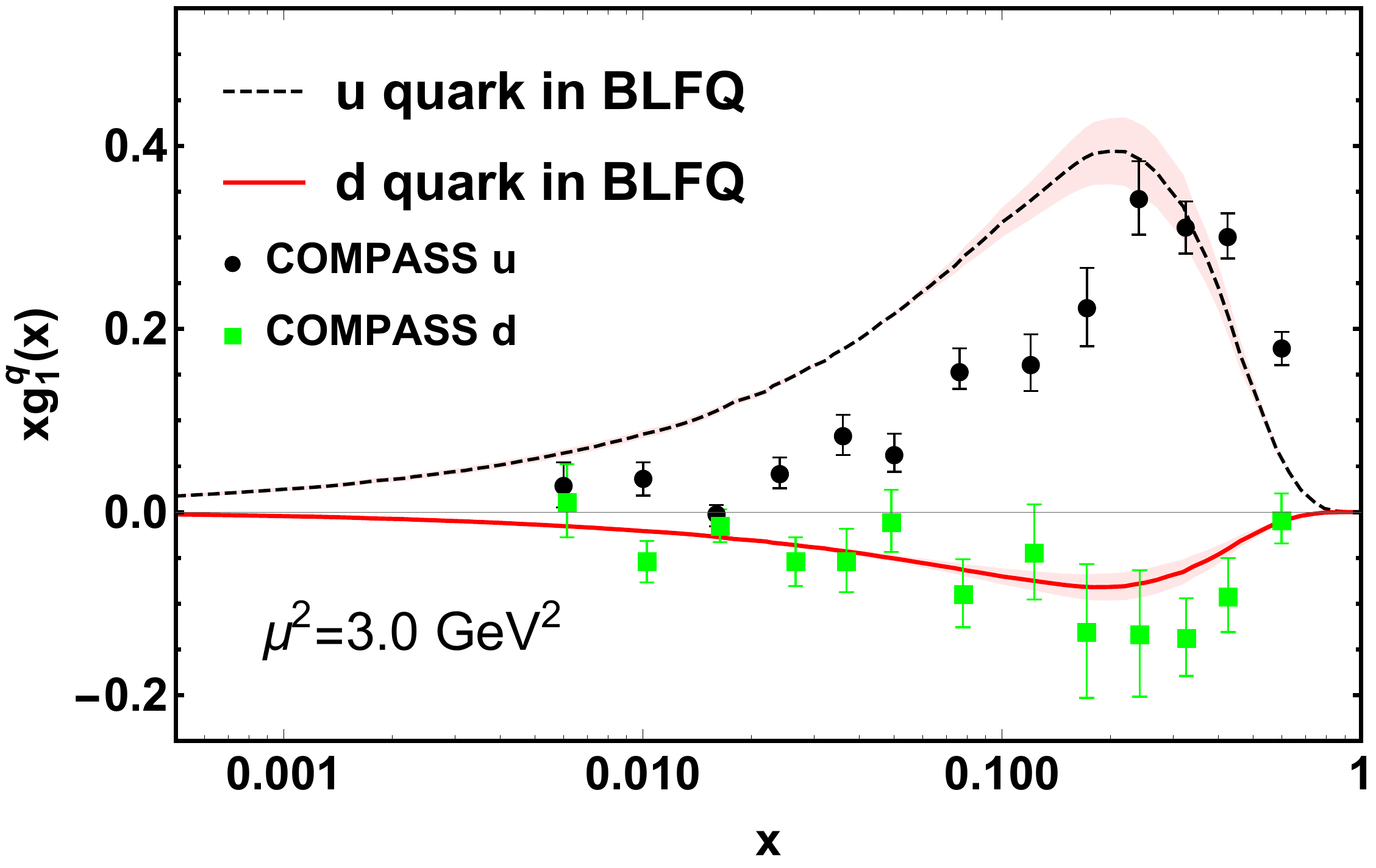}
\caption{The helicity PDFs for the up and down quarks in the proton. We compare BLFQ results (red solid line for down quark and black dashed line for up quark with uncertainties indicated by light red bands) with measured data from COMPASS Collaboration~\cite{Alekseev:2010ub}.}
\label{helicity_pdf}
\end{figure}
\subsubsection{Helicity PDFs and helicity asymmetries}
The helicity PDFs $g_1(x)$
reveal the spin structure of the nucleon from its quark and gluon constituents. The polarized PDFs describe the difference of the probability density between helicity-parallel
and helicity-anti-parallel quarks in the nucleon. With the LFWFs, the helicity distribution for the valence quarks is given by 
\begin{align}
g_1^q(x)=&\sum_{\{\lambda_i\}} \int \left[{\rm d}\mathcal{X} \,{\rm d}\mathcal{P}_\perp\right]\nonumber\\&\times \lambda_1 \Psi^{\uparrow *}_{\{x_i,\vec{p}_{i\perp},\lambda_i\}}\Psi^{\uparrow}_{\{x_i,\vec{p}_{i\perp},\lambda_i\}} \delta \left(x-x_1\right),
\end{align}
where $\lambda_1=1~(-1)$ for the struck quark helicity. The scale
evolution of the helicity PDFs is simulated by the
same scheme as we employed for the unpolarized PDFs. For
comparison with measurements, we evolve the helicity PDFs from the initial scale $\mu_0^2=0.195\pm 0.020$ GeV$^2$ to the relevant experimental scale $\mu^2=3$ GeV$^2$~~\cite{Alekseev:2010ub}. 

The helicity PDFs $g_1(x)$ are shown in Fig.~\ref{helicity_pdf}, at the scale $\mu^2=3$ GeV$^2$, for the up and down quarks in the proton. The error bands in our evolved PDFs are due to the spread in the initial scale $\mu_0^2=0.195\pm0.020$ GeV$^2$ and the uncertainties in the coupling constant, $\alpha_s=1.1\pm 0.1$.
 We find that our down quark helicity PDF agrees reasonably well with measured data from COMPASS~\cite{Alekseev:2010ub}. For the up quark, the $g_1(x)$ in our approach is however overestimated at low $x$, while it tends to agree with the data above $x\sim 0.25$ region. 

\begin{figure}[htbp]
\centering
\includegraphics[width=\columnwidth]{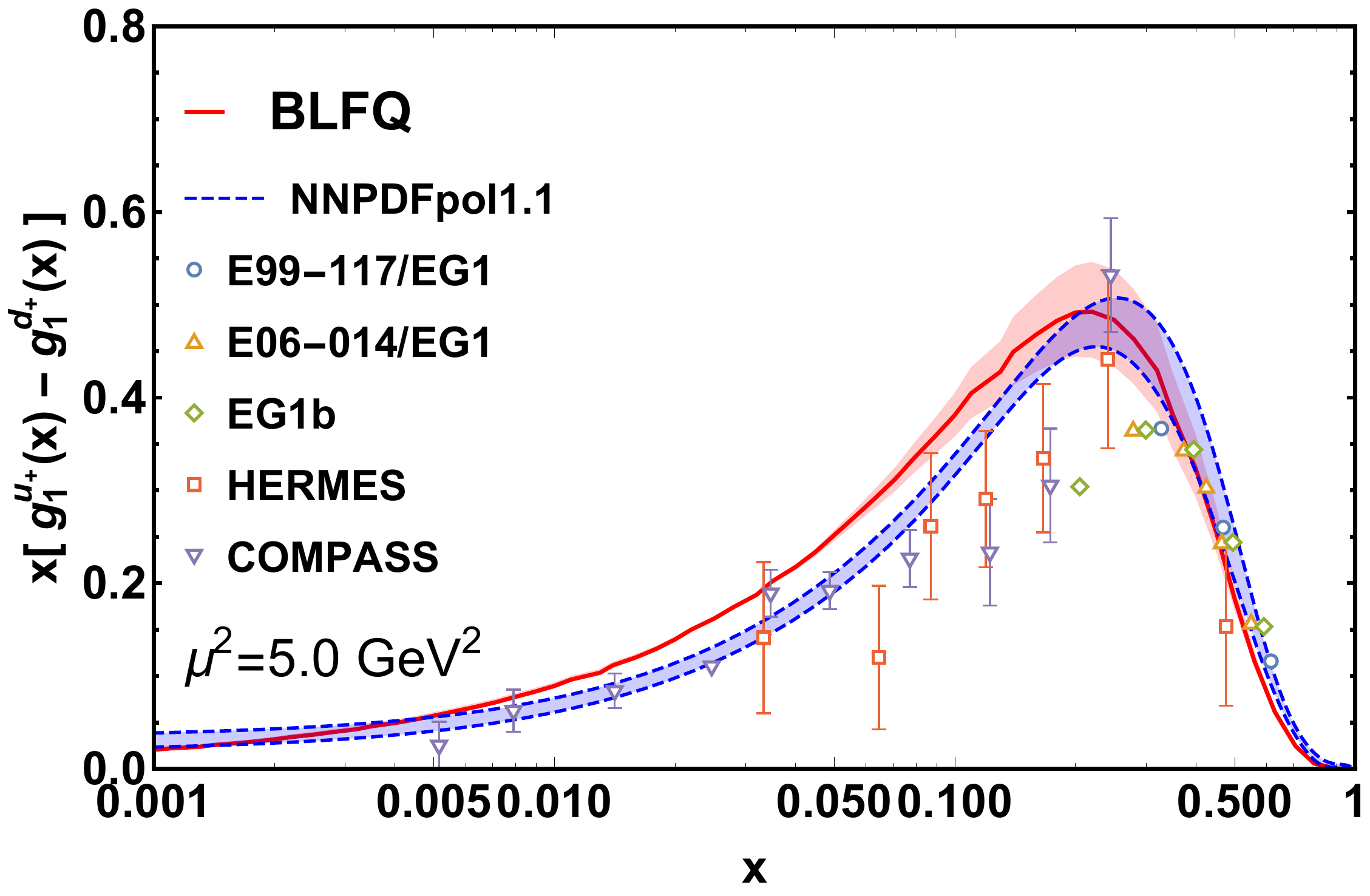}
\caption{Helicity PDFs of the isovector combination in the proton. The BLFQ result (red line with uncertainty indicated by the light red band) is compared with NNPDF global fit (blue band)~\cite{Nocera:2014gqa} and measured data~\cite{Zheng:2003un,Zheng:2004ce,Parno:2014xzb,Dharmawardane:2006zd,Airapetian:2003ct,Airapetian:2004zf,Alekseev:2010ub}.}
\label{difference_helicity_pdf}
\end{figure}
\begin{figure}[htbp]
\centering
\includegraphics[width=\columnwidth]{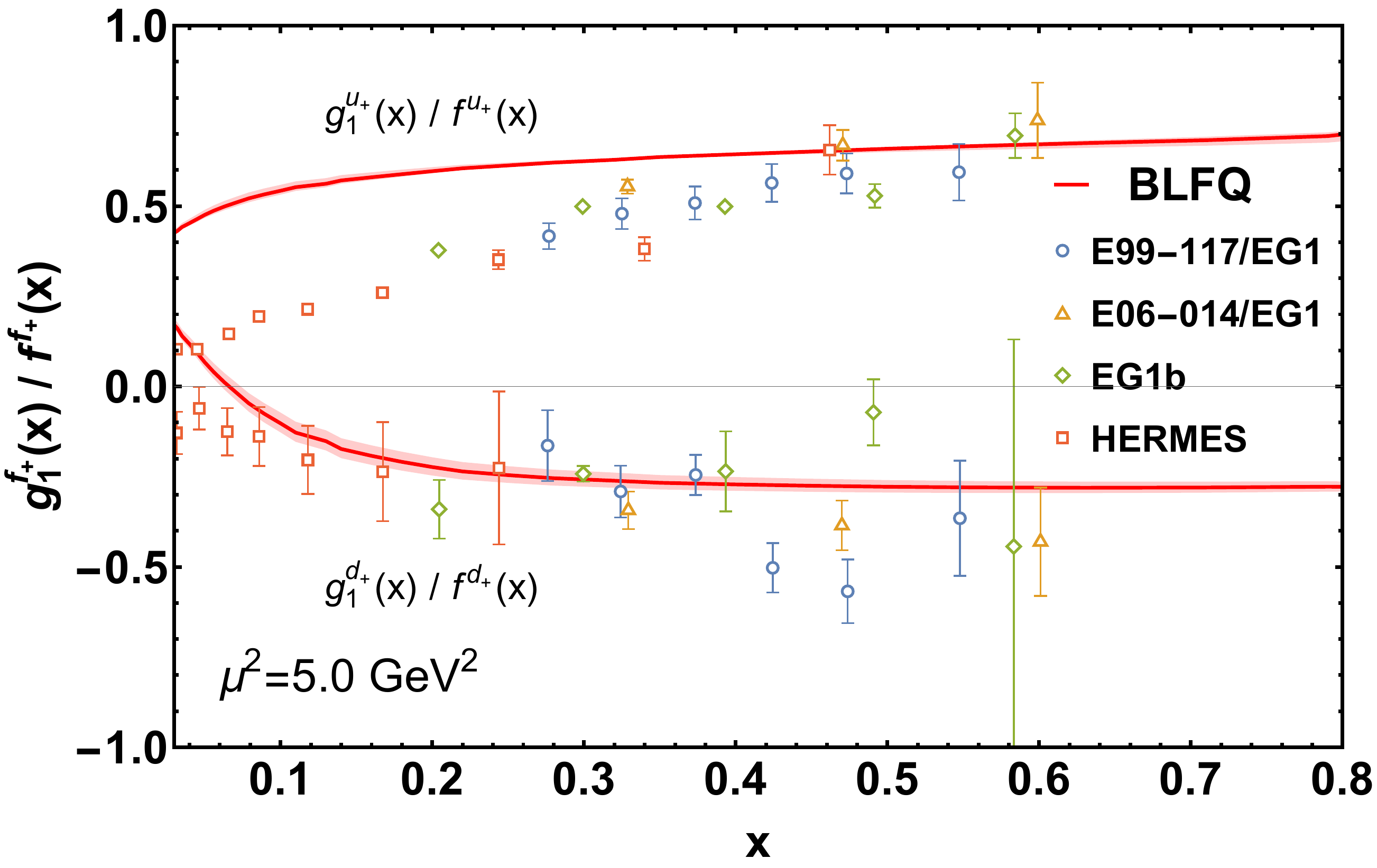}
\caption{Helicity asymmetry of $u+\bar{u}$ (upper curve/data) and  $d+\bar{d}$ (lower curve/data) in the proton. The BLFQ result (light red band) is compared with the measured data~\cite{Zheng:2003un,Zheng:2004ce,Parno:2014xzb,Airapetian:2003ct,Airapetian:2004zf}.}
\label{ratio_helicity_pdf}
\end{figure}

We also compute the isovector combination: $g^{u_+}_1(x)-g^{d_+}_1(x)$, where  $g^{q_+}_1(x)$ stands for $g^{q}_1(x)+g^{\bar{q}}_1(x)$. This is sensitive to the quark and the antiquark contributions. 
In Fig.~\ref{difference_helicity_pdf}, we compare our result for the isovector combination with the NNPDF global fit~\cite{Nocera:2014gqa} and the experimental data~\cite{Zheng:2003un,Zheng:2004ce,Parno:2014xzb,Dharmawardane:2006zd,Airapetian:2003ct,Airapetian:2004zf,Alekseev:2010ub} at the scale $\mu^2=5.0~\rm{GeV}^2$. Note that we directly compare the integrand of the first moment of the combination. As can be seen from the plot, our numerical result is reasonably consistent with the global fit and measurements. However, a deviation from the fit is also observed at $x < 0.2$ region. We emphasize that the sea quark distributions in our current treatment are solely generated from the QCD evolution and they are equal for the light flavors. Consequently, the sea quark contributions from the up and down flavors to $g^{u_+}_1(x)-g^{d_+}_1(x)$ cancel each other and effectively only the valence quark distributions contribute to the BLFQ results in Fig.~\ref{difference_helicity_pdf}. It has been noted that the sea quarks may have a noticeable contribution to the isovector combination distribution~\cite{Sufian:2016hwn}. To demonstrate the significance of the sea quarks in describing spin-related quantities, the higher Fock components, such as $|qqqq\bar{q}\rangle$, in the nucleon state  need to be included explicitly within BLFQ.

The first moment of the isovector distribution also relates to the axial charge $g_{\rm A}$, which is precisely determined by the neutron weak decay, $g_{\rm A}=1.2732\pm0.0023$~\cite{Tanabashi:2018oca}. As shown in Table~\ref{tab:AFF}, its value evaluated in our current approach is somewhat higher than the extracted data and this is consistent with the analysis of the isovector combination distribution discussed above.

Another very important observable is the helicity asymmetry, $g_1^q(x)/f_1^q(x)$, which is expected to increase with increasing $x$. In the limit $x\to 1$ the helicity asymmetry is predicted to approach $1$ by pQCD~\cite{Brodsky:1994kg,Avakian:2007xa}. Experimentally, the
expected increase of the asymmetry for the up quark is observed. However, for the down quark, the asymmetry is found to remain negative in the experimentally covered region of $x\lesssim 0.6$~\cite{Zheng:2003un,Zheng:2004ce,Parno:2014xzb,Dharmawardane:2006zd,Airapetian:2003ct,Airapetian:2004zf,Alekseev:2010ub} and the global analyses of the data extrapolated to large $x$ favor negative values of $g_1^d/f_1^d$ at $x\to 1$~\cite{deFlorian:2008mr,deFlorian:2009vb,Nocera:2014gqa,Jimenez-Delgado:2014xza,Ethier:2017zbq}, which is also supported by DSE calculations~\cite{Roberts:2013mja}. Thus, there is a tension with the pQCD constraints and the global analyses and DSE calculations in understanding the large-$x$ behavior of the polarized PDFs. Meanwhile, the large-$x$ region, which is dominated by the valence quarks and is thus much less affected by the sea quarks, will be tested in upcoming Jefferson Lab spin program~\cite{E12-06-110,E12-06-122}.
The helicity asymmetries in our BLFQ approach are shown in Fig.~\ref{ratio_helicity_pdf}. The BLFQ prediction shows a good agreement with the measurements~\cite{Zheng:2003un,Zheng:2004ce,Parno:2014xzb,Airapetian:2003ct,Airapetian:2004zf} for the down quark. Our result favors the global analyses and DSE calculations~\cite{Roberts:2013mja}. Note that $g_1^u(x)$ in our calculation deviates from the data in the low-$x$ domain (Fig.~\ref{helicity_pdf}). It is therefore not surprising that our asymmetry for the up quark also deviates from the experiments at low-$x$ region. 

\begin{figure*}[htbp]
\centering
\subfigure[]{
\begin{minipage}[t]{0.45\linewidth}
\centering
\includegraphics[width=\columnwidth]{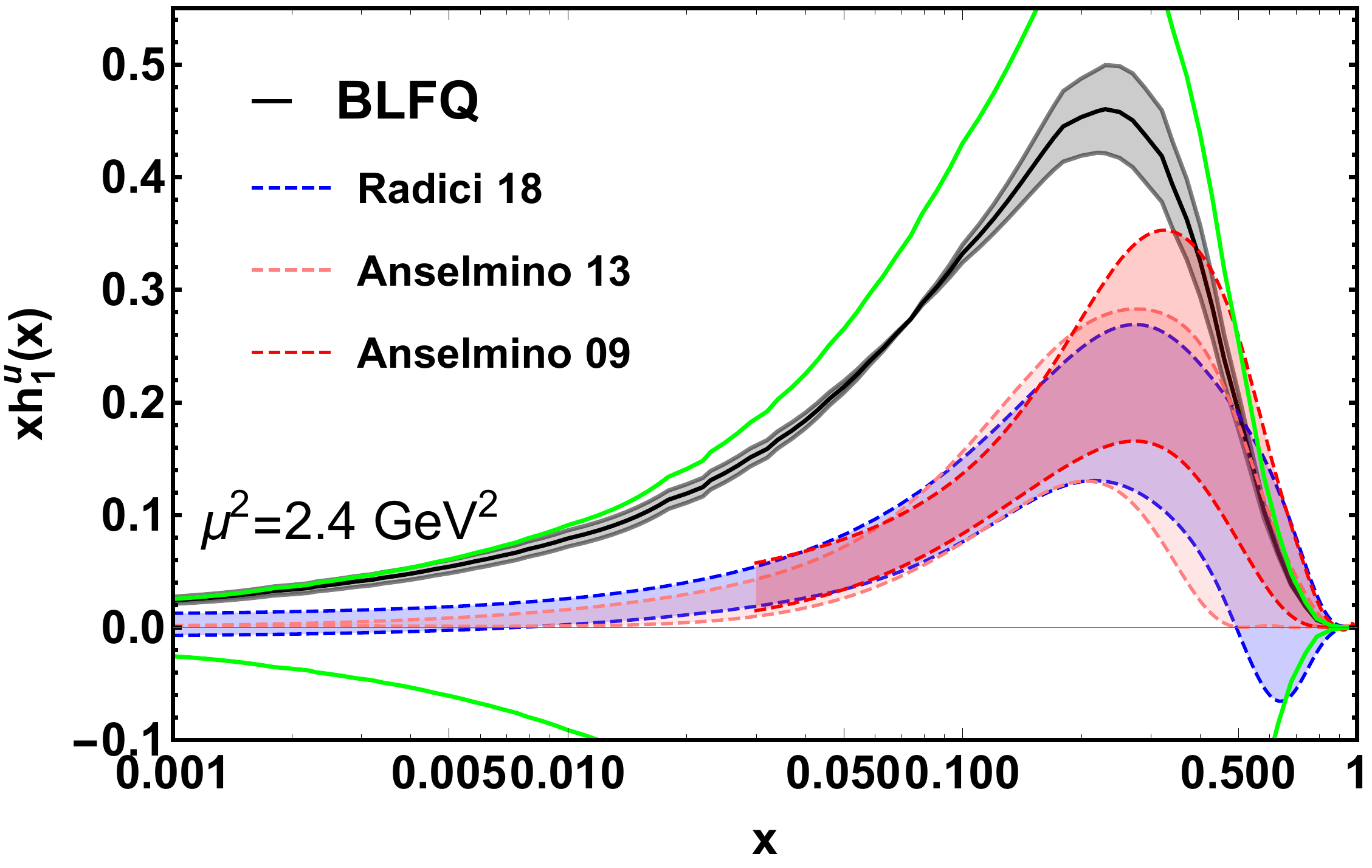}
\end{minipage}
\label{TPDF_D}
}
\subfigure[]{
\begin{minipage}[t]{0.45\linewidth}
\centering
\includegraphics[width=\columnwidth]{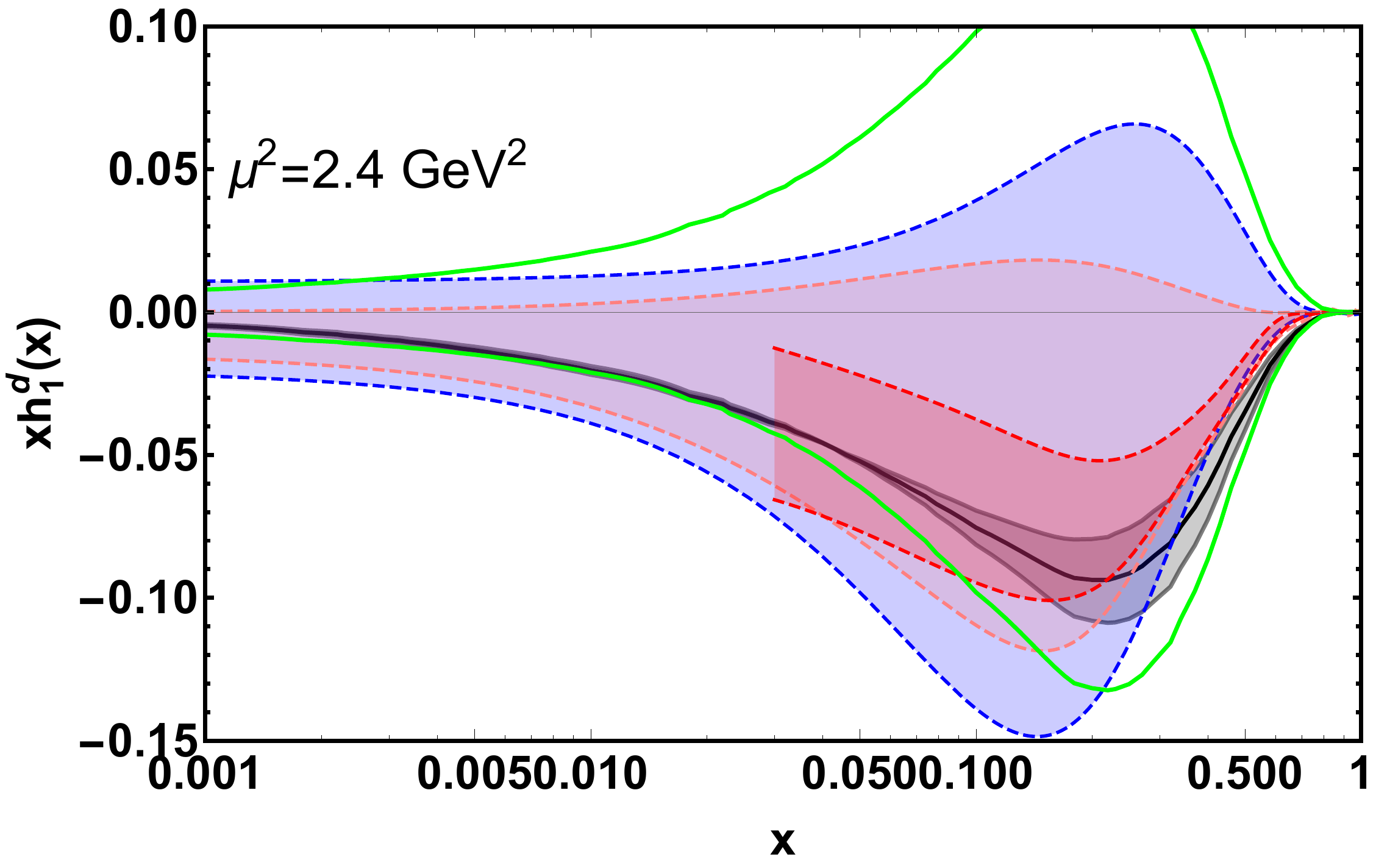}
\end{minipage}
\label{TPDF_U}
}
\caption{Quark transversity PDFs in the proton. (a) Up quark distribution; (b) down quark distribution. The BLFQ results (gray bands) are compared with the global fits~\cite{Radici:2018iag,Anselmino:2013vqa,Anselmino:2008jk}. The green lines correspond to the Soffer bound~\cite{Soffer:1994ww} multiplied by the momentum fraction $\frac{1}{2}|xf^f_1(x)+xg_1^f(x)|$.  }
\label{transversity_pdf}
\end{figure*}
\begin{figure*}[htbp]
\centering
\subfigure[]{
\begin{minipage}[t]{0.45\linewidth}
\centering
\includegraphics[width=\columnwidth]{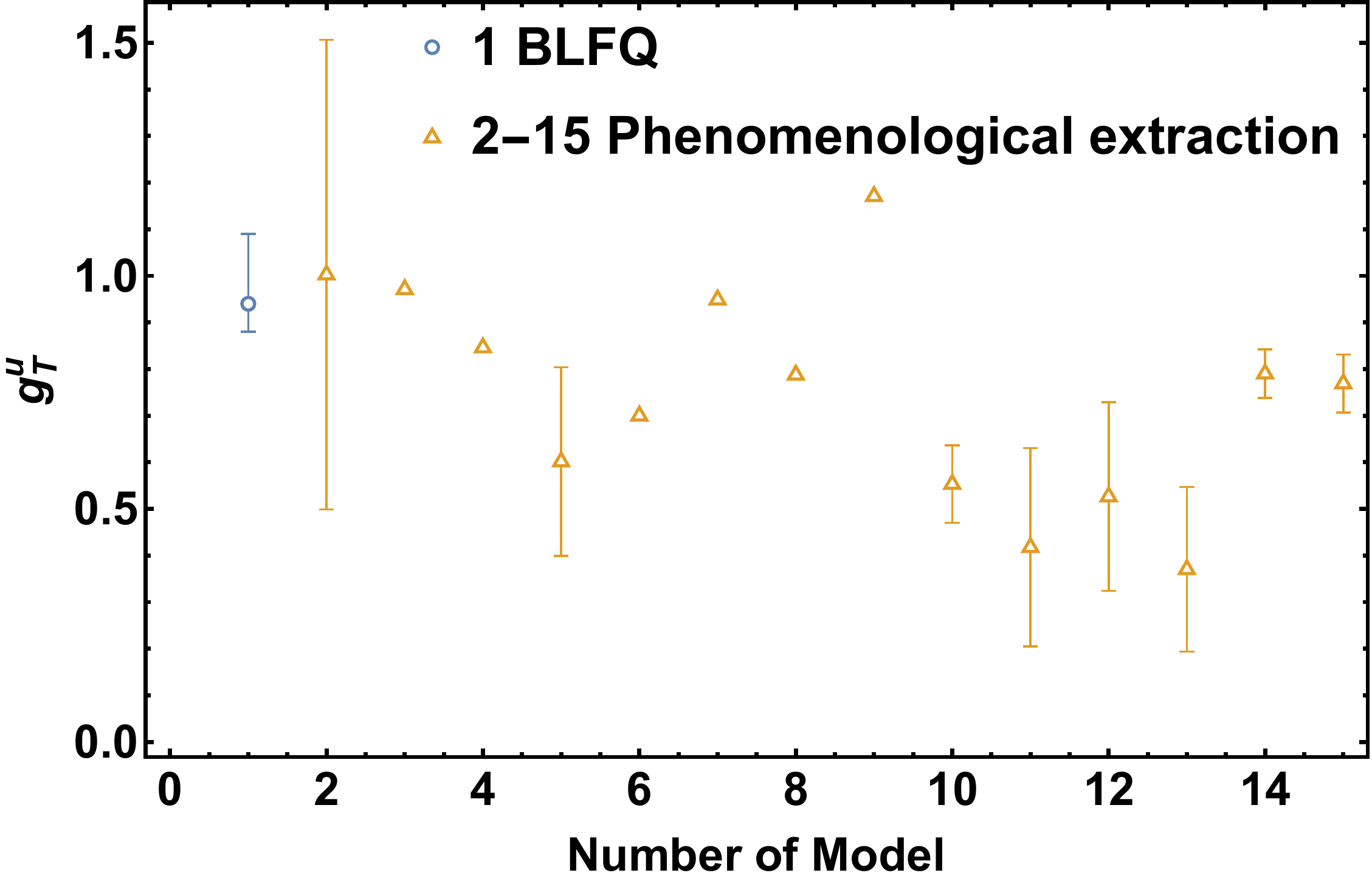}
\end{minipage}
\label{tensor_charge_flavor}
}
\subfigure[]{
\begin{minipage}[t]{0.45\linewidth}
\centering
\includegraphics[width=\columnwidth]{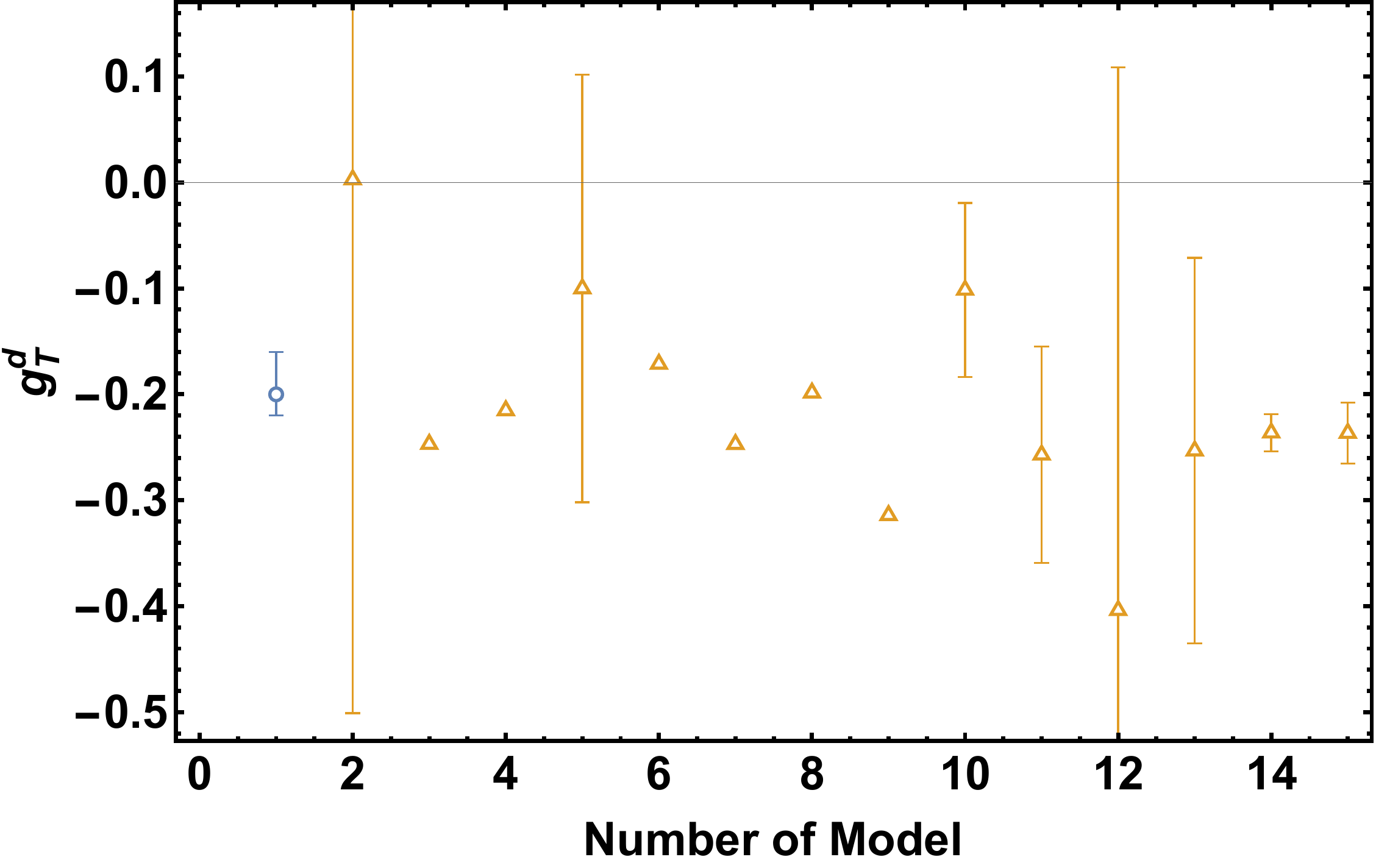}
\end{minipage}
\label{tensor_charge_total}
}
\caption{The quark tensor charges in the proton; (a) for the up quark; (b) for the down quark at  $\mu^2=2.4~\rm{GeV}^2$. The blue data (1) represent the BLFQ predictions. The other data points correspond to the phenomenological extractions of the tensor charges from different models: (2) He and Ji ($\mu^2\sim 1 ~\rm{GeV}^2$)~\cite{He:1994gz}, (3) Barone {\it et. al.} ($\mu^2\sim 25 ~\rm{GeV}^2$)~\cite{Barone:1996un}, (4) Schweitzer {\it et. al.} ($\mu^2\sim 0.6 ~\rm{GeV}^2$) \cite{Schweitzer:2001sr} (5) Gamberg {\it et. al.} ($\mu^2\sim 1 ~\rm{GeV}^2$)~\cite{Gamberg:2001qc}, (6) Pasquini {\it et. al.} 2005 ($\mu^2=10 ~\rm{GeV}^2$)~\cite{Pasquini:2005dk}, (7) Wakamatsu ($\mu^2=2.4 ~\rm{GeV}^2$)~\cite{Wakamatsu:2007nc}, (8) Pasquini {\it et. al.} 2007 ($\mu^2=10 ~\rm{GeV}^2$)~\cite{Pasquini:2006iv,Pincetti:2006hc}, (9) Lorc$\acute{\rm{e}}$ ($\mu^2=0.36 ~\rm{GeV}^2$)~\cite{Lorce:2007fa}, (10) Pitschmann {\it et. al.} ($\mu^2=4 ~\rm{GeV}^2$)~\cite{Pitschmann:2014jxa}, (11) Anselmino {\it et. al.} ($\mu^2=0.8 ~\rm{GeV}^2$)~\cite{Anselmino:2013vqa}, (12) Radici {\it et. al.} ($\mu^2=10 ~\rm{GeV}^2$)~\cite{Radici:2015mwa}, (13) Kang {\it et. al.} ($\mu^2 = 10 ~\rm{GeV}^2$)~\cite{Kang:2015msa}, (14) Abdel-Rehim {\it et. al.} ($\mu^2=4~\rm{GeV}^2$)~\cite{Abdel-Rehim:2015owa}, (15) Bhattacharya {\it et. al.} ($\mu^2=4 ~\rm{GeV}^2$)~\cite{Bhattacharya:2016zcn}. }
\label{tensor_charge}
\end{figure*}
\subsubsection{Transversity PDFs and tensor charge}
Quark transversity distributions describe the correlation of the transverse polarization of the nucleon with the transverse polarization of its
constituent quark. Recently, it has gained increasing attention because of the importance of a precise determination of its integral, the so-called tensor charge ($g_T$).  
In terms of the overlap of the LFWFs, the transversity PDFs read
\begin{align}
&h_1^q(x)=\sum_{\{\lambda_i\}} \int \left[{\rm d}\mathcal{X} \,{\rm d}\mathcal{P}_\perp\right] \nonumber\\&\times\left(\Psi^{\uparrow *}_{\{x^{}_i,\vec{p}^{}_{i\perp},\lambda^{\prime}_{i}\}}\Psi^{\downarrow}_{\{x_i,\vec{p}_{i\perp},\lambda_{i}\}} +(\uparrow \leftrightarrow \downarrow )\right)\delta(x-x_1),
\end{align}
where the quark helicities $\lambda_1^\prime=-\lambda_1$ for the struck quark and $\lambda^\prime_{2,3}=\lambda_{2,3}$ for the spectators. We compute $h^q_1(x)$ using our LFWFs given in Eq.~(\ref{wavefunctions}). We employ the same DGLAP equation with PDFs to evolve our transversity PDFs from our model scale $\mu_0^2$ to $\mu^2=2.4$ GeV$^2$. In Fig.~\ref{transversity_pdf}, we compare our predictions with the global analysis of the data of pion-pair production in DIS and in proton-proton
collisions with a transversely polarized proton by Radici {\it et.~al.} \cite{Radici:2018iag}, and with the global analysis of the data on azimuthal asymmetries in SIDIS, from the HERMES and COMPASS Collaborations, and  $e^+e^-$ data from the BELLE Collaboration by Anselmino {\it et.~al.} \cite{Anselmino:2008jk,Anselmino:2013vqa}. 
Our down quark transversity distribution is consistent with the global fits. The up quark $h^u_1$ PDF in our approach tends to overestimate the data in $x< 0.3$, but it shows reasonable agreement with those fits in the large-$x$ region. 
Furthermore, we find that our PDFs satisfy the Soffer bound~\cite{Soffer:1994ww}, which at an arbitrary scale $\mu$ is defined as
\begin{eqnarray}
|h_1^q(x,\mu^2)|\leq \frac{1}{2}|f^q_1(x,\mu^2)+g_1^q(x,\mu^2)|.
\end{eqnarray}
As can be seen from Fig.~\ref{transversity_pdf}, the transversity distributions $h^q_1$ for both the up and down quarks lie below the average of the unpolarized PDFs $f^q_1$ and helicity PDFs $g^q_1$.

The first moment of the transversity PDF gives the tensor charge:
\begin{eqnarray}
g_T^q = \int dx\, h_1^q(x,\mu^2).
\end{eqnarray}
Our results for $g^q_T$ are compared with extracted data as well as with lattice data, both at $\mu^2=2.4$ GeV$^2$, in Table~\ref{tab:TPDF}. Again we observe that BLFQ predicts the tensor charges for the down quark in good agreement with the global QCD analysis \cite{Anselmino:2013vqa} but it deviates from the extracted data for the up quark. On the other hand, our value of $g_T^u$ is closer to the recent lattice data \cite{Gupta:2018lvp}.  BLFQ results show compatibility with results from phenomenological models and lattice calculations~~\cite{He:1994gz,Barone:1996un,Gamberg:2001qc,Pasquini:2005dk,Wakamatsu:2007nc,Pasquini:2006iv,Pincetti:2006hc,Lorce:2007fa,Pitschmann:2014jxa,Anselmino:2013vqa,Radici:2015mwa,Kang:2015msa,Abdel-Rehim:2015owa,Bhattacharya:2016zcn} as can be seen from Fig.~\ref{tensor_charge}. 

We also observe that for both the up and down quark, we have $|g_T^q|<|g_A^q|$ as noticed in other theoretical calculations. 
We also provide the second moment of the difference between u and d quarks:
\begin{eqnarray}
\braket{x}_T^{u-d} = \int dx\, x\, (h_1^u(x,\mu^2)-h_1^d(x,\mu^2))\,,
\end{eqnarray}
at the scale $\mu^2=2.4~\rm{GeV}^2$ in the Table~\ref{tab:TPDF}. We find that the BLFQ prediction for $\braket{x}_T^{u-d}$ agrees reasonably well with the lattice data \cite{Alexandrou:2019ali}.
\begin{table}[htp]	
\centering
\caption{The tensor charge and second moment of transversity PDFs. Our results are compared with the extracted data~\cite{Anselmino:2013vqa} and recent lattice calculation~\cite{Gupta:2018lvp,Alexandrou:2019ali}.}
\label{tab:TPDF}
\begin{tabular}{|cccc|cc}
\hline\hline
Quantity  	   					~&~ BLFQ	          			~&~ Extracted data       				~&~ Lattice   	\\
\hline
$g_T^u $    					~&~ $0.94^{+0.06}_{-0.15}$	  	~&~ $0.39^{+0.18}_{-0.12}$  	    	~&~ $0.784(28)$	\\
$g_T^d $    					~&~ $-0.20^{+0.02}_{-0.04}$	  	~&~ $-0.25^{+0.30}_{-0.10}$  	   		~&~ $-0.204(11)$	\\
\hline
$\braket{x}_T^{u-d}$  	~&~ $0.229^{+0.019}_{-0.048}$	~&~ $-$  								~&~ $0.203(24)$	\\
\hline\hline
\end{tabular}
\end{table}

%
\subsection{Generalized parton distribution functions}
The GPDs are defined as off-forward matrix elements of the bilocal operator of light-front correlation functions. The unpolarized GPDs are parameterized as~\cite{Ji:1998pc}:
\begin{align}\label{gpd_definition}
&\int\frac{dz^-}{8\pi} e^{ixP^+z^-/2}\braket{P^{\prime},\Lambda^{\prime}|\bar{\Psi}(0)\gamma^+\Psi(z)|P,\Lambda}|_{z^+=\vec{z}_{\perp}=0}\nonumber\\
&=\frac{1}{2\bar{P}^+} \bar{u}(P^{\prime},\Lambda^{\prime})\Big[H^q(x,\zeta,t)\gamma^+
\nonumber\\ &\quad\quad\quad\quad\quad\quad\quad\quad + E^q(x,\zeta,t) \frac{i\sigma^{+j}q_j}{2M}\Big]u(P,\Lambda).
\end{align}
where the kinematical variables in the symmetric frame are
$
\bar{P}^\mu=(P^\mu + P^{\prime \mu})/2;~ q^\mu=P^{\prime \mu}-P^\mu;~ \zeta=-q^+/2\bar{P}^+;\nonumber
$
and $t=q^2$. For $\zeta=0$, $t=-\vec{q}_\perp^2=-Q^2$. 
We evaluate the GPDs in our model using the overlap representation of LFWFs. We consider the DGLAP region, i.e., $\zeta<x<1$
for our discussion. The particle number remains conserved in this kinematical region which corresponds to the situation where one removes a quark from the initial
nucleon with light-front longitudinal momentum $(x-\zeta)P^+$
and reinserts it into the final nucleon with longitudinal momentum $(x + \zeta )P^+$. In this paper, we concentrate only on the zero skewness limit, i.e., $\zeta=0$. The unpolarized GPDs in the diagonal ($3\to 3$) overlap representation, in terms of LFWFs, are given by 
%
%
\begin{align}
H^q(x,0,t)=& 
 \sum_{\{\lambda_i\}} \int \left[{\rm d}\mathcal{X} \,{\rm d}\mathcal{P}_\perp\right]\nonumber\\&\times \Psi^{\uparrow *}_{\{x^{\prime}_i,\vec{p}^{\prime}_{i\perp},\lambda_i\}}\Psi^{\uparrow}_{\{x_i,\vec{p}_{i\perp},\lambda_i\}} \delta(x-x_1) ;    \\
E^q(x,0,t)=& -\frac{2M}{(q^1-iq^2)}
 \sum_{\{\lambda_i\}} \int \left[{\rm d}\mathcal{X} \,{\rm d}\mathcal{P}_\perp\right]\nonumber\\&\times \Psi^{\uparrow *}_{\{x^{\prime}_i,\vec{p}^{\prime}_{i\perp},\lambda_i\}}\Psi^{\downarrow}_{\{x_i,\vec{p}_{i\perp},\lambda_i\}} \delta(x-x_1),
\end{align}
where the light-front momenta are ($i=1$) $x^{\prime}_1=x_1$; $\vec{p}^{\prime}_{1\perp}=\vec{p}_{1\perp}+(1-x_1)\vec{q}_{\perp}$ for the struck quark and $x^{\prime}_i={x_i}; ~\vec{p}^{\prime}_{i\perp}=\vec{p}_{i\perp}-{x_i} \vec{q}_{\perp}$ for the spectators.
\begin{figure*}[htbp]
\centering
\subfigure[]{
\begin{minipage}[t]{0.45\linewidth}
\centering
\includegraphics[width=\columnwidth]{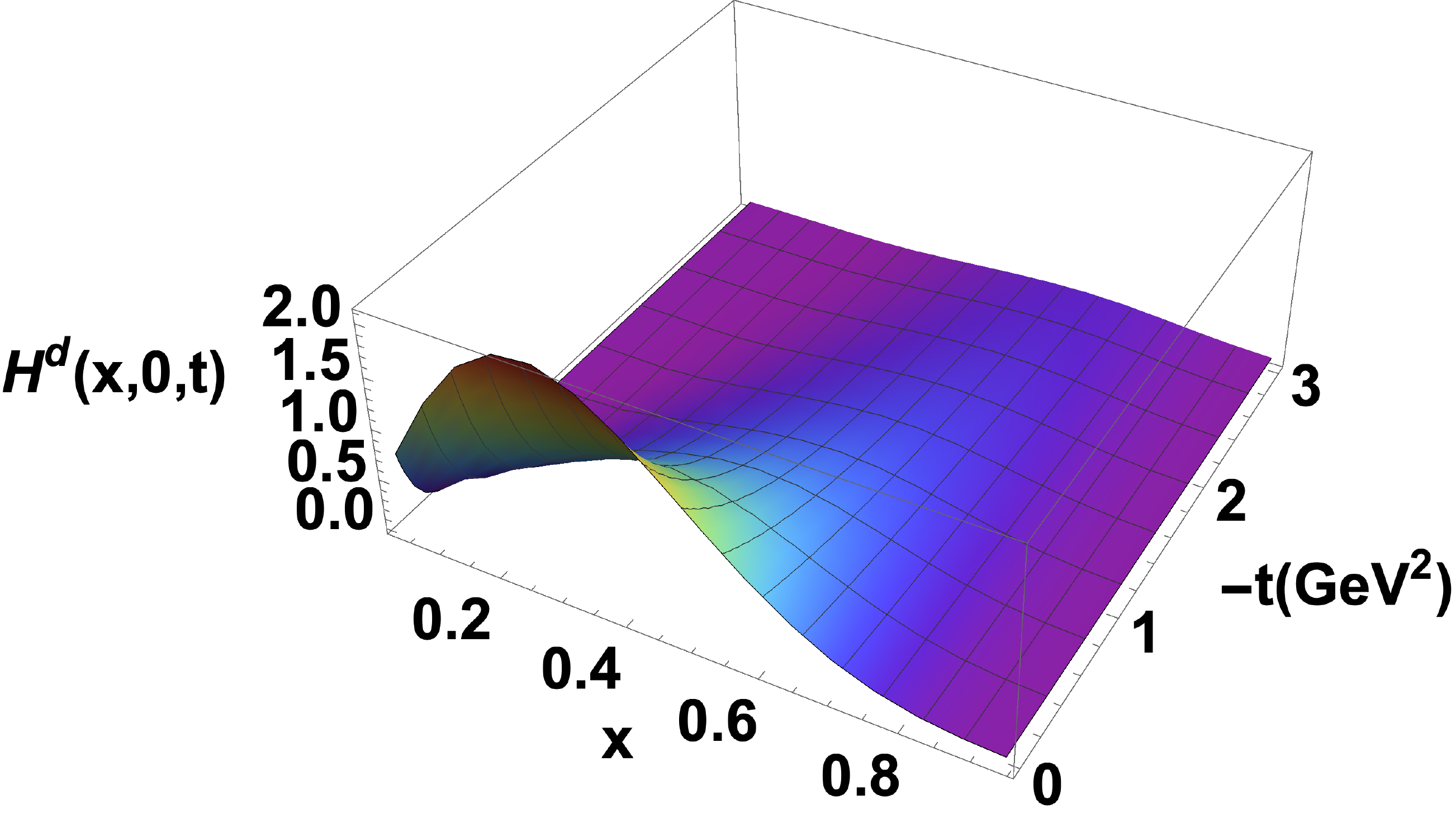}
\end{minipage}
\label{GPD_H_D}
}
\subfigure[]{
\begin{minipage}[t]{0.45\linewidth}
\centering
\includegraphics[width=\columnwidth]{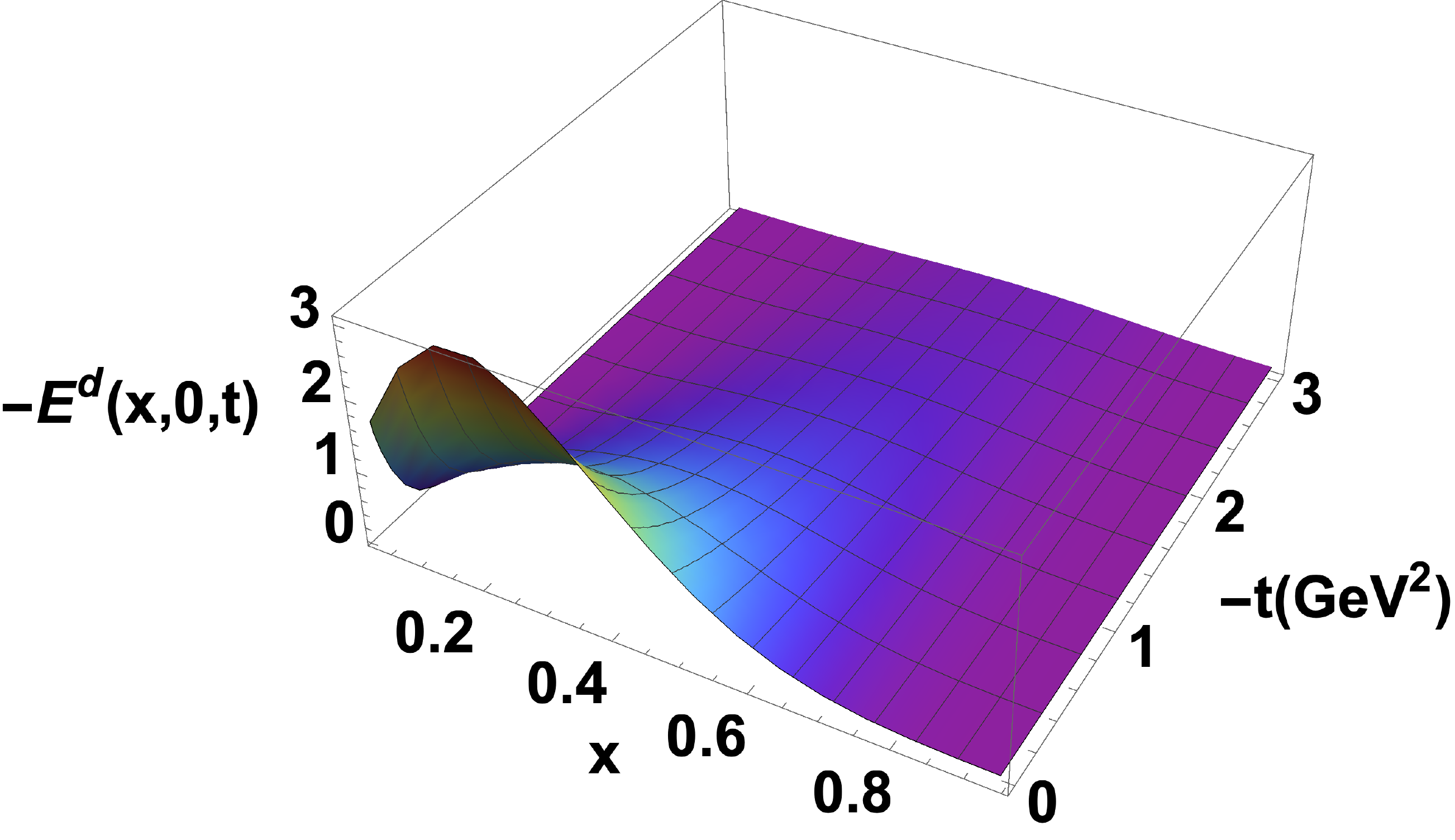}
\end{minipage}
\label{GPD_E_D}
}
\subfigure[]{
\begin{minipage}[t]{0.45\linewidth}
\centering
\includegraphics[width=\columnwidth]{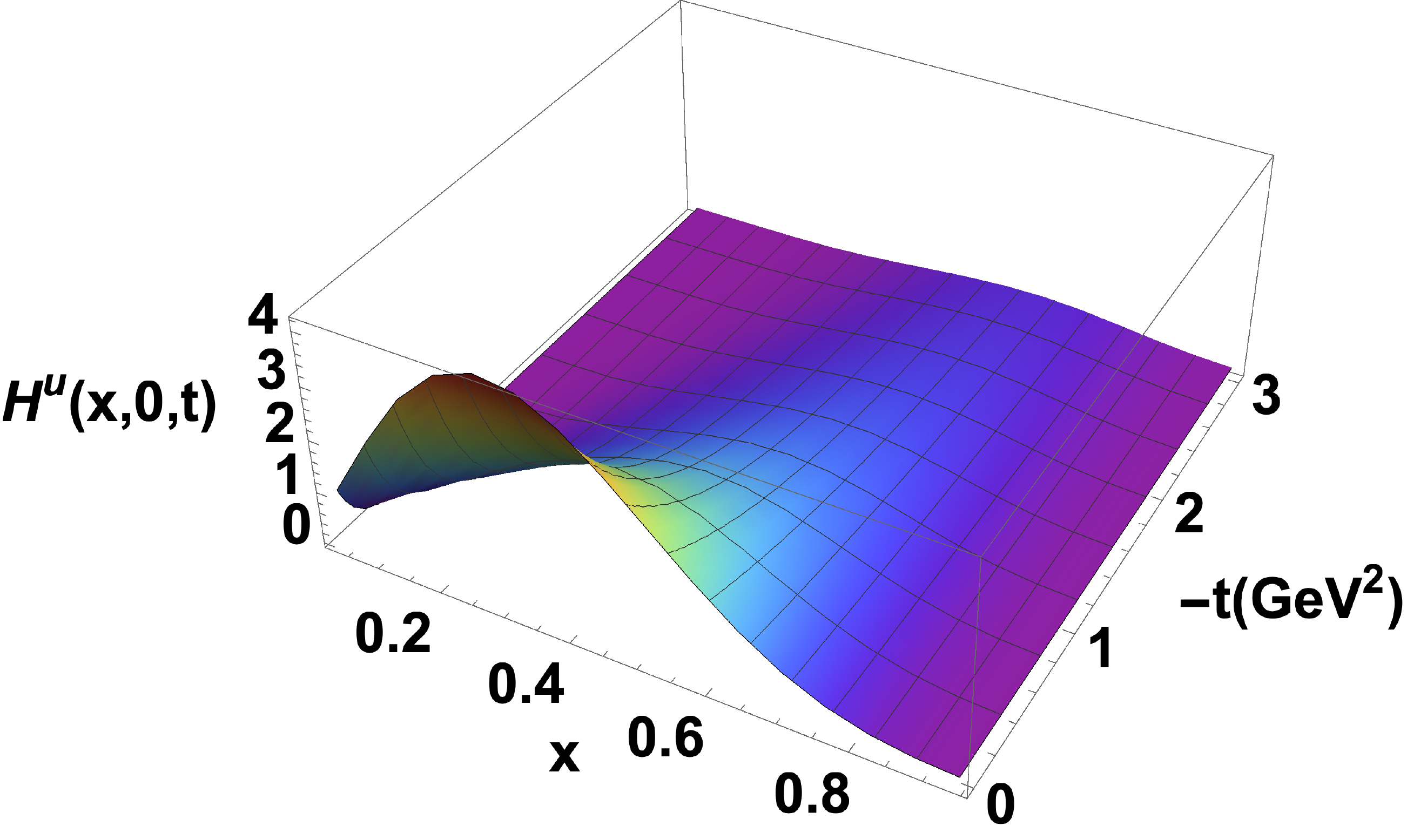}
\end{minipage}
\label{GPD_H_U}
}
\subfigure[]{
\begin{minipage}[t]{0.45\linewidth}
\centering
\includegraphics[width=\columnwidth]{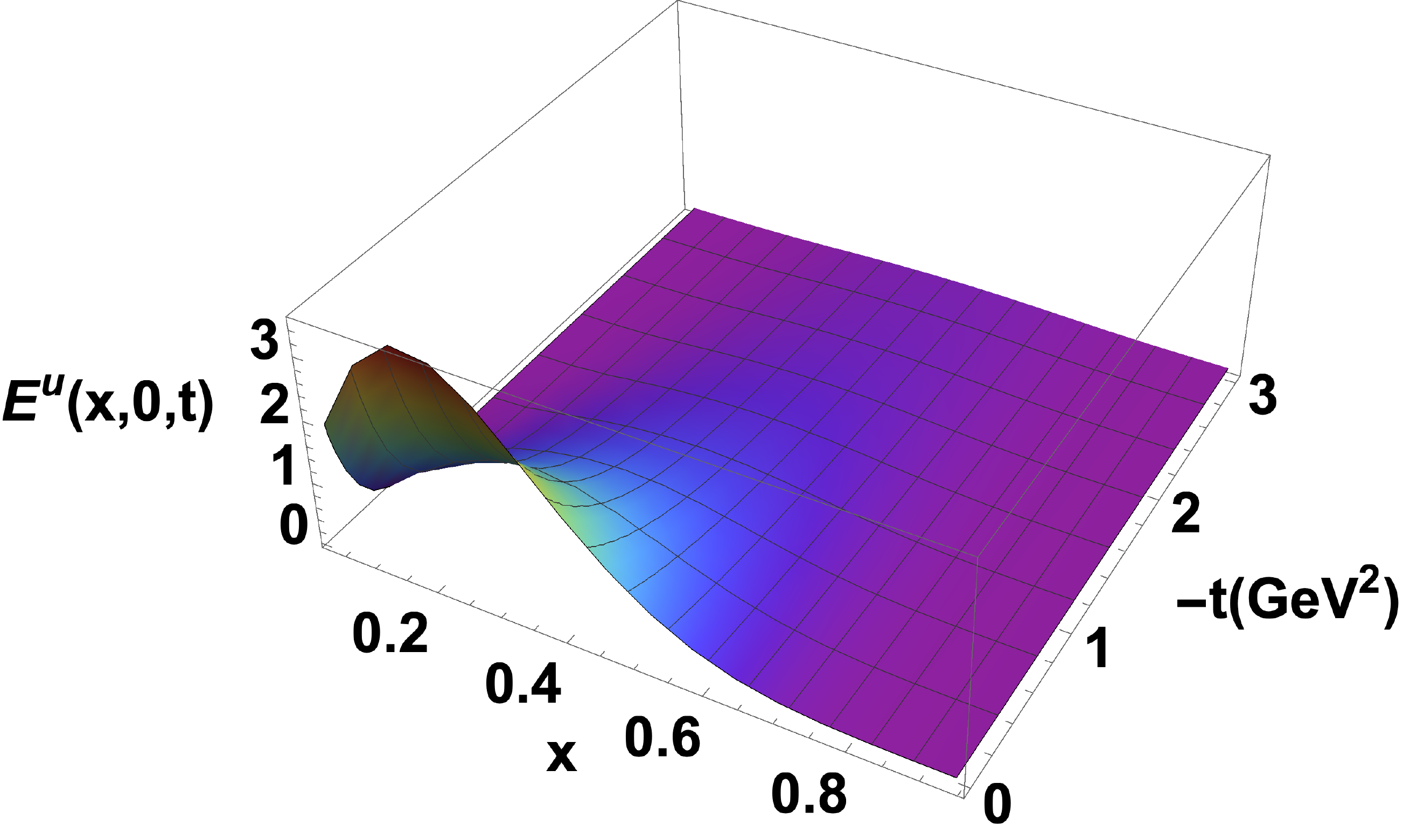}
\end{minipage}
\label{GPD_E_U}
}
\caption{The GPDs as functions of $x$ and $-t$ for the up and down quark in the proton at the model scale. Upper panels: for the down quark; lower panels: for the up quark. Note the sign change on the vertical axis of panel (b). }
\label{GPD_3D}
\end{figure*}
\begin{figure}[htbp]
\centering
\includegraphics[width=\columnwidth]{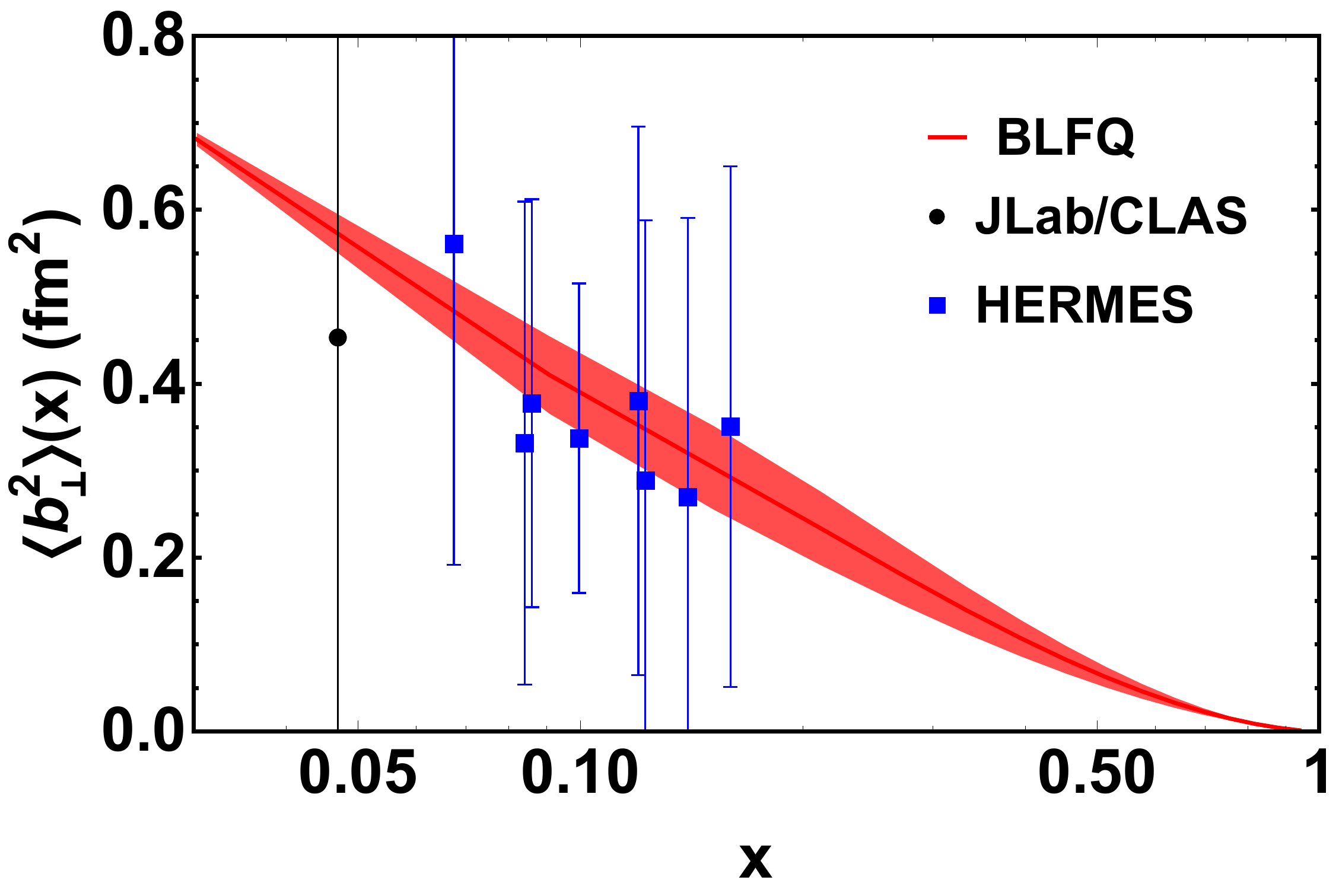}
\caption{$x$-dependence of  $\langle b^2_\perp \rangle$ for quarks in the proton. The line corresponds to the BLFQ prediction and the band indicates its uncertainty. The data points are taken from Ref.~\cite{Dupre:2016mai} }
\label{b2x}
\end{figure}
The GPDs $H^q(x,0,t)$ reduce to the unpolarized PDFs in the forward limit $(t\rightarrow 0)$ and satisfy the following normalization condition:
\begin{align}
&\int \,dx\, H^u(x,t=0)=F_1^u(0) =n_u,\nonumber\\
&\int \,dx\, H^d(x,t=0)= F_1^d(0)=n_d,
\end{align}
while the $E^q(x,0,0)$ provides the anomalous magnetic moments $(\kappa^q)$ of the quark after integrating over $x$:
\begin{align}
&\int \,dx\, E^u(x,t=0)=F_2^u(0) =\kappa_u,\nonumber\\
&\int \,dx\, E^d(x,t=0)= F_2^d(0)=\kappa_d.
\end{align}

The unpolarized GPDs, for zero skewness, as functions of $x$  and $-t$ for up and down quarks are shown in Fig.~\ref{GPD_3D}. We observe that with increasing momentum transfer $-t$, the peaks of the distributions shift towards larger $x$ and the corresponding magnitudes decrease as seen in various phenomenological models~\cite{Scopetta:2002xq,Boffi:2002yy,Ji:1997gm,Mondal:2015uha,Maji:2017ill,Vega,Vega2,CM1,deTeramond:2018ecg}. Except for the fact that the magnitude of the GPD $H$ for the up quark is larger than that for the down quark, the overall features are qualitatively similar for both quarks. Meanwhile, the GPD $E$ for both quarks has almost the same magnitude but with opposite sign.  The fall-offs of the unpolarized GPDs with increasing $x$ are similar for both the up and the down quarks.

The GPDs in transverse position space are defined by a two-dimensional Fourier transform with respect to the momentum transfer purely in the transverse direction ~\cite{Burkardt:2002hr,Diehl:2002he}:
\begin{align}
\mathcal{H}^q(x,0,b_{\perp})= \int \frac{{\rm d}^2\vec{ q}_{\perp}}{(2\pi)^2} e^{i\vec{ q}_{\perp}\cdot \vec{b}_{\perp}} H^q(x,t=-\vec{ q}_{\perp}^{~2}),
\end{align}
where $\vec{b}_{\perp}$ denotes the transverse impact parameter conjugate to the transverse momentum transfer $\vec{q}_{\perp}$ and $b_{\perp}=|\vec{b}_{\perp}|$. 
For zero skewness, $\vec{b}_{\perp}$ represents a measure of the transverse distance of the struck quark from the center of momentum of the nucleon and it follows the condition $\sum_i x_i \vec{b}_{\perp i}=0$, where the sum is over the index of quarks. The relative distance between the struck quark and the center of momentum of the spectators is given by ${ b_{\perp} /(1-x})$, which provides us with an estimate of the transverse size of the bound state \cite{diehl}.
The function $\mathcal{H}^q(x, b_{\perp})$ can be interpreted as the number density of quarks with longitudinal momentum  fraction $x$ at a given transverse distance $b_{\perp}$ 
in the nucleon~\cite{Burkardt:2000za}. We can  then
define the $x$ dependent squared radius of the quark density in the transverse plane as~\cite{Dupre:2016mai}:
\begin{align}
\langle b^2_\perp \rangle^q (x) = \frac{ \int d^2 {\vec{ b}_\perp} { b}^2_\perp \mathcal{H}^q(x, {b_{\perp}})}{\int d^2 {\vec{ b}_\perp}  \mathcal{H}^q(x, b_{\perp})},
\label{eq:cr5}
\end{align}
which can also be written through the GPD $H$ as:
\begin{equation}
\langle b^2_\perp \rangle^q (x)= - 4 \frac{\partial}{\partial \vec{ q}^{~2}_\perp} \ln H^q(x,0,-{\vec{ q}_\perp}^{~2}) \biggr| _{{\vec{ q}_\perp} = 0}.
\label{eq:crgpd}
\end{equation}

We show the $x$-dependent squared radius of the proton distributions, $\langle b^2_\perp \rangle (x) = 2 e_u \langle b^2_\perp \rangle^u (x)+e_d \langle b^2_\perp \rangle^d (x)$, in Fig.~\ref{b2x}, where we compare the BLFQ prediction with the available extracted data from the DVCS process within the range $0.05 \lesssim x \lesssim 0.2$~\cite{Dupre:2016mai}. $\langle b^2_\perp \rangle (x)$ describes the transverse
size of the nucleon and shows an increase of transverse radius with decreasing value of the quark momentum fraction $x$. As can be seen from Fig.~\ref{b2x}, our result for $\langle b^2_\perp \rangle (x)$  is found to be compatible with the extracted data. Following the reference~\cite{Dupre:2016mai}, we evaluate the proton's transverse squared radius
\begin{equation}
\braket{b_{\perp}^2} = \sum_q e_q \int_0^1 dx f^q(x) \braket{b_{\perp}^2}^q(x).
\end{equation}
In our approach, we obtain the squared radius of the proton, $\braket{b_{\perp}^2}=0.36\pm 0.04$ fm$^2$, about $20 \%$ below the experimental data~\cite{Dupre:2016mai}: $\braket{b_{\perp}^2}_{\rm exp}=0.43\pm 0.01$ fm$^2$.



\section{conclusion}\label{sec:con}
In this paper, we presented various observables of the nucleon within the theoretical framework of basis light-front quantization (BLFQ). We adopted
an effective light-front Hamiltonian incorporating confinement in both the transverse and the longitudinal direction and one gluon exchange interaction for the valence quarks
suitable for low-resolution properties. We obtained the nucleon LFWFs as the eigenvectors of this Hamiltonian by solving its mass eigenstates using BLFQ as a relativistic three-quark problem. We then employed the LFWFs to compute the nucleon electromagnetic and axial FFs, the PDFs for different quark polarizations, and the unpolarized GPDs. 

We found good agreement with the experimental data for the
electromagnetic and the axial FFs for the proton, whereas the electromagnetic FFs for the neutron were found to deviate from the data in the low $Q^2$ region. Comparing the flavor FFs with the extracted data, the up quark FFs were in reasonable agreement with the data but the down quark Pauli FF underestimated the data. 

We also computed the transverse charge and magnetization densities by employing the two dimensional Fourier transformations of the Dirac and Pauli FFs and found that the BLFQ results were reasonably consistent with the global parameterizations. We found excellent agreement with the data for the electromagnetic and axial radii of the proton as well as for the magnetic radius of the neutron, while our result for the neutron's charge radius deviates significantly from experiment in our current treatment.

We also calculated the unpolarized, helicity, and transversity PDFs at a low-resolution scale using our LFWFs. The PDFs at the higher scale relevant to global QCD analyses have been computed based on the NNLO DGLAP equations.
The unpolarized PDFs for both up and down quarks were in excellent agreement with the global fits by NNPDF, MMHT, and CTEQ Collaborations. The helicity and the transversity PDFs for the down quark also agreed well with results from the corresponding global fits or experimental data respectively. However, for the up quark, our results were too large in the region $x<0.3$, but tended to agree with the data in the large-$x$ domain. 

We further evaluated the helicity asymmetries for both quarks, which revealed reasonable agreement with the available data in relatively large $x$ domain. The small $x$ region is dominated by the sea quarks and to describe the contributions of the sea quarks in the asymmetries, the higher Fock components, such as $|qqqq\bar{q}\rangle$, should be included explicitly, which is a topic of future effort. The axial charge and the tensor charge also
showed reasonable agreement with the extracted data and with the lattice results.

We further employed our LFWFs to generate the unpolarized GPDs and found that the qualitative behavior of the GPDs in our approach was consistent with other phenomenological models. Comparing with the recently analyzed DVCS data, we also obtained a good description of the $x$-dependent squared radius of the quark density in the transverse plane. 

Our future plans include calculations of the polarized GPDs, such as $\tilde{H}(x,\zeta,t)$, $\tilde{E}(x,\zeta,t)$, $H_T (x,\zeta,t)$, and $E_T (x,\zeta,t)$, and a study of the orbital angular momentum distribution. We anticipate predicting the contribution from spin and orbital angular momentum of the quarks to the nucleon spin.

Since our LFWFs include all three active quarks’ spin, flavor, and three-dimensional spatial information on the same footing, we also plan to investigate other parton distribution functions, such as transverse momentum dependent parton distribution functions, Wigner distributions, and double parton distribution functions as well as mechanical properties of the nucleon. In other words, a detailed analysis of all the leading twist GPDs in the BLFQ approach will be reported in future studies. By establishing our approach for the nucleon in the leading Fock sector, we prepare a path for future systematic improvements that include the addition of higher Fock sectors with dynamical gluons and sea quarks. The presented results affirm the utility of our effective Hamiltonian approach within BLFQ and motivate its application to other hadrons. 

\section*{acknowledgments}
We thank Wei Zhu, Shuai Liu for many useful discussions. 
C. M. thank the Chinese Academy of Sciences Presidents International Fellowship Initiative for the support via Grants No. 2021PM0023.  X. Z. is supported by new faculty startup funding by the Institute of Modern Physics, Chinese Academy of Sciences, by Key Research Program of Frontier Sciences, Chinese Academy of Sciences, Grant No. ZDB-SLY-7020, by the Natural Science Foundation of Gansu Province, China, Grant No. 20JR10RA067 and by the Strategic Priority Research Program of the Chinese Academy of Sciences, Grant No. XDB34000000. J. P. V. is supported by the Department of Energy under Grants No. DE-FG02-87ER40371, and No. DE-SC0018223 (SciDAC4/NUCLEI). This work of  S. X., C. M., and X. Z. is also supported by the Strategic Priority Research Program of the Chinese Academy of Sciences, Grant No. XDB34000000. This research used resources of the National Energy Research Scientific Computing Center (NERSC), a U.S. Department of Energy Office of Science User Facility operated under Contract No. DE-AC02-05CH11231. A portion of the computational resources was also provided by Gansu Computing Center.

\end{document}